\documentclass[letterpaper,twocolumn,10pt]{article}
\usepackage{usenix-2020-09}

\usepackage{times,url,color,soul,xspace,enumitem}
\usepackage{xurl}
\usepackage{breakurl}
\usepackage{graphicx}
\usepackage{caption}
\usepackage{subcaption}
\usepackage{comment}
\usepackage{xspace}

\usepackage{booktabs}
\usepackage{tabularx}
\usepackage{multirow}
\usepackage{makecell}
\usepackage{threeparttable}

\usepackage{xcolor,colortbl}

\usepackage{tcolorbox}
\tcbuselibrary{skins}
\definecolor{BgBlue}{HTML}{336699}

\usepackage{tcolorbox}
\definecolor{C0}{RGB}{85,105,151}
\definecolor{C1}{RGB}{187,129,92}
\definecolor{C2}{RGB}{96,145,101}
\definecolor{C3}{RGB}{161,88,86}

\usepackage{tikz}
\usepackage{amsmath}
\usepackage{xcolor}
\usepackage[inline,draft,nomargin,index,author=]{fixme}

\usepackage[noabbrev,capitalize]{cleveref}
\crefformat{section}{Section #2#1#3}
\crefformat{subsection}{Section #2#1#3}
\crefformat{subsubsection}{Section #2#1#3}

\FXRegisterAuthor{nty}{anty}{\textcolor{orange}{Nty}}

\usepackage{todonotes}
\fxsetup{theme=color}

\usepackage{adjustbox}
\usepackage{tikz}
\usetikzlibrary{positioning,calc,fit,backgrounds}

\pgfdeclarelayer{bg0}
\pgfdeclarelayer{bg1}
\pgfdeclarelayer{bg2}
\pgfdeclarelayer{bg3}
\pgfdeclarelayer{foreground}
\pgfsetlayers{bg0,bg1,bg2,bg3,main,foreground}
\definecolor{lightblue}{HTML}{B1DDF0}
\definecolor{lightorange}{HTML}{FFE6CC}
\definecolor{lightred}{HTML}{FAD9D5}
\definecolor{lightgreen}{HTML}{B9E0A5}

\tikzset{
    title/.style={font=\fontsize{10pt}{10pt}\color{black!90}\bfseries,anchor=center,text centered,align=center},
    outerbox/.style={rectangle, draw=black!80, rounded corners,line width=.9pt, minimum size=0.5cm, minimum width=1cm,minimum height=2cm},
    box/.style={rectangle,font=\fontsize{10pt}{10pt}\color{black!90}\bfseries, draw=black!80, line width=.9, fill=white!100, line width=.5pt, minimum size=1cm, minimum width=2.5cm,minimum height=.5cm},
    dot/.style = {circle, fill, minimum size=3pt, inner sep=0pt, outer sep=0pt},
    fp/.style={rectangle,font=\fontsize{10pt}{10pt}\color{black!90}\bfseries, draw=black!80, line width=.9, fill=lightgreen, line width=.5pt, minimum size=1cm, minimum height=1cm, minimum width=1cm,align=center},
}

\newcommand{\plothdots}[1]{
    \node (dt#1) [dot] [below=.1cm and 0cm of #1] {};
    \node (dt1#1) [dot] [below=.1cm and 0cm of dt#1] {};
    \node (dt2#1) [dot] [below=.1cm and 0cm of dt1#1] {};
}

\newcommand{\plotvdots}[1]{
    \node (dt#1) [dot] [right=0cm and 0.6cm of #1] {};
    \node (dt1#1) [dot] [right=0cm and .2cm of dt#1] {};
    \node (dt2#1) [dot] [right=0cm and .2cm of dt1#1] {};
}

\newcommand{\block}[5]{
    \node (blk#1) [title] [#4] {Block #5};
    \node (pg1#1) [box,fill=gray!15] [below=.15cm and 0cm of blk#1] {Page 1};
    \node (pg2#1) [box,fill=gray!15] [below=-0.5pt and 0cm of pg1#1] {Page 2};
    \node (dt#1) [dot] [below=.1cm and 0cm of pg2#1] {};
    \node (dt1#1) [dot] [below=.1cm and 0cm of dt#1] {};
    \node (dt2#1) [dot] [below=.1cm and 0cm of dt1#1] {};
    \node (pgn#1) [box,fill=gray!15] [below=.1cm and 0cm of dt2#1] {Page $n$};
    \begin{pgfonlayer}{bg3}
        \node (#2) [outerbox,rounded corners=0,fill=lightorange] [fit=(blk#1) (pg1#1) (pg2#1) (dt#1) (dt1#1) (dt2#1) (pgn#1),#3] {};
    \end{pgfonlayer}
}

\newcommand{\plane}[2]{
    \block{0p#1}{blkbox0p#1}{}{}{0}
    \plothdots{blkbox0p#1}
    \block{mp#1}{blkboxmp#1}{below=0.1cm and 0cm of dt2blkbox0p#1}{below=0.1cm and 0cm of dt2blkbox0p#1}{$m$}
    \node (titlep#1) [title] [above=0.4cm and 0cm of blk0p#1]{Plane #1};
    \node (drp#1) [box,fill=lightblue] [below=.2cm and 0cm of blkboxmp#1] {Data Register};
    \node (crp#1) [box,fill=lightred] [below=.15cm and 0cm of drp#1] {Cache Register};
    \begin{pgfonlayer}{bg2}
        \node (p#2) [outerbox,fill=lightgreen] [fit=(titlep#1) (blkbox0p#1) (blkboxmp#1) (drp#1) (crp#1)] {};
    \end{pgfonlayer}
}

\usepackage{amssymb}
\usepackage{pifont}
\newcommand{\cmark}{\ding{51}}
\newcommand{\xmark}{\ding{55}}

\usepackage[acronym,automake,nonumberlist]{glossaries}
\makeglossaries

\setacronymstyle{long-short}

\newcommand{\as}[1]{\acrshort{#1}}

\newcommand{\af}[1]{\textit{\acrfull{#1}}} 

\newacronym{ssd}{SSD}{Solid State Drive}
\newacronym{hdd}{HDD}{Hard Disk Drive}
\newacronym{gc}{GC}{Garbage Collection}
\newacronym{ftl}{FTL}{Flash Translation Layer}
\newacronym{lba}{LBA}{Logical Block Address}
\newacronym{pba}{PBA}{Physical Block Address}
\newacronym{pbal}{PBA}{Page Boundary Alignment}
\newacronym{rps}{RPS}{Reader Pass-through Semaphore}
\newacronym{ecc}{ECC}{Error Correction Codes}
\newacronym{nvm}{NVM}{Non-Volatile Memory}
\newacronym{slc}{SLC}{Single-Level Cell}
\newacronym{mlc}{MLC}{Multi-Level Cell}
\newacronym{tlc}{TLC}{Triple-Level Cell}
\newacronym{qlc}{QLC}{Quad-Level Cell}
\newacronym{plc}{PLC}{Penta-Level Cell}
\newacronym{slr}{SLR}{Systematic Literature Review}
\newacronym{rsq}{RSQ}{Relevant Studies Query}
\newacronym{rlsq}{RLSQ}{Related Literature Studies Query}
\newacronym{scm}{SCM}{Storage Class Memory}
\newacronym{io}{I/O}{Input/Output}
\newacronym{iops}{IOPS}{\as{io} Operations per Second}
\newacronym{ram}{RAM}{Random Access Memory}
\newacronym{dram}{DRAM}{Dynamic \as{ram}}
\newacronym{sata}{SATA}{Serial Advanced Technology Attachment}
\newacronym{pcie}{PCIe}{\as{pci} Express}
\newacronym{pci}{PCI}{Peripheral Component Interconnect}
\newacronym{ahci}{AHCI}{Advanced Host Controller Interface}
\newacronym{nvme}{NVMe}{Non-Volatile Memory Express}
\newacronym{ops}{OPS}{Over-Provisioning Space}
\newacronym{wl}{WL}{Wear Leveling}
\newacronym{ocssd}{OCSSD}{Open-Channel \as{ssd}}
\newacronym{iot}{IoT}{Internet of Things}
\newacronym{scsi}{SCSI}{Small Computer System Interface}
\newacronym{sas}{SAS}{Serial Attached \as{scsi}}
\newacronym{ata}{ATA}{Advanced Technology Attachment}
\newacronym{hba}{HBA}{Host Bust Adapator}
\newacronym{sq}{SQ}{Submission Queue}
\newacronym{cq}{CQ}{Completion Queue}
\newacronym{fic}{FIC}{Flash Integration Challenge}
\newacronym{wa}{WA}{Write Amplification}
\newacronym{ra}{RA}{Read Amplification}
\newacronym{sa}{SA}{Space Amplification}
\newacronym{lfs}{LFS}{Log-Structured File System}
\newacronym{inode}{inode}{index node}
\newacronym{nat}{NAT}{Node Address Table}
\newacronym{lmu}{LMU}{Lazy Metadata Update}
\newacronym{libu}{LIBU}{Lazy Indirect Block Update}
\newacronym{sit}{SIT}{Segment Information Table}
\newacronym{ssa}{SSA}{Segment Summary Area}
\newacronym{cow}{CoW}{Copy on Write}
\newacronym{cp}{CP}{Checkpoint}
\newacronym{oob}{OOB}{Out-Of-Band}
\newacronym{lru}{LRU}{Least Recently Used}
\newacronym{cflru}{CFLRU}{Clean First Least Recently Used}
\newacronym{mru}{MRU}{Most Recently Used}
\newacronym{fab}{FAB}{Flash Aware Buffer}
\newacronym{wods}{WODS}{Write Optimized Data Structure}
\newacronym{caftl}{CAFTL}{Content Aware \as{ftl}}
\newacronym{rle}{RLE}{Run Length Encoding}
\newacronym{lz}{LZ}{Lempel-Ziv}
\newacronym{piobt}{PIO B-Tree}{Parallel I/O B-Tree}
\newacronym{l2p}{L2P}{Logical-to-Physical}
\newacronym{p2l}{P2L}{Physical-to-Logical}
\newacronym{ars}{ARS}{Adaptive Reserved Space}
\newacronym{pid}{PID}{Process Identifier}
\newacronym{mbf}{MBF}{Multiple Bloom Filters}
\newacronym{dac}{DAC}{Dynamic Data Clustering}
\newacronym{cat}{CAT}{Cost-Age-Time}
\newacronym{scj}{SCJ}{Segment Cleaning Journal}
\newacronym{qpsc}{QPSC}{Quasi Preemptive Segment Cleaning}
\newacronym{sac}{SAC}{Suspend Aware Cleaning}
\newacronym{fagc}{FaGC}{File-aware Garbage Collection}
\newacronym{uft}{UFT}{Update Frequency Table}
\newacronym{vbq}{VBQueue}{Valid Block Queue}
\newacronym{ar}{AR}{Address Remapping}
\newacronym{vba}{VBA}{Virtual Block Address}
\newacronym{v2l}{V2L}{Virtual-to-Logical}
\newacronym{v2p}{V2P}{Virtual-to-Physical}
\newacronym{rmipu}{RM-IPU}{Remap-Based In-Place-Updates}
\newacronym{fpc}{FPC}{File Access Pattern-Guided Compression}
\newacronym{sods}{SODS}{Space Optimized Data Structure}
\newacronym{peb}{PEB}{Physical Erase Block}
\newacronym{mlog}{mlog}{Minor Log}
\newacronym{rl}{RL}{Range Locking}
\newacronym{cpu}{CPU}{Central Processing Unit}
\newacronym{tlb}{TLB}{Translation Lookaside Buffer}
\newacronym{spdk}{SPDK}{Storage Performance Development Kit}
\newacronym{cfq}{CFQ}{Completely Fair Queuing}
\newacronym{fuse}{FUSE}{Filesystem in USErspace}
\newacronym{ipc}{IPC}{Inter-Process Communication}
\newacronym{fsp}{FSP}{File System as Processes}
\newacronym{vfs}{VFS}{Virtual File System}
\newacronym{amf}{AMF}{Application Managed Flash}
\newacronym{rfs}{RFS}{Refactored File System}
\newacronym{fua}{FUA}{Forced Unit Access}
\newacronym{dma}{DMA}{Direct Memory Access}
\newacronym{dnchain}{DN-Chain}{Data Node Chain}
\newacronym{zns}{ZNS}{Zoned Namespace}
\newacronym{at}{AT}{Age-Threshold}
\newacronym{cb}{CB}{Cost-Benefit}
\newacronym{sdf}{SDF}{Software-Defined Flash}
\newacronym{dbms}{DBMS}{Database Management System}
\newacronym{rqs}{RQs}{Research Questions}
\newacronym{rq}{RQ}{Research Question}
\newacronym{f2fs}{F2FS}{Flash-Friendly File System}
\newacronym{smr}{SMR}{Shingled Magnetic Recording}
\newacronym{zac}{ZAC}{Zoned Block Device ATA Command Set}
\newacronym{zbc}{ZBC}{Zoned Block Command}
\newacronym{wp}{WP}{Write Pointer}
\newacronym{mdts}{MDTS}{Maximum Data Transfer Size}
\newacronym{msf2fs}{msF2FS}{Multi-Streamed \as{f2fs}}
\newacronym{srr}{SRR}{Streamed Round-Robin}
\newacronym{spf}{SPF}{Stream Pinned Files}
\newacronym{amfs}{AMFS}{Application Managed File Streams}
\newacronym{ssab}{SSAB}{Stream Section Allocation Barrier}
\newacronym{ebpf}{eBPF}{extended Berkeley Packet Filter}
\newacronym{hpc}{HPC}{High Performance Computing}
\newacronym{vm}{VM}{Virtual Machine}
\newacronym{lbaf}{LBAF}{\as{lba} Format}
\newacronym{vfio}{VFIO}{Virtual Function \as{io}}
\newacronym{vfio-pci}{VFIO-PCI}{\as{vfio}-\as{pci}}
\newacronym{cmos}{CMOS}{Complementary Metal-Oxide Semiconductor}
\newacronym{bios}{BIOS}{Basic \as{io} System}
\newacronym{ssr}{SSR}{Slack Space Recycling}
\newacronym{wal}{WAL}{Write-Ahead Log}
\newacronym{lsmtree}{LSM-Tree}{Log-Structured Merge Tree}
\newacronym{sst}{SSTable}{Sorted String Table}
\newacronym{rm}{RM}{Research Methodology}
\newacronym{eu}{EU}{Execution Unit}
\newacronym{eus}{EUS}{Execution Unit Size}
\newacronym{zs}{ZS}{Zone Size}
\newacronym{lun}{LUN}{Logical Unit Number}
\newacronym{cdf}{CDF}{Cumulative Distribution Function}
\newacronym{liza}{LIZA}{Lifetime-based Zone Allocation}
\newacronym{lbads}{LBADS}{\as{lba} Data Size}
\newacronym{ict}{ICT}{Information and Communications Technology}
\newacronym{zslba}{ZSLBA}{Zone Start \as{lba}}
\newacronym{api}{API}{Application Programming Interface}
\newacronym{srq}{SRQ}{Survey Research Question}
\newacronym{eq}{EQ}{Evaluation Question}
\newacronym{fifo}{FIFO}{First in First out}

\begin{document}

\date{}


\title{\Large \bf A Survey on the Integration of NAND Flash Storage \\ in the Design of File Systems and the Host Storage Software Stack \\ { \normalsize Survey done: July 2022}} 
\author{
{\rm Nick Tehrany}\\
Delft University of Technology\\
n.a.tehrany@vu.nl
\and
{\rm Krijn Doekemeijer}\\
Vrije Universiteit Amsterdam\\
k.doekemeijer@vu.nl
\and
{\rm Animesh Trivedi}\\
Vrije Universiteit Amsterdam\\
a.trivedi@vu.nl
}

\maketitle

\begin{abstract}
    With the ever-increasing amount of data generate in the world, estimated to reach over 200 Zettabytes by 2025, pressure on efficient data storage systems is intensifying. The shift from \as{hdd} to flash-based \as{ssd} provides one of the most fundamental shifts in storage technology, increasing performance capabilities significantly. However, flash storage comes with different characteristics than prior \as{hdd} storage technology. Therefore, storage software was unsuitable for leveraging the capabilities of flash storage. As a result, a plethora of storage applications have been design to better integrate with flash storage and align with flash characteristics.

In this literature study we evaluate the effect the introduction of flash storage has had on the design of file systems, which providing one of the most essential mechanisms for managing persistent storage. We analyze the mechanisms for effectively managing flash storage, managing overheads of introduced design requirements, and leverage the capabilities of flash storage. Numerous methods have been adopted in file systems, however prominently revolve around similar design decisions, adhering to the flash hardware constrains, and limiting software intervention. Future design of storage software remains prominent with the constant growth in flash-based storage devices and interfaces, providing an increasing possibility to enhance flash integration in the host storage software stack.
\end{abstract}

\section{Introduction}\label{sec:introduction}
With the increasing amount of data, estimated to reach 200 Zettabytes by the year 2025~\cite{2020-Morgan-Data_Attack_Surface}, efficient storage systems are becoming imperative. A large contribution factor to increased data generation is the gain in popularity for big data~\cite{Khan2014BigDS,2019-Baig-Big_Data,2015-Abaker-Big_data_on_cloud} and cloud services~\cite{rimal2011architectural,antonopoulos2010cloud}. While there exist a plethora of different storage technologies, the most prevalent type is \af{hdd}~\cite{arpaci2018operating,2011-Deng-Disk_Drives_Future}, which are now largely being replaced by \af{ssd}~\cite{cornwell2012anatomy}. \as{hdd} is one of the cheapest forms of storage, however is limited in performance due to requiring on mechanical movement to access data on the disk. This results in high latency for random access patterns~\cite{ding2007diskseen,jiang2005dulo} and additionally increases power demand~\cite{2006-Feng-Disk_Energy,gurumurthi2003drpm}. While \as{ssd} is more expensive than \as{hdd}, it is becoming more affordable~\cite{mohan2012refresh} and provides increased performance over \as{hdd}~\cite{kasavajhala2011solid}, resulting in a growing adoption for enterprise businesses~\cite{daim2008forecasting,2008-Lee-Flash_in_enterprise_db}. 

One of the most fundamental mechanisms of storing and organizing data on \as{hdd}, \as{ssd}, and other storage technologies, is through the use of file systems, enabling the structural organization of data on persistent storage media. Building efficient and performant file systems for the evolving storage media technologies and progressing with future demands of data storage is of paramount importance. With \as{hdd} having been the prevailing storage technology for decades, file system and application design revolved around the intrinsic characteristics of these devices. In particular, aiming to limit access patterns to sequential accesses~\cite{chang2008bigtable,oneil1996log}, in an effort to minimize mechanical movement on the disk and thus optimize their performance.

The most widely adopted type of \as{ssd} is based on \textit{flash storage}, having different characteristics than traditional \as{hdd}. Performance of flash storage achieves several GB/s, with millions of \af{iops}~\cite{2022-samsung-zand,2022-intel-p}, and access latency as low as single digit $\mu$-second latency. However, flash storage has its own characteristics different from \as{hdd}. In particular, flash storage does not support in-place updates, requiring data to be erased at a larger unit in order to be written again. Additionally, the cost of erase operations is substantially higher than read and write operations~\cite{stoica2009evaluating,2022-intel-p}. In order to hide these constraints from host systems, flash \as{ssd} employs firmware, called the \af{ftl}, that exposes a sector-addressable interface. This allows \as{ssd} to be addressed in the same way as conventional \as{hdd}, requiring no changes in host software for accessing the different storage technologies.

While \as{ssd} and \as{hdd} utilize the same interfaces to be addressed, in order to exploit the increased performance benefits of flash storage, software must integrate with the characteristics of flash storage. Adapting software design to align with flash storage characteristics helps minimize \as{ftl} overheads to manage the flash storage. With the increasing adoption of flash \as{ssd} in enterprise, a plethora of applications and file systems have been proposed aiming at integrating software design with flash storage characteristics. In this survey we evaluate the changes in software, particularly in file systems, caused by integrating with flash storage characteristics, and how these changes have affected file system design. We additionally assess the future implications of evolving flash storage technologies to file system and software design. In order to evaluate the various work on flash storage implications for file system design, we devise three key \af{srq} that aim at analyzing past, current, and future trends.

\begin{itemize}[left=0.9cm]
    \item [\textbf{SRQ1.}] \textbf{What are the main challenges arising from NAND flash characteristics and its integration into file system design?} \\
        Flash storage has particular characteristics, such as sequential writing, no in-place updates, and requiring erasing of flash blocks. This \as{srq} aims at analyzing what particular challenges arise for storage software from the flash-specific constraints and resulting effects of on-device operations. Devising a list of key challenges provides the foundation based on which relevant work in this literature study is selected, and the final report is structured.
    \item[\textbf{SRQ2.}] \textbf{How has NAND flash storage influenced the design and development of file system and the storage software stack?} \\
        Using the identified challenges in \textbf{\as{srq}1}, this \as{srq} evaluates for each of the challenges, how file system design has changed to integrate with it. As file systems are commonly built on top of existing storage software layers, such as the Linux Block \as{io} layer, we include methods and mechanisms in the storage software stack particularly devised for file systems and flash storage integration. As a result, this \as{srq} evaluates how the depicted challenges are addressed throughout the various software stack layers, up to the file system.
    \item [\textbf{SRQ3.}] \textbf{How will NAND flash storage and newly introduced NAND flash-based storage devices and interfaces affect future file system design and development?} \\
        With a particular goal of this literature study being to evaluate the validity of data structures, algorithms, and mechanisms of flash, and understanding the applicability to \as{zns}, a newly arising storage technology, this research question furthermore aims at evaluating future challenges that may arise from new technology.
\end{itemize}

Furthermore, this literature study makes the following contributions:

\begin{itemize}
    \item We devise six key challenges for storage software arising from the integration of flash-based \as{ssd}, particularly focusing on leveraging its capabilities and enhancing device utilization.
    \item For each of the six devised flash integration challenges, we summarize the main methods of relevant work on dealing with and integration the particular challenge(s) into file system design.
    \item Based on the findings of this literature study, we present a discussion on the future applicability of the presented methods during this study, and evaluate the effects of newly arising flash-based \as{ssd} devices.
\end{itemize}

\section{Literate Study Research Methodology}\label{sec:lit_methodology}
Several methodologies for conducting literature studies exist such as an unguided traversal of the
literature~\cite{2018-Hegeman-Survey_Graph_Analysis} and using
\textit{snowballing}~\cite{Webster2002AnalyzingTP,2014-Wohlin-Snowballing}. However, we find that said methods lack
systematic mechanisms in their evaluation, where the search space of relevant literature with unguided is not clearly
defined and can become incomprehensibly large. Snowballing on the other hand allows limiting the search space by
evaluation going forward, studies that reference the seed paper, and backwards, studies that the seed paper references,
in the citations from a set of seed papers systematically. However, it can introduce possible bias from the seed paper
selection. Therefore, we utilize a combination of approaches and additionally apply the \af{slr} presented by Kitchenham
et al.~\cite{kitchenham2007guidelines}. The \as{slr} approach relies on three separate stages to construct a systematic
literature review, depicted in \cref{tab:SLR_process}. As we do not inclusively apply every possible stage of the
\as{slr} method, the table additionally depicts which methods we apply in this study (indicated with a \cmark), and
where relevant information can be found in this literature study.

\begin{table}[t]
    \crefformat{section}{\S#2#1#3}
    \crefformat{subsection}{\S#2#1#3}
    \centering
    \begin{tabular}{c c c}
        \hline
        & \textbf{Planning the review} & \\
        \hline
        \cmark & Identification of the need for a review & (\cref{sec:introduction}) \\
        \xmark & Commissioning a review & - \\
        \cmark & Specifying the research question(s) & (\cref{sec:introduction}) \\
        \cmark & Developing a review protocol & (\cref{sec:review_protocol}) \\
        \xmark & Evaluating the review protocol & - \\
        \hline \hline
        & \textbf{Conducting the review} & \\
        \hline
        \cmark & Identification of research & (\cref{sec:search_space_selection}) \\
        \cmark & Selection of primary studies & (\cref{sec:study_selection}) \\
        \xmark & Study quality assessment & - \\
        \cmark & Data extraction and monitoring & (\cref{sec:study_analysis}) \\
        \cmark & Data synthesis & (\cref{sec:study_analysis}) \\
        \hline \hline
        & \textbf{Reporting the review} & \\
        \hline
        \xmark & Specifying dissemination mechanism & - \\
        \cmark & Formatting the main report & - \\
        \xmark & Evaluating the report & - \\
        \hline
    \end{tabular}
    \crefformat{section}{Section #2#1#3}
    \crefformat{subsection}{Section #2#1#3}
    \caption{Outline of the Systematic Literature Review approach presented by Kitchenham et al.~\cite{kitchenham2007guidelines}, inspired by the structure of the literature study on Graph Analysis by Hegeman and Iosup~\cite{2018-Hegeman-Survey_Graph_Analysis}.}
    \label{tab:SLR_process}
\end{table}

With these three stages, clear protocol and process definitions are established prior to conducting the review
(discussed in \cref{sec:review_protocol}), limiting possible reviewer bias and additionally enhancing reproducibility.
The initial stage consists of planning the review, which includes establishing the need for this certain review and
developing the review protocol. The protocol encompasses the inclusion and exclusion criteria, as well as research
question definition, and the establishment of the review process. Based on an established protocol we conduct the
review, establish the selection of studies to evaluate, and proceed to extract and analyze the studies. Lastly, with all
collected data on selected studies, we format this survey to present the review in a comprehensible manner.

\subsection{Review Protocol}\label{sec:review_protocol}
\begin{figure}[t]
    \centering
    \begin{adjustbox}{width=0.5\textwidth}
    \begin{tikzpicture}[
            title/.style={font=\fontsize{13pt}{13pt}\color{black!90}\bfseries,anchor=center,text centered,align=center},
            innerbox/.style={rectangle,font=\fontsize{13pt}{13pt}, fill=lightblue, rounded corners, draw=black!60, line width=2pt, anchor=center, text centered, align=center,text width=2.9cm, minimum width=1.5cm, minimum height=1cm},
            box/.style={rectangle, font=\fontsize{14pt}{14pt}, fill=gray!5, draw=black!80, rounded corners,line width=1.5pt, minimum size=8cm, minimum width=4cm,minimum height=10cm},
            ]

            \node (spd) [title] {Search Space\\Selection};
            \node (sp) [innerbox,fill=lightorange] [below=0.4cm of spd.south] {\large Seed Paper Selection};
            \node (cwjs) [innerbox] [below=of sp.south] {\large Conference, Workshop, and Journal Selection};
            \node (ks) [innerbox] [below=of cwjs.south] {\large Keyword Selection};
            \node (qd) [innerbox] [below=of ks.south] {\large Query Definitions};

            \begin{pgfonlayer}{bg0}
                \node (spdbox) [box,fit=(spd) (sp) (cwjs) (ks) (qd)] {};
            \end{pgfonlayer}

            \node (sd) at (5,0) [title] {Study Selection\\};
            \node (qr) [innerbox] [below=of sd.south,right=1.8cm of sp.east] {\large Query Results Collection};
            \node (r) [innerbox] [below=of qr.south, right=1.8cm of cwjs.east] {\large Conference, Workshop, and Journal Collection};
            \node (ssc) [innerbox] [below=0.4cm of r.south, right=1.8cm of ks.east] {\large Snowballing Paper Collection};
            \node (sca) [innerbox] [below=of r.south, right=1.8cm of qd.east] {\large Selection Criteria Application};
            \begin{pgfonlayer}{bg0}
                \node (sdbox) [box,fit=(sd) (ssc) (qr) (r) (sca)] {};
            \end{pgfonlayer}

            \node (sa) at (10,0) [title] {Study Analysis\\};
            \node (de) [innerbox] [below=0.4cm of sa.south, right=1.8cm of qr.east] {\large Data Extraction};
            \node (ds) [innerbox] [below=of de.south, right=1.8cm of r.east] {\large Data Synthesis};
            \node (fr) [innerbox,fill=lightgreen] [below=of ds.south, right=1.8cm of ssc.east] {\large Formatting the Report};
            \node (dummy) [innerbox, draw=gray!5,fill=gray!5] [below=of fr.south,right=1.8cm of sca.east] {};
            \begin{pgfonlayer}{bg0}
                \node (sabox) [box,fit=(sa) (de) (ds) (fr) (dummy)] {};
            \end{pgfonlayer}

            \draw[->,line width=1.5pt] (sp.south) -- (cwjs.north);
            \draw[->,line width=1.5pt] (cwjs.south) -- (ks.north);
            \draw[->,line width=1.5pt] (ks.south) -- (qd.north);
            \draw[->,line width=1.5pt] (qd.east) -| +(1,0) |- (qr.west);
            \draw[->,line width=1.5pt] (qr.south) -- (r.north);
            \draw[->,line width=1.5pt] (r.south) -- (ssc.north);
            \draw[->,line width=1.5pt] (ssc.south) -- (sca.north);
            \draw[->,line width=1.5pt] (sca.east) -| +(1,0) |- (de.west);
            \draw[->,line width=1.5pt] (de.south) -- (ds.north);
            \draw[->,line width=1.5pt] (ds.south) -- (fr.north);
    \end{tikzpicture}
    \end{adjustbox}
    \caption{Review Protocol phases applied in this literature study.}
    \label{fig:review_protocol}
\end{figure}
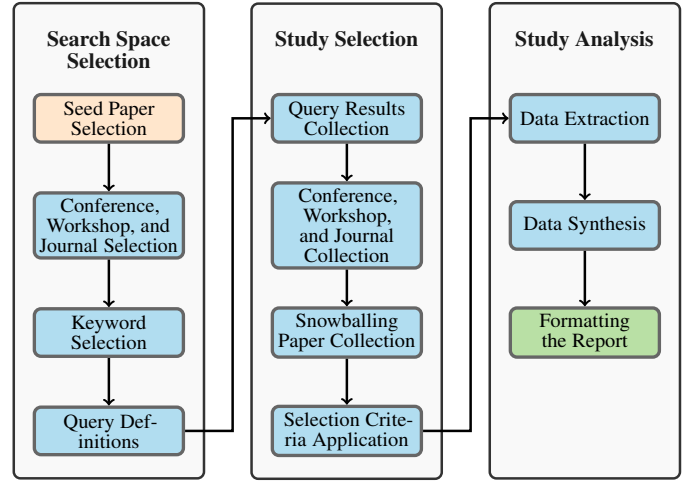

By making use of the \as{slr} process, we initially define a review protocol and review processes that we detail for this study. With emphasis on making this study reproducible, we provide detailed descriptions of each phase in the review process. A visual representation of the application of the review protocol phases is depicted in \cref{fig:review_protocol}. It revolves around three phases, firstly establishing the search space from which we extract relevant studies for this literature study. In the second phase we collect the studies and apply the defined selection criteria (explained in \cref{sec:study_selection}). Lastly, we analyze the collected relevant studies. The following sections explain each of the phases in detail, \cref{sec:search_space_selection} explains methods used for establishing the search space of literature. Next, \cref{sec:study_selection} depicts the selection criteria for extracting relevant studies from the defined search space. Lastly, \cref{sec:study_analysis} provides the methods of data extraction from studies and gives the organization of studies for this literature review.

\subsection{Search Space Selection}\label{sec:search_space_selection}
The first stage of our review protocol defines the definition of the search space from which relevant studies are extracted. We make use of several approaches for identifying relevant studies. Firstly, we apply the snowballing method on a set of seed papers. Seed papers selected for this literature review are depicted in \cref{tab:seed_papers}. With these seed papers, we analyze the studies from forward and backward citations of the seed paper. To further expand the search space of relevant studies, we examine the publications of numerous conferences, workshops, and journals which are focused on the area of systems and storage research. These conferences are analyzed in the range of 2010-2022, if present, as some are bi-annual or may not have been established in the given time range. The following venues are checked in this literature review: 

\begin{itemize}
    \item USENIX Annual Technical Conference (\textit{USENIX ATC})
    \item USENIX Conference on File and Storage Technologies (\textit{FAST})
    \item Networked Systems Design \& Implementation (\textit{NSDI})
    \item European Conference on Computer Systems (\textit{EuroSys})
    \item USENIX Symposium on Operating Systems Design and Implementation (\textit{OSDI})
    \item Symposium on Operating Systems Principles (\textit{SOSP})
    \item ACM International Systems and Storage Conference (\textit{SYSTOR})
    \item ACM Workshop on Hot Topics in Storage and File Systems (\textit{HotStorage})
    \item Architectural Support for Programming Languages and Operating Systems (\textit{ASPLOS})
    \item ACM Special Interest Group on Management of Data (\textit{SIGMOD})
    \item International Conference on Very Large Data Bases (\textit{VLDB})
    \item IEEE International System-on-Chip Conference (\textit{SOCC})
    \item International Conference on Distributed Computing Systems (\textit{ICDCS})
    \item ACM/IFIP Middleware Conference (\textit{Middleware})
    \item ACM Transactions on Storage (\textit{TOS})
    \item International Conference for High Performance Computing, Networking, Storage, and Analysis (\textit{SC})
    \item International Conference on Massive Storage Systems and Technology (\textit{MSST})
    \item IEEE International Conference on Computer Design (\textit{ICCD})
\end{itemize}

\begin{table}[!t]
    \centering
    \begin{tabular}{||c c c||}
        \hline
        Title & Venue & Publication Year \\
        \hline
        \hline
        F2FS~\cite{2015-Changman-f2fs} & FAST & 2016 \\
        JFFS~\cite{woodhouse2001jffs} & OLS & 2001 \\
        LogFS~\cite{engel2005logfs} & Linux Kongress & 2005 \\
        \hline
    \end{tabular}
    \caption{Seed papers used for this literature review. Titles are shortened.}
    \label{tab:seed_papers}
\end{table}

Lastly, we run individual queries on academic search engines for scholarly literature; Google Scholar, Semantic Scholar, and dblp. We utilize two types of queries; \af{rsq} for finding of relevant studies for this literature study, and \af{rlsq} for finding related work to this literature review. Related work encompasses surveys on file systems for flash, flash specific algorithms and data structures, and additional studies of flash related application and system integration. For each query, with each search engine, we analyze the 100 most relevant results (or less if there are fewer query results). The keyword queries for finding relevant studies and relevant related literature studies are, respectively:

\begin{itemize}[align=left,leftmargin=1.3cm]
    \item[\textbf{RSQ1.}] Flash File System
    \item[\textbf{RSQ2.}] NVM File System
    \item[\textbf{RSQ3.}] SSD File System
    \item[\textbf{RSQ4.}] File System \as{spdk}~\footnote{\af{spdk}~\cite{2017-Yang-SPDK} provides a number of tools and libraries for building high performant user-level storage software over NVMe, making it applicable to file system development on flash SSDs.}
\end{itemize}
\begin{itemize}[align=left,leftmargin=1.3cm]
    \item[\textbf{RLSQ1.}] Flash File System Survey
    \item[\textbf{RLSQ2.}] NVM File System Survey
    \item[\textbf{RLSQ3.}] SSD File System Survey
\end{itemize}

\noindent Finding of related surveys is not limited to the established queries, but additionally during the snowballing and synthesis of conference, workshop, and journal publications we identify related work based on the prior defined classifications. Inclusion of relevant studies outside of the time range from 2010-2022 is only applicable when the study is retrieved using snowballing from seed papers, or the study is present in one of the respective query results. The timing constraint is thus only applied to the extraction of relevant studies by analyzing publications at conferences, workshops, and journals.

\subsection{Study Selection Criteria}\label{sec:study_selection}
The second phase of the review protocol defines the study selection, which extracts the relevant studies from the defined search space with a set of established criteria. We define a specific Inclusion/Exclusion criteria, with which studies are selected for this literature review based on the appropriateness for the criteria. These criteria are based on the defined research questions and are aimed to narrow down the search space to a particular set of studies of interest in this review, and enforce only relevant work is included. While studies do not have to exclusively meet all the inclusion requirements to be included in this review, any if any of the exclusion criteria is present, the study is not included in this review.

\begin{itemize}
    \item[\textbf{I1.}] The work is novel.
    \item[\textbf{I2.}] The work designs a file system specifically for NAND flash storage.
    \item[\textbf{I3.}] The work adapts an existing file system to integrate with NAND flash storage.
    \item[\textbf{E1.}] The work designs or adapts a file system not specifically for NAND flash storage.
    \item[\textbf{E2.}] The work designs or adapts a hybrid/tiered file system that utilizes various storage technologies, not focusing file system design to NAND flash storage.
    \item[\textbf{E3.}] The work designs or adapts a file system which is evaluated on NAND flash storage, but not particularly built for NAND flash storage.
    \item[\textbf{E4.}] The work designs or adapts a file system for \as{ssd}, but not specifically for NAND flash storage.
\end{itemize}

While meeting inclusion requirements does not mean a study is guaranteed to be included, it is more likely to be included. In the case of exclusion criteria, exceptions are made in cases of using papers to establish background knowledge or building context, however they are not the core focus of the respective section where its content is discussed.

Furthermore, there has been a plethora of flash-based file systems which utilize hybrid/tiered storage devices, including NOR-based flash~\cite{2008-Chul-Hybrid_NOR_NAND_FS}, \af{scm}~\cite{2013-Sheng-NVMFS,2008-Youngwoo-PFFS,2010-Jaemin-Frash,2016-Mazumder-WSN_FS,2014-Li-PCM_NAND_Fusion}, byte addressable \gls{nvm}~\cite{2019-Lee-Parallel_LFS}, and methods that expose byte addressable NVRAM on SSDs with custom firmware for metadata placement~\cite{2021-Zhou-Remap_SSD,2017-Jin-Byte_Adressable_SSD}. Our focus is on block address storage, which is the most prevalent for storage. Additionally, combining of multiple storage technologies, such as \as{ssd} and \as{hdd}, commonly utilize \as{ssd} as a cache for the file system on the \as{hdd}, similarly shifting focus away from flash storage integration for the file system. For this reason we exclude such file system designs from the core study of this literature review. Similarly, file systems that are not designed specifically for flash storage, but have flash-friendly characteristics are also excluded. This mainly includes log-structured file systems that are intended for HDD, with a focus on writing sequentially to minimize arm movement on the disk, which coincidentally matches the sequential write requirement of flash.

\subsection{Study Analysis}\label{sec:study_analysis}
The last stage of the review protocol defines the extraction of data from the relevant studies, and establishing the final report. During evaluation of the different studies selected for this literature review, we disseminate the information presented based on their answering of the defined research questions. For this, we define the various key integration challenges of flash storage integration (\cref{sec:flash_challenges}), based on the hardware and software characteristics of flash storage. Using the defined integration challenges, we divide the contributions of the studies evaluated in this literature study into the respective challenge, and discuss its mechanisms for solving the particular flash integration challenge. These challenges are evaluated in \cref{sec:ssd_rw_asym,sec:gc,sec:io_amplification,sec:flash_parallelism,sec:wear_leveling,sec:io_sched,sec:failure_consistency}. Lastly, we discuss on the findings of the main findings from this literature study, followed by the relevant related literature studies for flash storage in \cref{sec:related_work}.

\subsection{Limitations}
Albeit this literature study being extensively defined and established through clear protocol definitions, there are several limitations that remain. Firstly, the search space selection is only based on studies that have scientific literature. Therefore, file systems for flash which may be in the mainline Linux kernel, are not guaranteed to be included in this survey, if no scientific literature on it exists. Secondly, the search space is limited to only scientific literature written in English. This additionally limits the inclusion of conferences, workshops, and journals to venues with proceedings in English. Thirdly, given that snowballing uses a manual selection of meeting inclusion/exclusion criteria, possible bias is inherent. Lastly, as the resulting queries on the selected literature search engines produces several hundred thousand results, and we select to evaluate the first 100 results, we rely on the sorting of results based on relevance that each search engine provides. While we utilize multiple search engines in an effort to avoid bias from each search engine, it does not completely eliminate it.

\section{Background}\label{chp:background}
The increasing adoption of flash-based \as{ssd}~\cite{daim2008forecasting,2008-Lee-Flash_in_enterprise_db}, due to its
higher performance capabilities compared to conventional \as{hdd}, and decreasing cost is making it a ubiquitous storage
media for storage systems. In this section we explain the high-level concepts of the construction of flash-based \as{ssd} (\cref{sec:bg_flash_storage,sec:building-ssd,sec:ssd-performance,sec:ssd-ftl}), technical details of the newly standardizes \as{zns} \as{ssd} (\cref{sec:zns-bg}), and how the de facto standard file system \as{f2fs} is designed for flash storage (\cref{sec:f2fs-bg}).

\subsection{Flash Storage Building Block: Flash Cells}\label{sec:bg_flash_storage}
Flash storage is based on \textit{flash cells} providing the persistent storage capabilities through programming of the cell and erasing it to clear its content. Each flash cell is constructed with a floating gate, which can hold an electrical charge, and a control gate, inside a \af{cmos} transistor~\cite{parat2015floating}. Using the electrical charge present inside  the flash cell, each cell is capable of representing a logical value (0 or 1). Such flash cells, capable of representing a single bit, are called \af{slc}. Data is written to flash cells by \textit{programming} the cell. Updating existing data in flash cells requires the cell to be \textit{erased}, followed by being programmed. In addition to \as{slc}, other types of flash cells are available, ranging from \af{mlc} (2 bits per flash cell) to \af{plc} (5 bits per cell)~\cite{2020-Ma-Flash_Memory_Bit_Errors}. To represent more bits with a single flash cell, the charge inside the flash cell is divided into a respective number of ranges in order to represent the number of bit combinations (e.g., 4 ranges to represent all combinations of 2 bits in \as{mlc}). 

Increasing the number of levels in flash cells however results in slower access time~\cite{arpaci2018operating}, increased wear on the flash cells and a lower lifetime~\cite{mohan2010learned,2016-Schroeder-Flash_Reliability}, as the cells erode over time with it being programmed. The wear on flash cells is accelerated with more bits being stored in the cell (i.e., increasing the cell level). For \as{slc} the number of program/erase cycles is typically 100K, and decreases to 10K or less for \as{mlc}~\cite{2016-Schroeder-Flash_Reliability}. As a result enterprise flash \as{ssd} most commonly employ \as{slc} for increased reliability and lifetime.

Similarly, the organization of flash cells dictates the resulting type of storage device. Flash cells can be organized
in the form of \textit{NOR} or \textit{NAND}, representing the connection architecture of the cells. With NOR flash, the
resulting flash is byte-addressable, whereas NAND flash provides a page-addressable structure, with a page representing
the unit of read and write. As a result NOR flash is commonly applied in \af{bios}~\cite{2012-Lai-NOR_BIOS}, however
NAND architecture is suitable for mass data storage, making it the primary architecture for flash \as{ssd}. Throughout
this literature study we focus purely on NAND flash, however for more information on flash cell architecture consult reports such as~\cite{bez2003introduction} and~\cite{compagnoni2017reviewing}.

\subsection{Building Mass Storage SSD with Flash Cells}\label{sec:building-ssd}
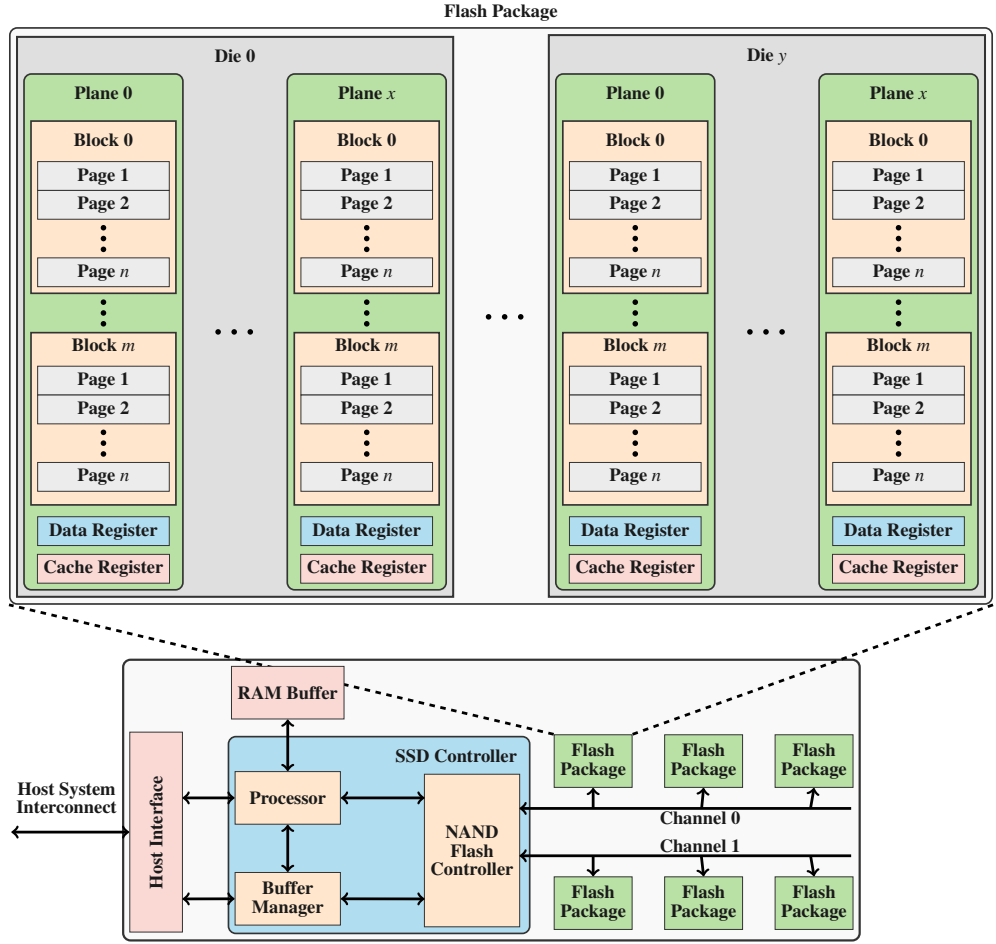
\begin{figure*}[t]
    \begin{center}
        \begin{adjustbox}{scale=0.7}
            \begin{tikzpicture}[]
                \plane{0}{0}
                \plotvdots{p0}
                \begin{scope}[shift={(5,0)}]
                    \plane{$x$}{x}
                \end{scope}
                \node (tdie0) [title] at (2.5cm,1.6cm) {Die 0};
                \begin{pgfonlayer}{bg1}
                    \node (die0) [outerbox,rounded corners=0,fill=gray!25,line width=1.1] [fit=(tdie0) (p0) (px)] {};
                \end{pgfonlayer}
                \plotvdots{die0}

                \begin{scope}[shift={(10.1,0)}]
                    \plane{0}{1}
                    \plotvdots{p1}
                    \begin{scope}[shift={(5,0)}]
                        \plane{$x$}{1x}
                    \end{scope}
                    \node (tdiex) [title] at (2.5cm,1.6cm) {Die $y$};
                    \begin{pgfonlayer}{bg1}
                        \node (diex) [outerbox,rounded corners=0,fill=gray!25,line width=1.1] [fit=(tdiex) (p1) (p1x)] {};
                    \end{pgfonlayer}
                \end{scope}
                    \begin{pgfonlayer}{bg0}
                        \node (fp) [outerbox,fill=gray!5] [fit=(die0) (diex)] {};
                    \end{pgfonlayer}
                \node (tfp) [title] [above=0cm and 1cm of fp] {Flash Package};

                \node (ssdct) [title] at (6.7,-11.7) {SSD Controller};
                \node (hi) [box,rotate=90,minimum height=1cm,minimum width=3.8cm,fill=lightred] at (1,-13.15) {Host Interface};
                \node (ram) [box,minimum width=2cm, minimum height=1cm,fill=lightred] at (3.5,-10.5) {RAM Buffer};
                \node (proc) [box,minimum width=2cm, minimum height=1cm,fill=lightorange] at (3.5,-12.5) {Processor};
                \node (bm) [box,minimum width=2cm, minimum height=1cm,align=center,fill=lightorange] [below=0.9cm and 0cm of proc] {Buffer\\Manager};
                \node (nfc) [box,minimum height=2.9cm,minimum width=1cm,align=center,fill=lightorange] at (7,-13.5) {NAND\\Flash\\Controller};
                \node (fp1) [fp] at (9.3,-11.8) {Flash\\Package};
                \node (fp2) [fp] [right=0cm and 0.6cm of fp1] {Flash\\Package};
                \node (fp3) [fp] [right=0cm and 0.6cm of fp2] {Flash\\Package};
                \node (fp4) [fp] at (9.3,-14.5) {Flash\\Package};
                \node (fp5) [fp] [right=0cm and 0.6cm of fp4] {Flash\\Package};
                \node (fp6) [fp] [right=0cm and 0.6cm of fp5] {Flash\\Package};

                \begin{pgfonlayer}{bg3}
                    \node (ssdc) [outerbox,rounded corners,fill=lightblue] [fit=(proc) (bm) (nfc) (ssdct)] {};
                \end{pgfonlayer}
                \begin{pgfonlayer}{bg2}
                    \node (ssdb) [outerbox,rounded corners,fill=gray!5,line width=1.2pt] [fit=(ram) (proc) (fp1) (fp2) (fp3) (fp4) (fp5) (fp6) (bm) (nfc) (hi) (ssdc)] {};
                \end{pgfonlayer}
                \node (ch1) [title] at (11.33,-12.87) {Channel 0};
                \node (ch2) [title] at (11.33,-13.4) {Channel 1};
                \node (hic) [title,align=center] at (-0.7,-12.5) {Host System\\Interconnect};

                \draw[<->,line width=1.5pt] (proc.north) -- (ram.south);
                \draw[<->,line width=1.5pt] (proc.south) -- (bm.north);
                \draw[<->,line width=1.5pt] (proc.east) -- (6.1,-12.5);
                \draw[<->,line width=1.5pt] (bm.east) -- (6.1,-14.42);
                \draw[<->,line width=1.5pt] (proc.west) -- (1.5,-12.5);
                \draw[<->,line width=1.5pt] (bm.west) -- (1.5,-14.42);
                \draw[<-,line width=1.5pt] (7.9,-13.6) -- (14.2,-13.6);
                \draw[<-,line width=1.5pt] (7.9,-12.7) -- (14.2,-12.7);
                \draw[->,line width=1.5pt] (9.3,-12.7) -- (fp1.south);
                \draw[->,line width=1.5pt] (11.36,-12.7) -- (fp2.south);
                \draw[->,line width=1.5pt] (13.428,-12.7) -- (fp3.south);
                \draw[->,line width=1.5pt] (9.3,-13.6) -- (fp4.north);
                \draw[->,line width=1.5pt] (11.3542,-13.6) -- (fp5.north);
                \draw[->,line width=1.5pt] (13.4295,-13.6) -- (fp6.north);
                \draw[<->,line width=1.5pt] (hi.north) -- (-1.75,-13.15);
                \begin{pgfonlayer}{bg2}
                    \draw[-,line width=1.5pt,dashed] (fp1.north east) -- (fp.south east);
                    \draw[-,line width=1.5pt,dashed] (fp1.north west) -- (fp.south west);
                \end{pgfonlayer}

            \end{tikzpicture}
        \end{adjustbox}
        \caption{Internal architecture of an SSD, with $n$ pages in a block, $m$ blocks in a plane, $x$ planes in a die, and $y$ dies in a flash package. The example SSD has two channels however any architecture with various numbers of channels is possible. Adapted from~\cite{2008-Agrawal-Design-Tradeoff-SSD,goossaertcoding}.}
        \label{fig:ssd_architecture}
    \end{center}
\end{figure*}

Utilizing the building block of flash cells, building mass storage devices is achieved by grouping together several flash cells into \textit{pages} (typically 8-16KiB in size), depicting the unit of access (read/write). A flash page additional has a small extra space, called the \af{oob} area, which is utilized for \af{ecc} and small metadata. As the operation for erasing of a flash page is costly~\cite{2009-Gupta-DFTL}, multiple pages are grouped into a \textit{block} (several MiB in size), which is the unit of erase. Increasing the erase unit from individual pages to a block helps amortize the erase overheads. Pages within a block have to be written sequentially and prior to a page being overwritten, the entire block must be erased. A number of blocks are then grouped into a \textit{plane}, of which multiple planes are combined into a \textit{die}~\footnote{\textit{Dies} are also referred to as \textit{chips}, we use the terms interchangeably and mean the same entity.}. Lastly, numerous chips are combined into a single flash package. \cref{fig:ssd_architecture} shows a visual representation of a flash-based \as{ssd} architecture.

In order to build full storage devices, multiple flash packages are packed together into a \as{ssd}. While a \as{ssd} can be constructed with any solid state technology such as Optane~\cite{2019-Wu-Unwritten_Contract_Optane,2020-Yang-Optane_3DXpoint}, we focus purely on NAND flash based \as{ssd}. The \as{ssd} contains a controller, which is responsible for processing requests, managing data buffers, and contains the NAND flash controller that manages the flash packages. An additional \af{ram} buffer, most commonly in the form of \af{dram}, is present on the device for maintaining address mappings. Lastly, a \as{ssd} contains the host interface that provides the means to connect the storage device to the host system over connection interfaces such as \af{sata} and \af{pcie}, and defining standards such as \af{ahci} and \af{nvme}~\cite{landsman2013ahci}. \as{nvme} is the interface specification particularly designed for fast \as{ssd} devices, capable of outperforming legacy protocols such as \as{sata} with 8x higher performance~\cite{xu2015performance}, due to its increased number of \as{io} queues to which requests can be issued in parallel. Similarly, \as{pcie} protocol achieves the highest throughput and lowest latency compared to \as{sata}~\cite{2013-Eshghi-SSD}. As a result, flash \as{ssd} are commonly connected to systems through \as{nvme} over \as{pci}.

\subsection{Increasing Flash SSD Performance}\label{sec:ssd-performance}
Given the architecture of flash \as{ssd}, the it contains a large degree of possible parallelism (i.e., multiple channels, flash packages, dies, and planes). Furthermore, based on a study on a Samsung \as{ssd}, the bandwidth of NAND flash \af{io} buses, connecting the flash to the flash controller, is limited to 32MB/s because of physical restrictions, and 40MB/s with interleaving in dies~\cite{2008-Agrawal-Design-Tradeoff-SSD}. A write operation to flash storage firstly writes the data to a data register, from which the data in the register is then programmed into the flash page. Because the programming operation takes longer than loading of data, these operations can be interleaved, such that while a page is being programmed, data of another write operation is loaded~\cite{kim2011parameter}. This avoids stalling whenever a page programming finishes and data needs to be loaded again, which is fundamentally similar to the concept of CPU pipelining~\cite{tanenbaum2016structured-pipelinining}. 

In order to increase performance, parallelism is utilized on the different flash entities in the \as{ssd}~\cite{chen2011essential}. Depending on the hardware configurations, the different types of parallelism are referred to as \textit{channel-level parallelism}, where requests are distributed individually across the flash channels, \textit{package-level parallelism} where flash packages on the same channel are accessed in parallel, \textit{chip-level parallelism} where flash packages with multiple chips access these in parallel, and \textit{plane-level parallelism} where the same operation can be run on multiple planes, on the same chip, in parallel. A commonly applied technique for enhanced parallelism is the utilization of \textit{clustered blocks}, where requests are issued to blocks in parallel to different chips and planes.

There are several additional performance optimizations for accessing flash storage. Other than accessing a flash entity in parallel and operation interleaving, flash packages can be accessed synchronously with a single request queue through \textit{ganging}~\cite{2008-Agrawal-Design-Tradeoff-SSD}. Flash packages within a gang share the same control bus from the flash controller, however can utilize different data buses, thus allowing a single command request queue with multiple data buses to provide the resulting data. Therefore, requests that require data from multiple pages can be split among a gang from a single request queue, and provide the data on different buses, avoiding the bottleneck of a single data bus. 

\subsection{Hiding Flash Management Idiosyncrasies on the SSD}\label{sec:ssd-ftl}
With the flash characteristic requiring sequential writing within blocks, and lacking support for in-place updates, flash \as{ssd} employs a \af{ftl} to hide these management idiosyncrasies, and provide seemingly in-place updates. As a result of the sequential write constraint, if data inside a flash page is updated, the page is simply marked as invalid and the new data is appended to a new a flash page, possibly in the same block. The \as{ftl} is responsible for managing flash page information on their validity and its mappings to \af{lba}, which is the mechanism for storage software to address the storage device. 

 Different implementations of a \as{ftl} can utilize different mapping levels, such as block- and page-level mappings. A
 naive \as{ftl} design is to maintain a fully associative mapping of each \af{lba} to every possible
 \af{pba}~\cite{2009-Gupta-DFTL}, referred to as \af{l2p} mapping, and inversely \af{p2l} mapping. These mappings are
 maintained in a \textit{mapping table}, which is kept in the \as{ram} of the flash device for faster accesses. For
 consistency the mapping table is also maintained in the persistent flash storage, where on startup time of the devices
 the mapping table is reconstructed in the device
 \as{ram}~\cite{chen2009understanding,2008-Agrawal-Design-Tradeoff-SSD}. With this, the \as{ssd} can provide seemingly
 in-place updates by simply writing data to a free page and invalidating the overwritten data. Further details on
 \as{ftl} mappings are not required for the remainder of this literature study, however for a detailed explanation of \as{ftl} mapping algorithms consult \cite{chung2009FTL_survey} and \cite{2009-Gupta-DFTL}.

As over time an increasing number of flash pages become invalid due to data updates, blocks contain valid and invalid data, requiring the \as{ftl} to run \af{gc}. During \as{gc}, the \as{ftl} selects a block from which it reads the still valid flash pages, writes these to an empty block, followed by erasing of the original block. The process of \as{gc} requires the device to provide an empty space that cannot be used as storage, but is only used for the \as{ftl} to move valid pages, referred to as the \af{ops}. If a device has fully utilized all its capacity, the \as{ftl} must be able to write out valid pages, which the overprovisioning space serves at. Therefore, \as{ssd} commonly ship with an overprovisioning space of 10-28\%, which is only usable by the \as{ftl}. In addition to \as{gc}, the \as{ftl} is responsible to ensure even wear across flash cells, as a flash cell has a limited number of program/erase cycles. This process is referred to as \af{wl}, where the \as{ftl} ensures the flash wears out evenly across the entire device, and no particular parts are burnt out faster than others.

\subsection{Zoned Namespace SSD}\label{sec:zns-bg}
While the \as{ftl} provides the seamless integration of flash \as{ssd} into storage systems, details about the flash
characteristics, such as the garbage collection unit, are commonly hidden from users, making reverse engineering of such information increasingly complex~\cite{2023-cluster-zns-study,2022-systor-ssd-study}. Furthermore, its unpredictable
performance as a result of \as{gc}~\cite{2015-Kim-SLO_complying_ssds,2014-yang-dont_stack_log_on_log} has shifted the
research community to new flash interfaces that expose flash management to the host system. Such interfaces allow for
better coordination between the storage device and the host software by decreasing \as{ftl} responsibility, and
increasing control for host software of the flash storage. Efforts for open flash \as{ssd} interfaces include
\af{sdf}~\cite{2014-Ouyang-SDF} and \af{ocssd}~\cite{bjorling2017lightnvm,picoli2020open}, which however failed to gain
large scale adoption due to the lack of standardization, resulting in device-specific implementations, and complex
interfaces, requiring software developers to have extensive knowledge of flash in order to be able to build
flash-specific software. 

The newest addition to opening flash \as{ssd} interfaces comes with arrival of \af{zns} \as{ssd}. The 2.0 base specification of \as{nvme}~\cite{2022-nvme-spec} (published in June 2021), establishes the standardization of \as{zns} with the concept of splitting the address space of the storage device into a number of \textit{zones}, which are independently addressable. This concept of representing the storage space with zones has previously been introduced with the addition of \as{smr} \as{hdd}s~\cite{Feldman2013ShingledMR,Gibson2011PrinciplesOO,Suresh2012ShingledMR}. These are a particular type of \as{hdd} that increase the storage density. Its concept of zones was included in the Linux kernel through the \af{zac} and \af{zbc} specifications~\cite{2014-ZBC,2015-ZAC}. Similar to \as{smr}, with \as{zns} zones have to be written sequentially, and must be reset prior to being overwritten, matching the internal characteristics of flash. 

\begin{figure}[!t]
    \center
    \includegraphics[width=\linewidth]{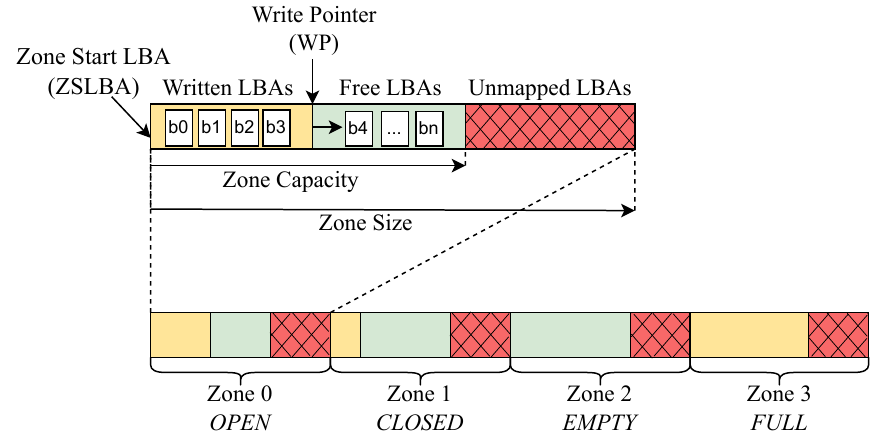}
    \caption{Layout of a ZNS SSD, depicting zone capacity, write pointer, and zone states associated to each zone.
    Adapted from~\cite{2022-Tehrany-Understanding_ZNS}.}
    \label{fig:ZNS_HW}
\end{figure}

\textbf{\as{zns} Interface.} \cref{fig:ZNS_HW} depicts a simplified layout of zones on a \as{zns} device, illustrating the management of the sequential write requirement within zones with a \af{wp} for each zone. The \as{wp} indicates the next \af{lba} to be written in the particular zone. The starting \as{lba} of a zone is represented by a \af{zslba}, identifying the first \as{lba} of each zone. For zone management, each zone has an associated state, such as \textit{FULL} and \textit{EMPTY}. 

In addition to the zone management, the \as{nvme} standardization of \as{zns} introduces three new concepts. Firstly, the specification details the zone capacity, which limits the addressable region within a zone. In order to integrate into the Linux kernel, the zone size must be a power of two value, since kernel operations rely on bit shifts for addressing. However, the addressable space in a zone may not be a power of two value, and can therefore be less than or equal to the zone size. Any \as{lba} beyond the zone capacity is not addressable. 

Secondly, the specification adds a limit on the number of concurrently active zones. Zone states indicate the state of the zone, where zones that are currently in an \textit{OPEN} or \textit{CLOSED} state are considered to be active. As the device must allocate resources for active zones, such as write buffers, it enforces a limit on concurrently active resources. Lastly, \as{zns} introduces a new \textit{zone-append} command, which instead of requiring the host to manage I/Os, such that they adhere to the sequential write constraint within a zone, allows the host to issue write I/Os to a zone without specifying the \as{lba}. The device handles the write and returns the address at which the data is written. 

The \textit{zone append} command is particularly beneficial with large queue depths (submitting numerous asynchronous
write \as{io} requests), which is not possible with write commands, as these must be issued at consecutive addresses and
writes can be reordered in the block layer of the Linux kernel or inside the storage device. In order to adhere to the
write constraint in a zone without relying on the zone append command, the \texttt{mq-deadline} scheduler within the
Linux kernel must be enabled. The \texttt{mq-deadline} scheduler holds back \as{io}s and only submits a single \as{io}
at a time to the \as{zns} device. Furthermore, this allows to merge \as{io} requests in the scheduler, enhancing
performance by issuing a smaller number of larger \as{io} requests. Evaluations on the performance of the different
schedulers show the benefits of merging \as{io} requests with larger \as{io}s of $\ge16$KiB being required to saturate
the device bandwidth~\cite{2022-Tehrany-Understanding_ZNS,2020-nvmsa-zns-implications}. 

\subsection{F2FS: Flash-Friendly File System}\label{sec:f2fs-bg}
A ubiquitous approach of managing persistent storage devices is with file systems, providing the familiar file and
directory interface for organizing storage. The lack of in-place updates on flash pages, enforcing sequential writes,
makes \af{lfs}~\cite{1992-Rosenblum-LFS,2006-Konishi-Linux_LFS,seltzer1993implementation} a suitable file system design.
\as{lfs} revolves around writing data as a log, appending new data sequentially on the storage device. In this section,
we describe the de facto standard \as{lfs} for flash-based storage devices, \af{f2fs}~\cite{2015-Changman-f2fs}, a
plethora of file systems base their design on the foundations presented by \as{f2fs}. 

\subsubsection{F2FS Data Layout}
\begin{figure*}[!t] 
    \centering
    \includegraphics[width=\linewidth]{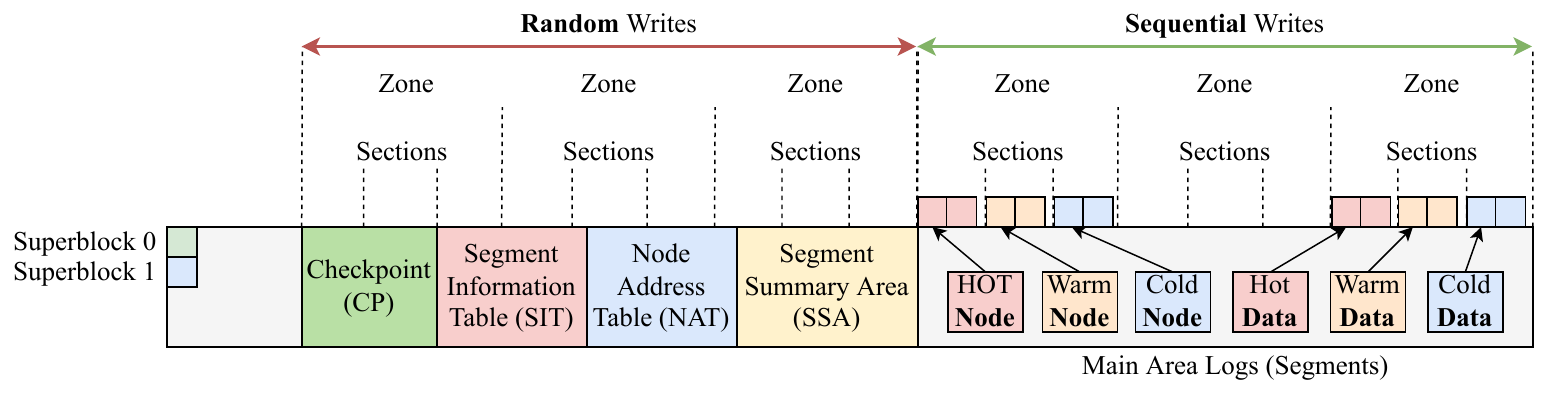}
    \caption{On-device layout of \as{f2fs} data. Adapted from F2FS~\cite{2015-Changman-f2fs}.}
    \label{fig:f2fs-layout}
\end{figure*}

Internally, \as{f2fs} utilizes a data allocation unit referred to as a \textit{block}, of 4KiB, in which blocks are allocated on the log. Consecutive blocks are collected into a 2MiB \textit{segment}, of which one or multiple segments are further grouped into a \textit{section}, that are combined into a \textit{zone}. \cref{fig:f2fs-layout} shows the layout of segments, sections, and zones for the data logs (on the right half of the figure). The left half of figure \cref{fig:f2fs-layout} shows the metadata structures in \as{f2fs} to manage the file system, consisting of a \af{cp}, \af{sit}, \af{nat}, and \af{ssa}. We now explain each of these data structures.

\textbf{Checkpoint.} \as{f2fs} utilizes \textit{checkpointing}, in which all essential metadata for the file system is stored to provide recovery in the case of system failure. A checkpoint is periodically generated, or explicitly triggered, and persists all metadata information from memory. In the case of a system crash or power loss, the file system can recover the state from the latest checkpoint, referred to as \textit{roll-back recovery}, since the latest changes which are not in the checkpoint are reverted. In order to recover the latest changes, the host must call \textit{fsync()} to ensure that metadata and data are flushed from memory to the device. This recovery is referred \textit{roll-forward recovery} since it recovers the state past the latest checkpoint. F2FS can only guarantee roll-forward recovery with \textit{fsync()}.

\textbf{Segment Information Table.} With the data allocation in \as{f2fs} being organized in 2MiB segments on the log, the \as{sit} maintains information on each of the segments. It maintains bitmaps for each segment to indicate valid and invalid blocks (blocks that have been overwritten).

\begin{figure}[!t] 
    \centering
    \includegraphics[width=0.5\textwidth]{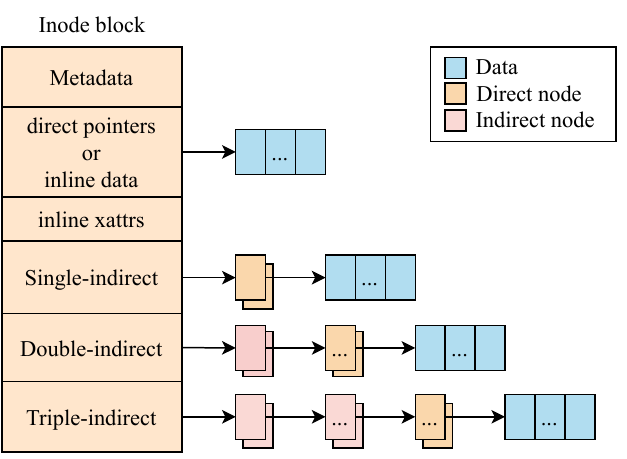}
    \caption{The \as{inode} structure of \as{f2fs}. Retrieved from F2FS~\cite{2015-Changman-f2fs}.}
    \label{fig:f2fs-inode}
\end{figure}

\textbf{Node Address Table.} Similar to other file systems, a file in \as{f2fs} is managed through an \af{inode}, and contains all the file information, including the file specific metadata on creation time, access permissions, file name, and more. \cref{fig:f2fs-inode} shows the inode of F2FS. For identifying the data blocks of the associated file, \as{inode}s contain a fixed number of \textit{pointers} to the addresses of the file data. Since an \as{inode} is allocated in a block (4KiB), they can often contain inline data of the file, if the file data fits in the available space of the \as{inode}. If data for a file is updated, a new block is allocated on the log for the new data, followed by an update to the \as{inode}, in order to modify the pointer to the data to point to the newly written data. However, this allocates a new block for the \as{inode} of the file, requiring the metadata to track the inode locations to similarly be updated. These changes continue propagating to the parent nodes, resulting in a high increase of required metadata updates. Therefore, \as{f2fs} utilizes the \as{nat}, to maintain an identifier of each node and its corresponding block address. Upon an update of a node, only the block address in the \as{nat} is modified to depict the new block address of the node. Finding a node address then checks the \as{nat} entry for the respective node identifier.


\textbf{Segment Summary Area.} In addition to the segment information in the \as{sit}, \as{f2fs} tracks information such as the owner of segments in the \as{ssa}. Furthermore, the \as{ssa} provides a cache for frequently accessed \as{nat} and \as{sit} information. 

\textbf{Main Area Logs.} The right half of \cref{fig:f2fs-layout} shows the layout of the main area logs, to which data and \as{inode}s are written. \as{f2fs} utilizes multiple concurrently writable logs to enhance data grouping and performance, with 3 logs to which nodes are written, and 3 logs for data. An essential mechanism for limiting the required \as{gc}, which \as{f2fs} must run to free space on the logs, comes from efficient \textit{data grouping}. With data grouping, data that has a similar lifetime is grouped together. Given that data has a similar lifetime, it is likely to be updated within close proximity. Therefore, when \as{gc} is run, fewer valid blocks are present, as the data with the same lifetime has likely been updated, reducing the amount of valid data that must be moved by the \as{gc} process. 

To support data grouping \as{f2fs} utilizes the three types of lifetime classes (hot/warm/cold), which are separated into the three different logs for node and data. The lifetime of data can be explicitly set for each file by an application through the passing of a \textit{lifetime hint} with the \texttt{fcntl()} function. The Linux kernel provides a total of 5 different lifetime hints, which \as{f2fs} reduces to the three lifetime hints it utilizes. If a lifetime hint for a file is not set, \as{f2fs} either assigns the default warm lifetime classification, or assigns a lifetime classification based on the file type. With the extension of a file (e.g., \texttt{.txt}, \texttt{.pdf}), \as{f2fs} identifies which files are likely to be updated in the future. Multi-media files (e.g., \texttt{.mp4}, \texttt{.gif}, \texttt{.png}) are less likely to be updated and are directly classified as cold data.

\subsubsection{F2FS Garbage Collection}\label{sec:f2fs-gc-bg}
As \as{f2fs} is log-structured, over time it contains valid and invalid blocks, similar to the \as{ftl}, and must therefore also run \as{gc}. In \as{f2fs} the process of \as{gc} is done at the unit of a section, where valid blocks in all the segments of the section are read and written to a free space, prior to all the segments in the section being freed. In \as{f2fs} \as{gc} is referred to as \textit{cleaning}, and is run periodically (called \textit{background cleaning}) or when free space for writing is needed (called \textit{foreground cleaning}). The foreground cleaning utilizes a \textit{greedy} approach for finding the section to garbage collect, which results in the largest amount of space being freed by erasing the section. Background cleaning on the other hand utilizes a \textit{cost-benefit} method that considers the required data blocks to be moved during the \as{gc} of a section, and the resulting free space that is generated by the erasing of the section.

Given that during \as{f2fs} \as{gc} block addresses are modified, as the cleaning moves still valid blocks to free space, the metadata is not directly updated to depict these changes, in order to provide recovery. Therefore, \as{f2fs} is required to create a checkpoint after each \as{gc} call. Similarly, discard commands, issued after \as{gc} calls to delete data from the flash \as{ssd}, can only be issued after a checkpoint, such that in the case of a necessary recovery, the prior checkpoint still points to existing data that has not been discarded. Once a new checkpoint has been written a discard command can be issued.

\subsubsection{Aligning F2FS Data Layout with the FTL}
\begin{figure*}[!t]
    \centering
    \subfloat[]{\includegraphics[width=0.4\textwidth]{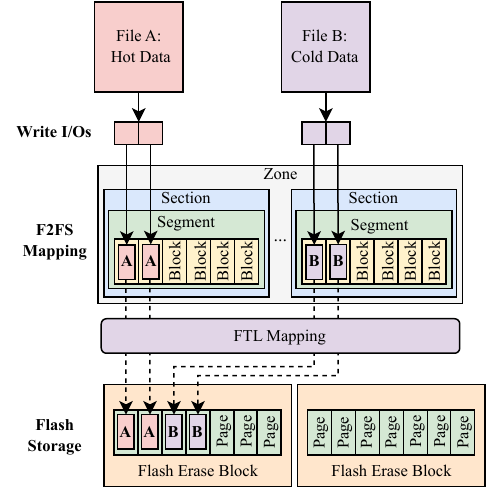}\label{fig:f2fs-zone-blind-allocation}}\hfill
    \subfloat[]{\includegraphics[width=0.59\textwidth]{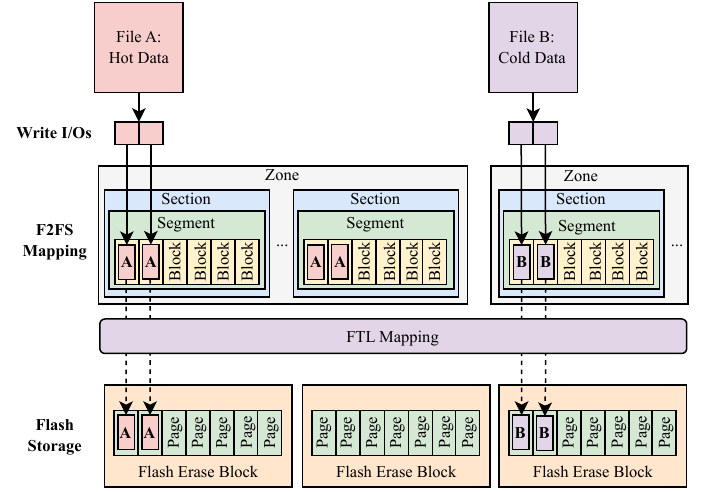}\label{fig:f2fs-zone-aware-allocation}}
    \caption{(a) \textit{zone-blind allocation} suffering from inadequate physical data separation, (b) \textit{zone-aware allocation} utilizing larger zones to ensure physical data separation.}
    \label{fig:f2fs-erase-unit-align}
\end{figure*}

To ensure the data grouping of \as{f2fs} is similarly depicted by the mapping of data to flash pages in the \as{ftl}, \as{f2fs} utilizes the separation provided by sections and zones. The goal of sections is to align the \as{f2fs} allocation with the \as{gc} unit of the underlying \as{ftl}. Therefore, the \as{f2fs} \as{gc} occurring at the unit of a section, matches the \as{ftl} \as{gc}. Zones are utilized to avoid sections in different zones to be mapped into the same on-device erase unit by the \as{ftl}. With a mapping that results in different sections of different lifetimes (e.g., hot and cold data) being in written to the same erase unit on the flash \as{ssd}, as is illustrated in \cref{fig:f2fs-zone-blind-allocation}, data grouping is violated, furthermore resulting in possible \as{gc} overheads if only the hot data is updated while the cold data remains valid. This allocation of inadequately separating sections is referred to as \textit{zone-blind allocation}. 

With \textit{zone-aware allocation}, illustrated in \cref{fig:f2fs-zone-aware-allocation}, the zone serves the purpose to provide a large enough separation between particular sections, such that the \as{ftl} similarly separates the written flash pages for the different file data into different erase units. As a result, only data of similar lifetime is within the same erase unit, resulting in reduction of \as{gc} overheads. As the erase unit of flash \as{ssd} is commonly hidden from users, the default \as{f2fs} configuration utilizes a single segment in each section, and a single section in a zone. However, if the flash storage device characteristics are known, these values can be configured.

\subsection{Summary}
The foundational building block of flash \as{ssd} is based on the \textit{flash cell}. However, the characteristics of flash cells, and their utilization to construct mass storage flash \as{ssd} introduce flash management idiosyncrasies, such as having to erase flash prior to updating the data. Due to the resulting lack of in-place updates for flash storage, flash \as{ssd} employ firmware called the \as{ftl}, that hides the complexity of managing the flash, which however introduces garbage collection overheads. File system design, particularly \as{f2fs}, has therefore focused on utilizing flash-friendly data structures and mechanisms to integrate with flash storage. The hiding of flash management idiosyncrasies, specifically the process of \as{gc}, however results in unpredictable performance and high tail latency~\cite{2013-Dean-tail_at_scale,2015-Kim-SLO_complying_ssds,2014-yang-dont_stack_log_on_log}. Therefore, interfaces that expose flash storage characteristics are appearing, with \as{zns} being the first standardized effort. The \textit{zone} interface of \as{zns} eliminates the flash \as{ssd} \as{gc}, moving the responsibility of \as{gc} to the storage software on the host (e.g., the file system).

\section{Flash Storage Integrations}
As there are various methods for integrating flash into storage devices, in addition building full \as{ssd} devices, ranging from directly attaching the flash chip to the motherboard, as is common with embedded mobile and \as{iot} devices, or custom integrations of flash chips, we evaluate the file systems based on their level of integration. Given that a different integration exposes a different interface, the possibility to enhance particular operations is highly dependent on the integration.

\begin{figure*}[!t]
    \centering
    \subfloat[]{\includegraphics[width=0.3\textwidth]{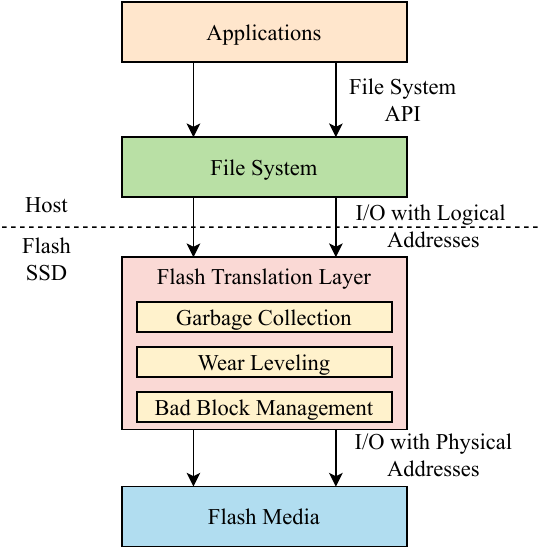}\label{fig:flash_integration_ssd}}\hfill
    \subfloat[]{\includegraphics[width=0.3\textwidth]{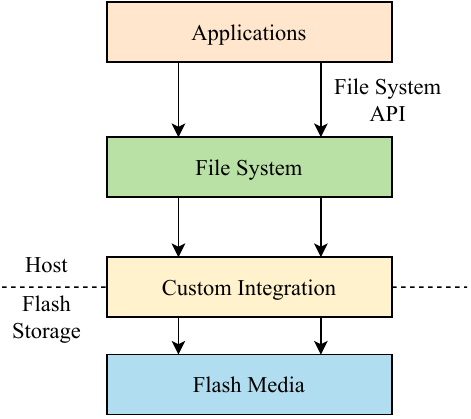}\label{fig:flash_integration_custom}}\hfill
    \subfloat[]{\includegraphics[width=0.3\textwidth]{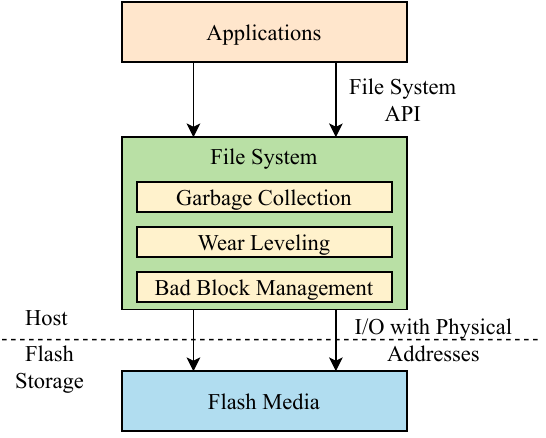}\label{fig:flash_integration_embedded}}
    \caption{Integration of flash storage into host systems with (a) showing a conventional \as{ssd} with a \as{ftl} on the storage device, (b) a custom flash integration, through interfaces such as \as{ocssd}, custom \as{ftl}, custom device drivers, and multi-stream SSD and (c) flash storage on embedded systems with direct management for flash from the file system.}
    \label{fig:flash_integrations}
\end{figure*}

Therefore, throughout this literature study, we divide the relevant work based on the type of flash integration. \cref{fig:flash_integrations} shows three integration levels for flash, where \cref{fig:flash_integration_ssd} depicts the conventional integration with a \as{ssd}. \cref{fig:flash_integration_custom} shows a custom integration of flash storage for devices such as \af{ocssd}~\cite{2019-Lu-Sync_IO_OCSSD,bjorling2017lightnvm}, multi-stream SSD~\cite{bhimani2017enhancing,kang2014multi}, and \af{sdf}~\cite{2014-Ouyang-SDF}. The main benefit of these types of integration is that the flash characteristics are no longer hidden behind the device, giving the host an increasing level of storage control. \as{ocssd} is a type of \as{ssd} that exposes the device geometry to the host, allowing the host to manage device parallelism and allocation. While such a device allows increased data management for the host software, it comes at increased complexity for managing the device constraints. 

Lastly, \cref{fig:flash_integration_embedded} shows flash integration at the embedded level, such as is commonly used in
mobile devices and \as{iot} devices. In embedded flash configuration the flash chip is commonly directly attached to the
motherboard, giving the host system full control over the underlying flash storage. Throughout this literature study, we group file system design and mechanisms based on these three integration levels, as different levels of integration allow different degrees of flash management and ranging possibility for flash integration.


\section{Challenges of Flash Storage Integration}\label{sec:flash_challenges} 
While \as{ssd} uses the same block interface that is used with \as{hdd}, flash has different characteristics that software must account for to better integrate flash storage. This section details the challenges that arise from integrating flash storage into systems, providing the guidelines along which we dictate the bottom up view of changes in the host storage software stack, up to the file system. We define the challenges to account for the characteristics of flash storage, as well as enhance its integration into host systems. In the case of \as{ssd} devices, these challenges often largely depend on the underlying \as{ftl}, as it is making the final decision, independent of what data placement the host implements, however aiding the \as{ftl} can increase the performance. Embedded devices provide a higher level of host data placement by eliminating the \as{ftl} and directly attaching flash chips to the motherboard. Each challenge is assigned a specific identifier with the \af{fic}, in order to refer back to the specific challenge throughout this literature review. \cref{tab:flash-challenges} summarizes the 6 key challenges arising from flash storage devices.

\begin{table*}[!t]
    \centering
    \begin{tabular}{||p{1cm}|p{35mm}|p{95mm}||}
        \hline 
        ID & Flash Integration Challenge & Description \\
        \hline
        \hline
        \textbf{\as{fic} 1} & Asymmetric Read and Write Performance & Write operations require more time than read operations~\cite{stoica2009evaluating,2022-intel-p,chen2009understanding,2021-osdi-zns+,2017-parity-stream}\\
        \hline
        \textbf{\as{fic} 2} & Garbage Collection & The lack of in-place updates results in flash storage running garbage collection to free space and clear invalid pages. \\
        \hline
        \textbf{\as{fic} 3} & I/O Amplification & The lack of in-place and the required garbage collection introduce write amplification, writing more flash pages than the size of the I/O issued by the host. \\
        \hline
        \textbf{\as{fic} 4} & Flash Parallelism & The architecture of flash utilizes a high degree of parallelism (channels/chips/planes) to be utilized to enhance performance. \\
        \hline
        \textbf{\as{fic} 5} & Wear Leveling & Limited lifetime of flash cells requires careful consideration during writes to ensure flash is worn out evenly across the storage space. \\
        \hline
        \textbf{\as{fic} 6} & I/O Management & Optimizations on the I/O requests, such as merging, aims at leveraging the flash storage capabilities and reducing I/O latency. \\
        \hline
    \end{tabular}
    \caption{Overview of the challenges arising from integrating flash storage. The identifier corresponds to the respective \af{fic} referred to throughout this literature study.}
    \label{tab:flash-challenges}
\end{table*}

\noindent \textbf{\as{fic}1: Asymmetric Read and Write Performance.} On flash storage write operations require more time than read operations~\cite{stoica2009evaluating,2022-intel-p,chen2009understanding,2017-parity-stream,2021-osdi-zns+}, making it important for software to limit write operations. Particularly, frequent small writes that are smaller than the allocation unit, referred to as \textit{microwrites} incur significant performance penalties, and should be avoided where possible. Similarly, methods for enhancing the write performance are important to account for the lower write performance, compared to read performance.

\noindent \textbf{\as{fic}2: Garbage Collection (GC).} While the \as{ftl} hides the flash access constraints from host applications, providing seemingly in-place data updates, it adds the cost of performing garbage collection to free up space. \as{gc} overheads have unpredictable performance penalties for the host system~\cite{2015-Kim-SLO_complying_ssds,2014-yang-dont_stack_log_on_log}, resulting in large tail latency~\cite{2013-Dean-tail_at_scale}. Dealing with, and aiming to minimize required garbage collection for the flash device is a key challenge in integrating flash storage.

\noindent \textbf{\as{fic}3: \as{io} Amplification.} Due to the characteristics of flash avoiding in-place updates of flash pages, writes often encounter \af{wa}. With this the amount of data that is written on the flash storage is larger than the write that is issued by the host system. For example a 4KiB issued write may increase to 16KiB being written on the device, due to possible garbage collection requiring to copy data, resulting in a \as{wa} factor of 4x. \as{wa} furthermore adds to an increase in wear on the flash cells~\cite{2013-Lu-SSD_WA_Lifetime}. \af{ra} similarly is caused by requiring to read a larger amount of data than is issued in the read \as{io} request. \as{ra} most commonly happens when reading metadata in order to locate data, thus requiring an additional read of metadata on top of the request read \as{io} request for the data. This is most often inevitable, as all data requires metadata for management, however this should be kept to a minimum at application-level. Furthermore, minimization of \as{wa} is more important than \as{ra}, since write requests have a higher latency than read requests, and writing has a more significant impact on the flash storage, resulting in increased flash wear. While read requests also incur wear on the flash cell, called read disturbance~\cite{2015-Liu-Read_leveling}, it is not as significant as for write requests. 



\noindent \textbf{\as{fic}4: Flash Parallelism.} With the various possible levels of parallelism on flash storage devices (discussed in \cref{sec:ssd-performance}), exploiting of the various possibilities requires software design consideration to aligning with these. Although the \as{io} scheduling of on-device parallelism, such for channel-level parallelism, is responsibility of the \as{ftl} (on devices at \as{ssd} integration level), the \as{ftl} implements particular parallelism, given that the host \as{io} requests aligning with the possibility of parallelizing the request, such as with large enough \as{io}s to stripe across channels and dies. Embedded device and custom flash integrations have more possibility to manage flash device parallelism at the host software level.

\noindent \textbf{\as{fic}5: Wear Leveling (WL).} Given limited program/erase cycles for flash cells, even wear over the entire device is required to ensures that no specific areas of the device are burnt out faster than others. Similar to flash parallelism, this largely depends on the flash integration level, as the \as{ftl} at the \as{ssd} integration level ensures \as{wl}, however embedded flash integration and custom flash integration is required to place more significance on ensuring even wear across the flash cells. Strongly related to prior flash integration challenges, wear is commonly a result of \as{gc}, which in turn increases the \as{io} amplification, and particularly the  \as{wa} and \as{ra}~\cite{2014-Desnoyers-Analytic_Models_SSD,2009-Hu-WA_SSD}. 

\begin{figure}[!t]
    \centering
    \includegraphics[width=\linewidth]{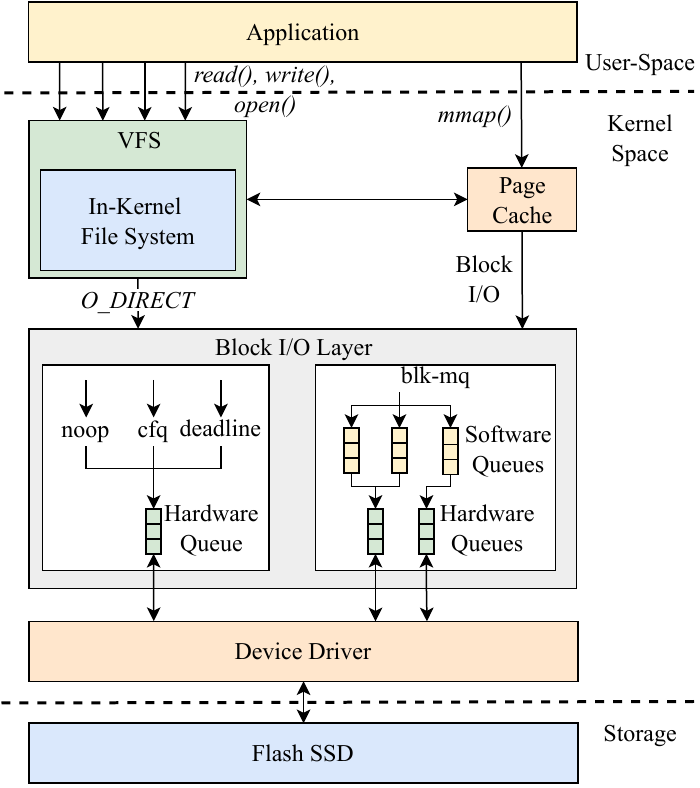}
    \caption{Visual overview of the storage stack in the Linux kernel. Adapted from~\cite{mishra2017BigDataStorage,fischer2017linux}.}
    \label{fig:storage_stack}
\end{figure}

\noindent \textbf{\as{fic}6: \as{io} Management.} As \as{ssd} ships with integrated firmware to expose the flash storage as a block addressable storage device, integration into the current software stack is seamless. \cref{fig:storage_stack} shows the integration of a flash \as{ssd} into the Linux kernel storage stack. Since flash storage devices are significantly faster than prior storage technologies, such as \as{hdd}, the storage software stack becomes the dominating factor in \as{io} latency~\cite{2010-Caulfield-Moneta,2012-Caulfield-Fast_User_Space_Access}. One particular optimization for performance of \as{io} requests to flash storage devices is provided by an \as{io} scheduler, deciding when to issue \as{io} requests to the storage device. As is visible in \cref{fig:storage_stack}, the block \as{io} layer implements various schedulers with different functionality. Providing a ranging degree of optimizations for \as{io} requests, such as varying scheduling policies and merging of \as{io} requests, or possible reordering, specific configurations are favorable to increase performance with flash storage. Particularly the utilization of multiple queues, with multiple software and hardware dispatch queues (visible in the \texttt{blk-mq} configuration of the block \as{io} layer), allows better exploitation of flash storage capabilities, and avoids certain Linux kernel overheads. Furthermore, evaluating mechanisms that reduce the latency of \as{io} operations, and particularly write \as{io} operations.

\subsection{Flash Integration Organization}
\begin{table}[!t]
    \centering
    \begin{tabular}{||p{35mm}|p{38mm}||}
        \hline 
        Integration Level & File Systems \\
        \hline
        \hline
        \as{ssd} Flash Integration & \cite{2015-Changman-f2fs,2021-Gwak-SCJ,2012-Min-SFS,2014-Lu-ReconFS,2020-Tu-URFS,2019-Yoshimura-EvFS,2022-Jiao-BetrFS,jannen2015betrfs,huang2013improve,2019-Liu-fs_as_process,2018-Kannan-DevFS,2021-Liao-Max,2015-Kang-SpanFS,2022-Oh-exF2FS,rodeh2013btrfs} \\
        \hline
        Custom Flash Integration & \cite{kawaguchi1995flash,josephson2011direct,2020-Wu-DualFS,2018-Rho-Fstream,dubeyko2019ssdfs,lee2014refactored,2019-Lu-Sync_IO_OCSSD,2016-Lee-AMF,2018-Yoo-OrcFS,2016-Zhang-ParaFS,2021-Qin-Atomic_Writes} \\
        \hline
        Embedded Flash \newline Integration & \cite{2006-Lim-NAND_fs,hunter2008brief,2009-Sungjin-FlexFS,2008-Jung-ScaleFFS,woodhouse2001jffs,park2013enffis,nahill2015flogfs,schildt2012contiki,2015-Park-Suspend_Aware_cleaning-F2FS,2009-Zuck-NANDFS,2009-Park-Multimedia_NAND_Fs,2007-Hyun-LeCramFS,2021-Ji-F2FS_compression,2011-Park-Multi_NAND,manning2010yaffs,aleph2001yaffs,manning2002yaffs,engel2005logfs,2008-Kim-DFFS,park2006flash,2020-Yang-F2FS_Framentation,2022-Lee-F2FS_Address_Remapping,2011-Lim-DeFFS,2014-Hyunchan-O1FS,2022-Zhang-ELOFS,2020-Zhang-LOFFS,kim2006mnfs,2009-Tsiftes-Coffee_FS} \\
        \hline
    \end{tabular}
    \caption{Classification on the flash integration level utilised for the file systems evaluated in this literature study.}
    \label{tab:integration_levels}
\end{table}

With the different possible levels for integration of flash storage (recall \cref{fig:flash_integrations}), and while mechanisms for solving flash integration challenges are frequently applicable at various integration levels, several of the mechanisms we present require deeper integration of flash storage, levering increased control, in order to be implemented. For instance, the incorporation of the various levels of on-device parallelism is not directly possible at the \as{ssd} integration level, as the \as{ftl} hides the parallelism on the physical device from the host system. The custom and embedded level flash integration provide the host with more possibility to manage these. In order to separate the possibility of mechanisms to be implemented with a particular flash integration level, \cref{tab:integration_levels} provides a classification of each evaluated file system in this literature study to the respective integration level. During the evaluation we present the different mechanisms to solve a \as{fic}, and indicate which file systems utilize these. Therefore, when considering the feasibility of a mechanism for a particular flash integration consult this table to see its applicability. Note that file systems are not limited to the classification we provide, as for instance file systems designed for \as{ssd} flash integration also work on some embedded flash integration. However, we utilize only a single classification for each file system to avoid confusion. Exceptions are made only in specific cases where an existing file system is adapted for a different flash integration level.

We divide the discussion of mechanism by the \as{fic} for which the evaluated study presents a novel solution. This implies that mechanisms that solve multiple \as{fic} are discussed in detail in the first section they appear in, however are also mentioned in all latter sections for the \as{fic} that the mechanism solves. Therefore, each \as{fic} section contains a table of the respective mechanisms presented to solve that particular \as{fic}, along with a reference to the corresponding section of its detailed discussion.

\section{\as{fic}-1: Asymmetric Read and Write Performance}\label{sec:ssd_rw_asym}
Given that read and write performance on flash storage is asymmetric~\cite{stoica2009evaluating,2022-intel-p,chen2009understanding}, this section provides the various mechanisms to handle the asymmetric performance. \cref{tab:rw_asym} shows the various methods discussed in this section, for dealing with asymmetric performance, and how to increase file system performance. Such methods and mechanisms include data structures particularly optimized for characteristics of flash storage, efficient methods of data caching, and effective organization of data on the storage device.

\begin{table}[!t]
    \centering
    \crefformat{section}{\S#2#1#3}
    \crefformat{subsection}{\S#2#1#3}
    \crefformat{subsubsection}{\S#2#1#3}
    \begin{tabular}{||p{50mm}|p{25mm}||}
        \hline 
        Mechanism & File Systems \\
        \hline
        \hline
        Write Optimized Data Structures (\cref{sec:wods}) & \cite{2020-Tu-URFS,hunter2008brief,rodeh2013btrfs,2022-Jiao-BetrFS,dubeyko2019ssdfs} \\
        \hline
        Write Buffering (\cref{sec:write_buffering}) & \cite{park2006cflru,jo2006fab,park2013enffis,2018-Kannan-DevFS,josephson2011direct,2019-Lu-Sync_IO_OCSSD,2010-Josephson-DFS,2011-Park-Multi_NAND} \\
        \hline
        Deduplication (\cref{sec:deduplication}) & \cite{huang2013improve,dubeyko2019ssdfs,2011-Lim-DeFFS} \\
        \hline
        Compression (\cref{sec:compression}) & \cite{woodhouse2001jffs,dubeyko2019ssdfs,2021-Ji-F2FS_compression,2007-Hyun-LeCramFS,ning2011design} \\
        \hline
        Delta-Encoding (\cref{sec:delta_encoding}) & \cite{huang2013improve,dubeyko2019ssdfs} \\
        \hline
        Virtualization (\cref{sec:virtualization}) & \cite{2009-Zuck-NANDFS,2022-Lee-F2FS_Address_Remapping} \\
        \hline
        Flash Dual Mode Switching (\cref{sec:flash_dual_mode}) & \cite{2020-Wu-DualFS,2009-Sungjin-FlexFS} \\
        \hline
    \end{tabular}
    \crefformat{section}{Section #2#1#3}
    \crefformat{subsection}{Section #2#1#3}
    \crefformat{subsubsection}{Section #2#1#3}
    \caption{Mechanisms for file systems to deal with \textbf{\as{fic}1}, asymmetric read and write performance of flash storage, and the respective file systems that implement a particular mechanism.}
    \label{tab:rw_asym}
\end{table}

\subsection{Write Optimized Data Structures}\label{sec:wods}
The characteristic of flash having lower write than read performance requires that data structures for flash are write optimized. Such data structures which are optimized for write operations are referred to as \af{wods}~\cite{bender2015introduction}. With the addition of the missing support for in-place updates on flash, the best suiting data structure for flash-based file systems is a log-based structure. As a result, all file systems discussed in this section are \as{lfs}. The nature of a \as{lfs} being append-only writes accounts for the lower write performance on flash, which matches the write updates to the operations more optimal for flash, which are smaller fine-fine grained updates~\cite{2013-Lu-Flash_Lifetime_Reduce_With_WA} in a log structured fashion~\cite{2015-Changman-f2fs}. 

However, while the log provides increased write performance, metadata is scattered throughout the log, requiring a full scan of the log to locate metadata. Therefore, file systems commonly employ tree-based data structures for metadata, decreasing the worst case time complexity from $\mathcal{O}(n)$ to $\mathcal{O}(\log{}n)$. B-tree is a commonly used data structure for storage systems, including databases~\cite{comer1979ubiquitous,graefe2011modern} and file systems~\cite{rodeh2013btrfs}. Nodes in a B-tree can be larger than in conventional binary search trees, allowing the node to align to a unit of the underlying storage. B-trees furthermore are maximizing the breadth of the tree, instead of its height, in order to minimize the \as{io} requests in order to locate data, since each traversal to a child node requires an \as{io} request. The tree itself is sorted and self-balancing, making the worst case complexity for search relative to the tree height, however also requiring to balance the tree. The B+tree further reduce required \as{io} by only having data in the leaf nodes, such that higher nodes only contain keys, increasing the number of keys that fit inside a block. Leaf nodes are linked in a linked list, for faster node traversal, which in turn speeds up searching. 

In order to write optimize these trees, $\text{B}^\varepsilon$-trees~\cite{bender2015introduction,brodal2003lower} adapt the node structure of the B-tree to include a buffer to which updates to its children nodes are written. As node updates are initially written in memory, this allows to gather a larger number of small writes, encode these in the added buffer of the respective nodes, and write these as a larger unit, avoiding frequent small updates to nodes. The most recent version of BetrFS~\cite{2022-Jiao-BetrFS} (published in 2022) implements such a $\text{B}^\varepsilon$-tree as the metadata storage for indexing data. It is implemented in the Linux kernel as a key-value store, based on TokuDB~\cite{tokudb}, which exposes a key-value addressable interface. The benefit of this $\text{B}^\varepsilon$-tree is that the nodes have a larger (2-4MiB) sequentially written log, batching updates into larger units and thus avoiding small updates. Initially all updates go into the root node message log, which when full gets flushed to its child nodes. While \as{wods} provide optimized write performance by batching updates, read requests require reading an entire node (2-4MiB), causing small read requests to suffer from significant read amplification.

SSDFS~\cite{dubeyko2019ssdfs} adapts the tree design to utilize hybrid nodes, since the node allocation uses blocks, which are several KB in size, and may not directly be filled directly. Therefore, hybrid nodes in the tree adapt the size of the node, such that if a hybrid node is allocated and filled, it allocates an additional hybrid node, which upon being filled is merged with the first hybrid node and becomes a leaf node. This allows to reduce write amplification (solving \textbf{\as{fic}3}) if a node is not filled enough. An additional optimization to B-trees is the \af{piobt}~\cite{2011-Roh-Bp_tree_optimizations}, which implements a parallel flash-optimized B-Tree variant (solving \textbf{\as{fic}4}), utilizing a larger \as{io} granularity to exploit package-level parallelism, maintain a high number of outstanding \as{io} requests to utilize the channel-level parallelism, and avoid mixed read and write operations.

\subsection{Write Buffering}\label{sec:write_buffering}
In addition to \as{wods} optimizing write operations file systems need to also avoid small writes, referred to as \textit{microwrites}. Particularly, as file systems write in units of blocks, which commonly are 4KiB, small writes require a full unit to be filled. The majority of file systems provide the possibility for inline data in inodes, such that the inode data, which is written regardless, has a small capacity to include data. However, this still requires that for small files inodes are directly written to flash. Therefore, several schemes that involve buffering and caching of data in memory, before flushing to the flash storage, provide increased write performance, and additionally increase overall performance.

While buffering of \as{io} prior to flushing to flash storage provides performance gains, buffers are often limited in size and are much smaller than the persistent storage. Furthermore, in addition to buffering write requests for new data, accessing or updating existing data is also cached in the buffer. Therefore, the caches utilize effective methods that minimize the cache misses by optimizing the eviction policy. \af{lru}~\cite{tanenbaum2016structured-LRU} is a common caching policy that maintains the items in the cache in the order of usage, where the least recently used item is selected for eviction. This mechanism is extended by DFS~\cite{2010-Josephson-DFS,josephson2011direct}, which utilizes \textit{lazy \as{lru}}, which does not insert accessed data into the buffer upon a cache miss, but rather inserts the data into the buffer only on an additional cache. Requiring of two cache misses implies that the cache only contains data that is frequently accessed, instead of caching all data from a cache miss.

\af{cflru}~\cite{park2006cflru} is another extension on the \as{lru} algorithm, which splits the buffer into two segments, one as the working region in which recently accessed pages reside, which are managed in \af{mru} fashion, therefore depicting the frequently accessed pages. The second region, called the clean-first region contains the less commonly accessed pages in \as{lru} fashion. On eviction (e.g., when writing a new page and freeing space in the buffer), it first attempts to evict a clean page, rather than a dirty page from the clean-first region, as this does not require a flush of the dirty data to the flash storage, and only resorts to evicting dirty pages as last resort. \af{fab}~\cite{jo2006fab} optimizes the caching policy to align with the flash characteristics by organizing pages in the cache in a larger unit, called \textit{block}, and upon eviction flushes entire blocks to the flash storage, issuing larger \as{io}s and aligning better to the flash erase unit. 

Similarly, NAFS~\cite{2011-Park-Multi_NAND} implements a \textit{double list cache}, containing a clean list with only clean pages for caching of data, and a second dirty list for writing of modified pages and to prefetch of pages based on the access patters, only containing dirty pages. The benefit of two separate lists is that firstly it allows caching write operations and avoid small writes to the flash device, and secondly it prefetches data into the clean list in order to minimize cache misses. EnFFiS~\cite{park2013enffis} presents the \textit{dirty-last cache} policy, which considers the flash characteristics by utilizing a \textit{delay region} and a \textit{replacement region}, which correspond to a multiple of the flash blocks. By using a multiple of the flash blocks, the data written is sequentialized and written to the flash as a larger unit, where dirty pages are initially moved from the replacement region to the delay region, before being flushed to the flash. This buffering in the delay region allows collecting more dirty pages, improving performance and avoiding smaller writes to flash. 

StageFS~\cite{2019-Lu-Sync_IO_OCSSD} likewise utilizes two stages, where write operations are initially written into the first stage, called the \textit{file system staging area}. The issued writes to the staging area are completed in a \textit{persistence-efficient} way, which utilizes a log to account for asymmetric flash performance for the staging area. Subsequently, writes are then regrouped, based on the file system structure and hot/cold identification, and are written to the second stage, based on the group assignment. The staging area allows writing synchronous \as{io} directly to the staging log with optimal flash write characteristics, lowering completion latency, followed by better grouping of data when writing from the staging to the second stage file system area.

In an effort to limit the required memory for data caching, DevFS~\cite{2018-Kannan-DevFS} proposes the \textit{reverse caching} mechanism for effectively managing host memory and device memory. Reverse caching aims at keeping only active files in the device memory, which is limited in size, and upon closing of a file migrates the metadata to the host memory. Reopening a file migrates it back to the device memory, and consistency of metadata is not violated as any actively modified metadata is in the device memory, and only inactive metadata is in the host memory. To efficiently utilize reverse caching, DevFS uses a host-memory cache that is able to use \af{dma} to move metadata between itself and the device memory, additionally minimizing host overheads.

\subsection{Reducing Write Traffic}\label{sec:reduce_write_traffic}
In order to avoid the slower write performance of flash storage another mechanisms aims at minimizing the amount of data that is written to the storage device. Whenever data is updated on the flash storage, updating all metadata and writing the new data incurs a significant amount of write traffic, in addition to causing \as{wa}. Therefore, reducing the amount of data being written through mechanisms such as deduplication, compression, and delta-encoding helps at avoiding the increased \as{wa}.

\subsubsection{Deduplication}\label{sec:deduplication}
File systems commonly contain duplicate data, from backups or archival copies made and general work. Therefore, file systems often aim to avoid creating duplicates of existing data, which is referred to as \textit{deduplication}. Evaluations show that in high-performance computing centers on average about 20-30\%, with peaks of 70\%, of the stored data can be removed through deduplication~\cite{meister2012study}. Deduplication avoids writing the same data multiple times, which in turn helps reduce the write traffic and minimizes write and space amplification (solving \textbf{\as{fic}2}) and prolong the device lifetime. Effective deduplication relies on hash functions, which provide a deterministic output, called the \textit{digest} or commonly referred to as the \textit{fingerprint} of the data, based on which duplicates can be identified. The utilized hash functions for deduplication are \textit{one way hash functions}~\cite{merkle1989one}, which calculate a fingerprint of the data, or digest, but given the digest or fingerprint, the data cannot be generated. Given data from a file for instance, hashing the data provides a digest, and if the same data is to be written again, the same digest is generated. By identifying if a digest already exists in the file system, can it identify if the data is being duplicated, and avoid writing the duplicate by making the metadata of the newly written data point to the existing data. 

In order to apply deduplication on file system data, two methods exits. \textbf{(1)} Deduplication with fixed-sized chunks, where hashes are generated based on the entire file or a pre-determined chunk size, and \textbf{(2)} variable-sized chunks where the chunk size that is hashed depends on the file and content. The effectiveness of deduplication with fixed-sized chunks is limited and highly dependent on the chunk size and modification sequences in the chunks~\cite{2011-Lim-DeFFS}, as for example a small change in a large file, where the chunk size is the entire file, results in a completely different hash, even if much of the data is duplicated. While variable-sized chunks reduces the configuration dependability it is significantly more complex to implement. DeFFS~\cite{2011-Lim-DeFFS} implements a duplicate elimination algorithm that reduces the complexity of identifying duplicates. As comparing duplicates of every byte is not possible, the algorithm finds the smallest modified region in a file adapting the chunk size, which is initially assigned the default value that corresponds to the flash page size.

The idea of eliminating duplication through fingerprints is not new for \as{ssd}, as existing \as{ftl} implementations implement such mechanism. \af{caftl}~\cite{chen2011caftl} generates collision-free fingerprints of write requests and maintains fingerprints of all flash resident data in order to avoid writing the same data twice. Flash Saver~\cite{li2012flash} is a similar architecture running between the SSD and the file system, which manages the file system \as{io} requests and ensures deduplication using SHA-1 fingerprints~\cite{wang2005finding}. A plethora of similar hash-based deduplication mechanisms exist~\cite{fu2011aa,ha2013deduplication,kim2012deduplication,xia2014combining}, which may not be particular to file systems, but can be adopted by a wide range of storage systems. File systems with deduplication are CSA-FS~\cite{huang2013improve}, which implements deduplication by calculating the MD5 digest~\cite{rivest1992md5} (MD5 is a hash function applied to the input) and looks up the result in a hash table which provides the corresponding LBA of the block. If it already exists and the user requested a new file with it, the new inode simply uses the given LBA from the hash table, instead of duplicating the data again. Similarly, SSDFS~\cite{dubeyko2019ssdfs} maintains fingerprints with its metadata B-Tree, and DeFFS~\cite{2011-Lim-DeFFS} stores hash keys with all inodes for their data to avoid duplicates. 

\subsubsection{Compression}\label{sec:compression}
An effective method for reducing the required storage space for data is to utilize \textit{compression}. While there exist lossy compression algorithms~\cite{iverson2012fast,kavitha2016survey}, which discard parts of the data, and lossless compression algorithms~\cite{kavitha2016survey,sharma2010compression}, which maintain all data, lossy compression is very application dependent, and in the case of file systems we assume a lossless data storage and therefore focus on lossless compression algorithms. Lossless compression methods rely on three methods. \textbf{(1)} \af{rle} which minimizes size by taking the repeating characters, referred to as the run, and replacing them with a 2 byte sequence of the number of repetitions of the character, called the run count, followed by the replaced character. For example the sequence of ``aaabbccc`` is represented as ``a3b2c3`` with \as{rle}. \textbf{(2)} \af{lz}~\cite{ziv1977universal,ziv1978compression} which utilizes a dictionary to replace strings with their dictionary value. Variations of the \as{lz} algorithm exist, such as LZ77~\cite{ziv1977universal}, LZ78~\cite{ziv1978compression}, and LZW~\cite{welch1984technique}. The dictionary is constructed at the compression time and used a decompression time~\cite{kapoor2013review}. A well-known compression algorithm GZIP~\cite{deutsch1996gzip} is based on two \as{lz} algorithms. \textbf{(3)} Huffman Coding~\cite{huffman1952method} which creates a full binary tree based the frequency of occurring characters and generating code of the character and the frequency. 

JFFS~\cite{woodhouse2001jffs} and several other file systems~\cite{woodhouse2001jffs,dubeyko2019ssdfs,2021-Ji-F2FS_compression,2007-Hyun-LeCramFS,goyal2005energy,cramfs,ning2011design} provide a data compression, however metadata compression has shown to cause decompression overheads~\cite{kang2022pr}. Applying compression at the file system level allows reducing the utilized storage space, especially on flash this reduces the space and write amplification factors (solving \textbf{\as{fic}3}), and in turn prolonging the flash lifetime as shown by Li et al.~\cite{2015-Li-SSD_lifetime_compression} on the benefits of different compression algorithms for NAND flash lifetime. SSDFS~\cite{dubeyko2019ssdfs} utilizes LZO compression~\cite{oberhumer2008lzo}, a library for data compression with LZ algorithms, for data and metadata. Given that different data benefits from different compression mechanisms, where for instance data is less frequently read can be compressed more efficiently than frequently read data, which may require more simplistic decompression to minimize overheads, or embed metadata in the compressed data to avoid decrypting data and metadata separately. Adaptive compression selection is employed in an extension of F2FS~\cite{2021-Ji-F2FS_compression}, where \af{fpc} selects the compression method for files based on how compressible they are.

\begin{figure}[!t]
    \centering
    \includegraphics[width=0.45\textwidth]{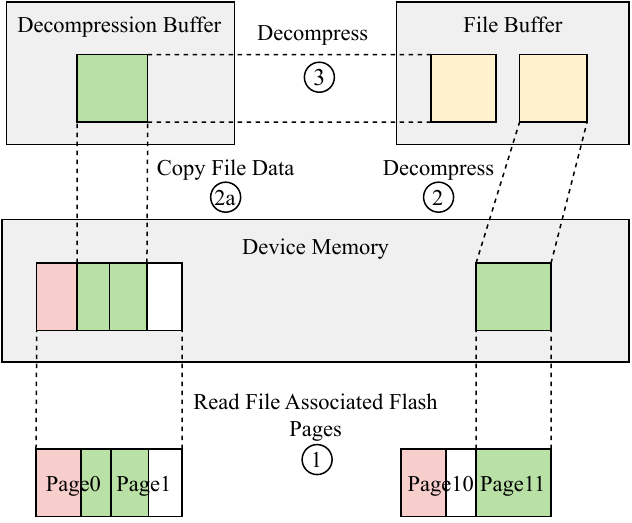}
    \caption{The benefit of \af{pbal} during decompression arises from the elimination of copying associated file data.}
    \label{fig:pba_decompression}
\end{figure}

An issue of compression is that it must be aligned to the flash unit, in order to avoid unnecessary read amplification and required data copying. \cref{fig:pba_decompression} shows that if the compressed data is not aligned to the flash unit of a page, such that compressed data can extend across pages, the decompression process becomes more complex. This is due to first having to read all the associated flash pages into the device memory. However, if compressed data can extend across flash pages, this implies that non-associated data can be included. Therefore, only the associated data must be copied to a \textit{decompression buffer} (step 2a), from which the data can be decompressed (step 3). LeCramFS~\cite{2007-Hyun-LeCramFS} (short for less crammed FS) implements a read-only file system with compression for embedded devices, that avoids this additional copy of data by using \as{pbal}. Any compressed data will not extend across page boundaries, implying that no additional pages with non-associated data are read, and the data can be decompressed directly (step 2). Therefore, it avoids the additional copy of data and reduces the read amplification (solving \textbf{\as{fic}}3). In order to avoid wasting space if the page is not fully utilized from a single compression, LeCramFS extends the compression implementation with a \textit{partial compression}, which splits the data into parts that fit into the available space. These parts are then compressed separately, allowing to utilize all available space.

\subsubsection{Delta-Encoding}\label{sec:delta_encoding}
Similar to deduplication, delta-encoding lowers the write traffic, further limiting the space and write amplification (solving \textbf{\as{fic}2}) by instead of writing out entire data when updated, with delta-encoding the new data is compared to the old data and only the differences, called the \textit{delta}, are written. This is especially beneficial in the case of small changes in large files, where it avoids writing the entire file again and writes only the changes in the data. While delta-encoding provides a similar space reduction to compression, particular file characteristics of what is being encoded can affect resulting performance. For instance, data sets that have a significant redundancy across them, such as email data sets, require significantly less space than being compressed~\cite{douglis2003application}. However, non-textual data, such as pdf documents, will have large deltas for even small changes, where compression is likely to be a better choice. Therefore, the application of delta-encoding depends on the data characteristics and what type of content is being encoded, in order to achieve better utilization of delta-encoding compared to simpler compression. CSA-FS~\cite{huang2013improve} implements delta encoding on its file system metadata (superblock, descriptor, bitmaps, and tables), allowing minor updates such as the access time of metadata and bitmaps to only write the new changes. Similarly, SSDFS~\cite{dubeyko2019ssdfs} uses delta-encoding for user data.

\subsubsection{Virtualization}\label{sec:virtualization}
\begin{figure}[!t]
    \centering
    \includegraphics[width=0.45\textwidth]{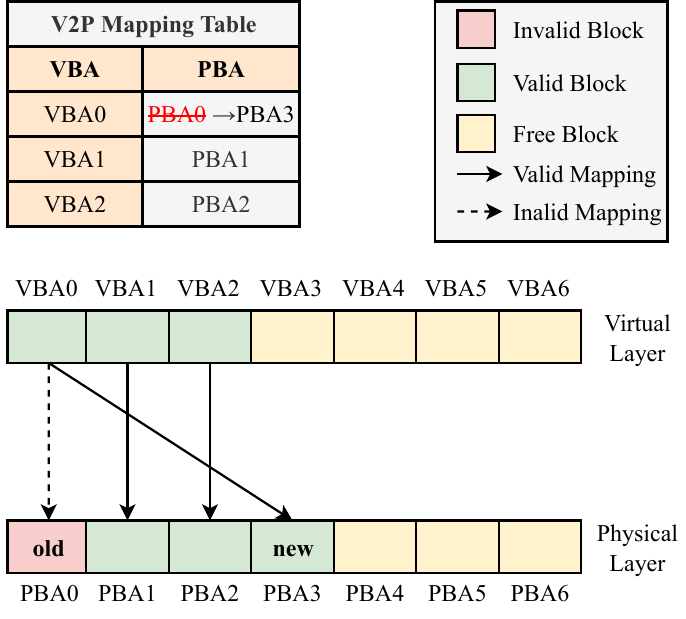}
    \caption{A simplified example scenario of virtualization on top of the physical storage.}
    \label{fig:virtualization}
\end{figure}

As Butler Lampson mentioned in his famous quote from 1972 (which was originally stated by David Wheeler), ``Any problem in computer science can be solved with another level of indirection``~\cite{lampson1993principles}. By adding a layer on flash storage the write traffic can be reduced as a result of avoiding metadata update propagation (as with the wandering tree problem~\cite{bityutskiy2005jffs3}). Adding a virtual layer is conceptually identical to \af{lba} on top of \as{pba}. \cref{fig:virtualization} shows an example of how virtualization maintains the same \af{vba} when file data is updated, eliminating a need for metadata updates. Note, for simplicity we assume the virtual layer to be on top of the physical layer, providing \as{pba}s, however it can be directly on the physical layer of the flash storage, depending on the flash integration level, and would therefore be using \af{l2p} mappings. In the illustrated scenario, three blocks are mapped to \as{pba}0, \as{pba}1, and \as{pba}2, respectively. Assuming these are part of a file, metadata points to these respective virtual block addresses. If the first block of the file is modified, a new data block is written at \as{pba}3. With virtualization, the \as{vba} remains unchanged, only the \as{v2p} mapping table is modified to depict the new mapping of \as{pba}3. As a result, file metadata remains unchanged, avoiding write amplification for updating file metadata. 

A similar mechanism is implemented in NANDFS~\cite{2009-Zuck-NANDFS}, using a \textit{sequencing layer} that implements the block allocation as immutable storage on the logical layer. Therefore, when file system data is updated, it cannot overwrite and simply marks the data as obsolete, and the new updated data is written in a newly allocated block. The sequencing layer implements the \as{l2p} mapping,  such that the \as{lba} of the new data is the same as the old \as{lba}, avoiding the updating of data addresses in the file system metadata. This implies that the file system does less writing by eliminating metadata updates, which in turn also reduces the required \as{gc}.

The concept of virtualization can similarly be applied through \af{ar}. With \as{ar} only the \as{l2p} mapping table of the storage device is modified, which similar to virtualization avoids rewriting of metadata. \as{ar} is an effective mechanism for solving additional overheads in file systems, such as file system garbage collection~\cite{kang2018address}, journaling~\cite{choi2009jftl,kang2018address,weiss2015anvil}, and data duplication~\cite{2021-Zhou-Remap_SSD}. Lee et al.~\cite{2022-Lee-F2FS_Address_Remapping} propose \af{rmipu}, an address remapping scheme implemented in F2FS to solve issues of out-of-place updates. Instead of applying \as{ar} into a single file system operation, \as{rmipu} includes all write operations to utilize \as{ar}. All updated data is first stored in the log, followed by \as{ar} to update data pointers, thus avoiding metadata updates for file overwrite operations. File write operations to new files are appended to the log as usual, and the contiguity in \as{lba}s for files is maintained.

\subsection{Flash Dual Mode Switching}\label{sec:flash_dual_mode}
Some modern flash devices allow the host to switch the cell level of underlying flash blocks (e.g., switching \as{mlc} to \as{slc})~\cite{2020-Wu-DualFS,li2019accelerating}. This provides the benefit that a lower cell level, representing fewer bits, has a lower read, write, and erase latency, thus providing the flash with a higher performance than with a larger cell level configuration~\cite{2020-Wu-DualFS}. This switching between flash modes is referred to as \textit{flash dual mode}, which however comes at the cost of being able to store less data in the same amount of flash, as the block is only able to store half the data in the example of switching from \as{mlc} to \as{slc}. DualFS~\cite{2020-Wu-DualFS} utilizes the flash dual mode feature to provide a dynamically sized \as{slc} area, alongside the remaining \as{mlc} area, to accelerate the performance of critical \as{io} requests. By evaluating the \as{io} queue depth of \as{io} requests, DualFS determines the criticality of the incoming request, and maps it to the \as{slc} area for increased performance. It further profiles incoming request based on the hotness and allocates hot data into the \as{slc} mode for lower request latency.

FlexFS~\cite{2009-Sungjin-FlexFS} implements a similar mechanism for increased performance on the \as{slc} area. The drawback of a design that utilizes a \as{slc} area for incoming write \as{io}, is that data is required to be moved from the \as{slc} area to the \as{mlc} area as it has a lower write lifetime. With the \as{slc} are being used for critical \as{io} requests that require lower completion latency, it introduces \textit{data migration overheads}. To solve this, FlexFS implements several migration techniques that aim to hide the overheads for the data migration from the host. The first technique revolves around \textit{background migration}, which pushes the migration to happen when the system is idle. The second technique is \textit{dynamic allocation}, which writes non-critical requests to the \as{mlc} area, saving on flash degradation in the \as{slc} area, as well as avoiding future data migration. The dynamic allocator functions based on measurements of prior system idle times from \as{io} requests, to predict current idle time, and if sufficient idle time is predicted in order to complete data migration, the data is written to the \as{slc}, and otherwise part of the data, depending on required migration and idle time, is written to the \as{mlc} area. The last technique, \textit{locality-aware data management}, takes into account the hotness of data, and the dynamic allocator attempts to migrate only cold data from the \as{slc} area.

\subsection{Summary}
With the write performance of flash storage being lower than its read performance, particular attention is paid to designing effective mechanisms and methods for achieving faster write performance. Effective caching methods, buffering data before writing, allow to reduce write latency. Similarly, mechanisms that reduce the write traffic to the device, such as deduplication, delta-encoding, and virtualization allow are effective for dealing with the lower write performance of flash storage. Lastly, the possibility of flash-dual mode switching allows to change the flash cell level to provide lower latency write request completion for critical write request.

\section{\as{fic}-2: Garbage Collection}\label{sec:gc}
A significant challenge of flash storage is \as{gc} overheads having unpredictable performance penalties for the host system~\cite{2015-Kim-SLO_complying_ssds,2014-yang-dont_stack_log_on_log}, resulting in large tail latency~\cite{2013-Dean-tail_at_scale}. Dealing with, and aiming to minimize required garbage collection for the flash device is a key challenge in integrating flash storage. Naturally, as flash storage does not provide in-place updates, data is written in a log-based fashion, sequentially in the flash blocks. Therefore, over time as data is overwritten, the blocks contain an increasing number of invalid pages that must be erased to free space. However, as the block also contains valid data, and the erase unit is a block, the still valid pages are moved to a new block, such that the old black can be erased. 

\begin{figure}[!t]
    \centering
    \includegraphics[width=\linewidth]{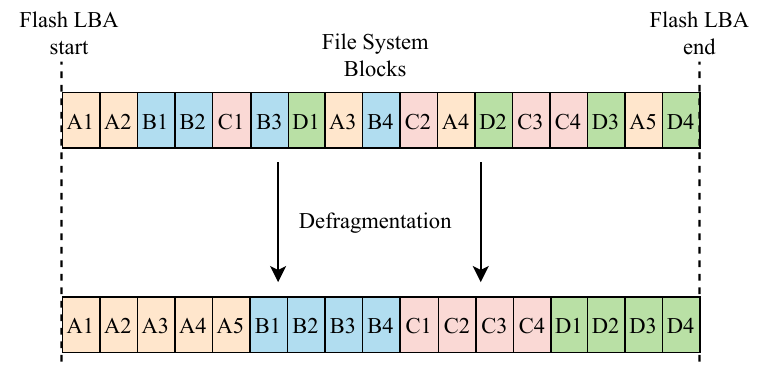}
    \caption{Illustration of single file fragmentation on the storage over time, as parts of file data are overwritten and, due to the \as{lfs} design new data is appended at the head of the log.}
    \label{fig:fragmentation}
\end{figure}

In a \as{lfs} updates to file data are written at the head of the log, resulting in the parts of the file that are updated to be located in a different flash block than the parts of the original file data. Furthermore, \as{gc} causes file data to moved around the storage space as well, resulting in scattering of parts of files, referred to as \textit{fragmentation}. Figure~\ref{fig:fragmentation} shows the resulting fragmentation for several files that occurs over time from file updates and \as{gc}. Fragmentation results in increased read time~\cite{chen2009understanding}, due to files being in non-contiguous regions, and introduces increased garbage collection overheads due to the failed grouping of data. Ji et al.~\cite{2016-Ji-Framgentation_empirical_study} show an empirical study on fragmentation in mobile devices, identifying that fragmentation introduces performance degradation due to increased \as{io} requests, and it further produces increased pressure on host caching. The effects on caching are due to increased difficulty of prefetching data, since it no longer is in contiguous physical ranges, but scattered throughout the physical space. Therefore, prefetching cannot bring correct data into the caches, resulting in an increase in cache misses. 

In addition to increased \as{io} requests for reading and rising cache pressure, increased garbage collection is caused when frequently modified files or file fragments, referred to as the \textit{hot data}, are in the same block as rarely accessed files, referred to as the \textit{cold data}. When hot data is modified, its flash pages are invalidated. Once enough flash pages in a block are invalidated, it can be erased. However, if it is co-located with cold data, the cold must be copied to a free space, as it is still valid. If this cold data is then again co-located with hot data, the modifications of the hot data cause the block to be cleared during \as{gc}, which requires the cold data to be moved again. Therefore, co-locating hot and cold data in the same physical erase unit results in significant \as{gc} increase due to the unnecessary moving of the cold data.

\begin{table}[!t]
    \centering
    \crefformat{section}{\S#2#1#3}
    \crefformat{subsection}{\S#2#1#3}
    \crefformat{subsubsection}{\S#2#1#3}
    \begin{tabular}{||p{50mm}|p{25mm}||}
        \hline 
        Mechanism & File Systems \\
        \hline
        \hline
         \cellcolor{lightgreen!75} Reducing Write Traffic (\cref{sec:reduce_write_traffic}) & \cellcolor{lightgreen!75}\cite{huang2013improve,2011-Lim-DeFFS,woodhouse2001jffs,dubeyko2019ssdfs,2021-Ji-F2FS_compression,2007-Hyun-LeCramFS,ning2011design,2009-Zuck-NANDFS,2022-Lee-F2FS_Address_Remapping} \\
        \hline
        Aligning the Allocation Unit (\cref{sec:align_alloc}) & \cite{2020-Tu-URFS,2009-Park-Multimedia_NAND_Fs} \\
        \hline
        Data Grouping (\cref{sec:data_grouping}) & \cite{2006-Lim-NAND_fs,2022-Zhang-ELOFS,2020-Zhang-LOFFS,2015-Changman-f2fs,2018-Rho-Fstream} \\
        \hline
        \as{gc} Policies (\cref{sec:gc_policy}) & \cite{2012-Min-SFS,2018-Yoo-OrcFS,2021-Gwak-SCJ,2015-Park-Suspend_Aware_cleaning-F2FS} \\
        \hline
        Coordinating the Software Stack Layers (\cref{sec:software_stack}) & \cite{2020-Wu-DualFS,2016-Zhang-ParaFS,2016-Lee-AMF,lee2014refactored} \\
        \hline
    \end{tabular}
    \crefformat{section}{Section #2#1#3}
    \crefformat{subsection}{Section #2#1#3}
    \crefformat{subsubsection}{Section #2#1#3}
    \caption{Mechanisms for file systems to deal with \as{gc} overheads from flash storage, and the respective file systems that implement a particular mechanism. Green highlighted table cells depict previously discussed mechanisms with their respective section.}
    \label{tab:gc}
\end{table}

Reducing the write traffic to the storage device, an effective method to handle the asymmetric flash performance (see \cref{sec:reduce_write_traffic}), is a solution to minimize fragmentation and reduce possible future \as{gc} overheads. However, additional mechanisms are required for effective \as{gc} management. With fragmentation being a significant contribution to increased garbage collection, avoiding it is a key objective to reducing garbage collection overheads. Fragmentation is classified into three different types~\cite{2007-Sato-defrag}. Firstly, \textit{single file fragmentation}, where data in a single file is dispersed over the storage (as is shown in Figure~\ref{fig:fragmentation}). Secondly, \textit{relevant file fragmentation}, where files that are relevant to each other and should be grouped together are split over the storage, such as co-locating hot and cold data in the same erase unit. Lastly, \textit{free space fragmentation}, where the file system has a large amount of small free space, because of deletion of dispersed small files. The cause of fragmentation occurring over time is referred to as \textit{file system aging}~\cite{smith1997file}. While several tools exist that implement \textit{defragmentation}~\cite{hahn2017improving,f2fs_defrag_tool,park2021fragpicker}, additional mechanism can be utilized to avoid fragmentation and \as{gc} overheads. \cref{tab:gc} depicts the mechanisms for file systems to deal with and minimize garbage collection overheads. The process of countering the different types of fragmentation is commonly referred to as \textit{storage gardening}~\cite{2020-Kesavan-Fragmentation,2019-Kesavan-Storage_Gardening}, and for file system development various aging tools exist in order to generate real-world file system workloads and simulate file system aging~\cite{2019-Conway-FS_aging,2018-Kadekodi-Geriatrix}. 

\subsection{Aligning the Allocation Unit}\label{sec:align_alloc}
\as{gc} is a result of having to move valid blocks in the erase unit to a free space, in order erase the flash block. While data grouping allows to align the validity inside the block, such that blocks are likely to be updated within close proximity, multiple files may be co-located in the same block. Therefore, a similar method is to align the allocation unit of data blocks for a single file to the erase unit, resulting in only a single file being located in a block. Such a mechanism is implemented in URFS~\cite{2020-Tu-URFS}, which aligns the data allocation unit for large files to the flash erase unit, allowing files to be erased as a single unit, limiting required \as{gc}. However, as the erase unit of flash can be several hundreds MBs, resulting in significant over-allocation for small files, it makes such a mechanism only beneficial with large files. NAMU~\cite{2009-Park-Multimedia_NAND_Fs} similarly showcases a file system that aligns its content with the requirement of large files. Focused on the multimedia domain, where files have the particular characteristics of rarely being modified, and if removed all file data blocks are erased in one unit, \as{gc} in NAMU is done at the granularity of a file. In addition to improving on \as{gc}, the memory requirements for mapping tables are also minimized. For generic file systems that vary in file characteristics, \af{ars}~\cite{2020-Yang-F2FS_Framentation} minimizes the issue of over-allocating space by allocating in a smaller unit of 2MB (a single segment). File data is written to the space until it is exhausted, upon which a new segment is allocated. While this does not map entire files to the flash erase unit, it allows writing of file data sequentially for each segment, eliminating fragmentation to a degree. Therefore, the resulting reduction in fragmentation provides benefits in reducing the amount of data that is required to be moved during \as{gc}.

\subsection{Data Grouping}\label{sec:data_grouping}
A key circumvention method for fragmentation relies on grouping of related data. Most commonly this is applied in the type grouping data by its access and modification frequency into hot and cold data, however other less commonly used groupings based on \textit{death-time prediction} exist~\cite{2021-Chakraborttii-DT_LBA}. Commonly more classifications than simply hot and cold are utilized for more effective grouping. A plethora of methods for grouping data in such a way have been proposed, which we split by its application type into several groups.

\subsubsection{Data Type Grouping}
Common write patterns in storage systems follow a bi-modal distribution, where many very small write requests and a numerous very large write requests are issued~\cite{chang2008hybrid}. This stems from the fact that small changes are caused by metadata updates, which occur the most frequently, whereas large changes are file updates. Given that metadata is more likely to be updated frequently, separating metadata from data improves on required garbage collection. Specifically, since metadata is often updated even if the data is not updated, for example in scenarios where the file attributes (access time, permissions, etc.) are updated, or the file is moved. Therefore, based on the request size, the data can be classified to be metadata and be grouped accordingly. F2FS also groups based on the data type, where metadata is considered to be always hot data, which is implemented by similar file systems~\cite{2006-Lim-NAND_fs}.

Dividing of file system data is similarly applied in ELOFS~\cite{2022-Zhang-ELOFS,2020-Zhang-LOFFS}, which splits the flash storage two partitions, where a directory partition contains the data of directory entries, and a data partition contains the file system data, which is compacted with the inodes. Jung et al.~\cite{jung2010process} propose the addition of classifying data based on the \af{pid}, as a process is likely to generate similar access patterns and data types throughout its lifetime, classifying by the \as{pid} allows to indirectly infer a data type. A similar data type separation is implemented by Fstream~\cite{2018-Rho-Fstream}, for which the authors modify ext4~\cite{cao2007ext4} and xfs~\cite{sweeney1996scalability} to map different operations to different streams on a stream \as{ssd}. Ext4Stream, the modified ext4 to support streams, maps different metadata operations to different streams, including the journal writes for consistency, the inode writes, as well as different streams for the directory blocks and the bitmaps (inode and block). Furthermore, it utilizes different streams that can be created for different files and for different file extensions. The goal of such streams is to map particular files, such as \texttt{LOG} files (temporary hot data) for key-value stores for example to a particular stream, separating its access patterns from that of other files and file system data. Similarly, the modified XFStream utilizes different streams for the log, inodes, and specific files.

\subsubsection{Dynamic Grouping}\label{sec:dynamic_grouping}
While grouping is an effective method for minimizing \as{gc}, it commonly relies on a static definition on classification targets for the number of hot/cold degrees to classify to (e.g, hot/warm/cold). Shafaei et al.~\cite{2016-Shafaei-WA_extent_temp} identify that the majority of hot/cold data grouping methods fail to account for the accuracy in the hot/cold grouping mechanism, as well as relying on an individual classification of each LBA, making the management of increasingly larger flash storage difficult. Therefore, Shafaei et al.~\cite{2016-Shafaei-WA_extent_temp} propose an extent-based temperature identification mechanism. It is based on the density stream clustering problem~\cite{jia2008grid,chen2007density,forestiero2013single,isaksson2012sostream}, which is a common approach of classification in artificial intelligence and stream processing, however has not been applied to storage before. The density-based stream clustering groups data in a one dimensional space as the data arrives, hence its applicability for stream processing. 

Applying this method to storage, the one-dimensional space is the range of \as{lba}s, and extent-based clustering splits the available space into a number of extents to group by. Initially, the entire space is a single extent and as writes occur the extent is split into smaller extents with different classifications. Over time as more writes are issued, extents are expanded and merged (merging of extents with the same classification). Such a grouping allows a more detailed grouping due to the increase in classification targets, compared to binary hot/cold grouping. However, an evaluation by Yang and Zhu~\cite{2015-Yang-algebric_WA_modeling} on a configurable garbage collection policy, where the number of hotness classification targets is evaluated, shows that various hotness classification targets can significantly increase the write amplification during garbage collection. 

\subsubsection{\as{lba} Hotness Classification}\label{sec:lba_classification}
Different to grouping data based on its type, hotness can be classified on the \as{lba} based on its access frequency. The naive approach at modeling hotness for each \as{lba} is with table-based classification model~\cite{hsieh2005efficient}. This however comes at a high overhead cost, as an entry for each LBA is needed, which becomes increasingly expensive as flash storage grows.  Therefore, a more non-trivial method is based on two-level \as{lru} classification~\cite{chang2002adaptive}, with two LRU lists. Upon an initial \as{lba} access, the \as{lba} is stored in the first list, and a subsequent access moves it to the next list, which is referred to as the \textit{hot list}. Therefore, if a \as{lba} is in the hot list, it is considered to be frequently accessed. 

A different approach is implemented by \af{mbf}~\cite{jagmohan2010write,park2011hot}, which uses bloom filters to identify if a \as{lba} is hot. Bloom filters rely on a hash function that, given an input such as the \as{lba}, provide an output which is mapped to a bit array, and sets the bit to true. Therefore, if a \as{lba} is accessed, applying the hash function sets the respective bit in the array to true, and checking if a \as{lba} is hot simply applies the hash function and checks if the bit is set. However, depending on the length of the array, multiple \as{lba}s can map to the same bit location, as proven by the pigeonhole principle~\cite{ajtai1994complexity}, resulting in false positive hotness classifications for a \as{lba}. To avoid frequent false positive classifications, \as{mbf} utilizes multiple bloom filters at the same time. With multiple bloom filters, the same amount of arrays exist, applying all bloom filter hash functions to the \as{lba}, and setting the respective bit in each of the arrays. As a result, collisions on all bloom filter hash functions are less likely, minimizing the possibility for false positives.

Similar to \as{mbf}, Kuo et al.~\cite{kuo2006configurability} present a hot data identification method by using multiple hash functions and a hash table. Upon a write, the \as{lba} is hashed by multiple hash functions, and a counter for each hash function is incremented in a hash table. To check if a \as{lba} is hot, the \as{lba} is hashed and a configured $H$ most significant bits of the resulting hash table indicate if the \as{lba} is hot if they are non-zero, as the counter is increased on accesses the most significant bits are only non-zero if the \as{lba} is frequently accessed. Multiple hash functions are used for the same reason multiple bloom filters are used in \as{mbf}, to avoid false positive classifications. Lee and Kim~\cite{2013-Lee-data_grouping_empirical_study} provide a study into comparing performance of two-level \as{lru}, \as{mbf}, and \af{dac}, which are similar to the density-based stream clustering by Shafaei et al.~\cite{2016-Shafaei-WA_extent_temp} (discussed in the prior \cref{sec:dynamic_grouping}). The authors show that on the evaluated synthetic workloads \as{dac} provides the highest reduction in write amplification factor, which in turn leads to a decrease in \as{gc} overheads.

Unlike all prior approaches basing classification on the access frequency directly to identify hotness, Chakraborttii and Litz~\cite{2021-Chakraborttii-DT_LBA} propose a temporal convolutional network that predicts the \textit{death-time} of a \as{lba}, based on modification history. This allows to more optimally group data based on the death-time of individual \as{lba}, which has been shown to be an effective grouping mechanism~\cite{2017-He-SSD-Unwritten-Contract}. Grouping related death-time LBA reduces the required garbage collection, as blocks containing \as{lba}s with similar death-times are erased together, which in turn reduces the write amplification (solving \textbf{\as{fic}3}).

\subsection{Garbage Collection Policies}\label{sec:gc_policy}
While data grouping provides benefits of co-locating data based on their update likelihood, an additional essential part of garbage collection is the policy of victim selection for segments to clean. Since during garbage collection a segment to clean is required to be selected, where all still valid data in the segment is moved to a free space, selecting a victim becomes non-trivial. With the importance of data grouping with hotness for effective \as{gc}, conventional \as{gc} policies, such as greedy and cost-benefit lack their inclusion. SFS~\cite{2012-Min-SFS} proposes the \textit{cost-hotness} policy to account for the hotness of segments instead of the segment age, better incorporating the data grouping into victim selection. The cost-hotness is calculated as
\begin{equation*}
    \text{cost-hotness}=\frac{\text{free space generated}}{\text{cost}*\text{segment hotness}}=\frac{(1-u)*\text{age}}{2u*h}
\end{equation*}
where the cost considers reading and writing the valid blocks (equivalent to $2u$) with the segment hotness. A further \as{gc} policy that is based on the cost-benefit policy, is the \af{cat} policy~\cite{chiang1999cleaning}. It extends cost-benefit by including an erase count for each block, improving wear leveling (solving \textbf{\as{fic}5}).

The majority of \as{gc} policies have a fixed algorithm, limiting configuration possibility. The \textit{$d$-Choice} algorithm~\cite{van2013mean} is a configurable \as{gc} policy that combines greedy selection with random selection. The tunable parameter $d$ defines the number of blocks to be selected randomly out of the $N$ total blocks. Therefore, configuring $d=1$ results in fully random victim selection, as a single block is randomly selected from the total blocks, providing effective wear leveling through randomness~\cite{2015-Yang-algebric_WA_modeling} (solving \textbf{\as{fic}5}). Configuring $d\rightarrow \infty$ selects a larger subset of blocks to use greedy selection on, such that $d=N$ is equivalent to fully greedy victim selection, allowing to provide the lowest cleaning latency. Configuration of $d$ thus allows to define the tradeoff between wear leveling and performance. 

An evaluation by Yang and Zhu~\cite{2015-Yang-algebric_WA_modeling} of the algorithm shows the significance of the number of hotness classification targets that are utilized, and the configuration of the $d$ parameter, where various hotness classification targets can significantly increase the \as{wa} during \as{gc}. Similarly, \af{fagc}~\cite{yan2014efficient} is a \as{gc} policy that maintains an \af{uft} for each \as{lba} of a file, in order to group valid pages in the victim block based on the access frequency when these are copied to a new block during \as{gc}. This is similar to grouping hot and cold data, but at the level of \as{gc} for each \as{lba} in a file. SFS~\cite{2012-Min-SFS} implements a \as{gc} policy that accounts for data grouping, with a lower overhead of having to maintain an \as{uft} for each file. It maintains a hotness classification for each block, and combines blocks with k-means clustering~\cite{hartigan1979algorithm}, into groups with similar hotness classification. 

The \as{gc} cost of foreground cleaning in F2FS can take up to several seconds~\cite{2018-Yoo-OrcFS}, as it is not a preemptive task. Implementing preemptive scheduling in kernel file systems is challenging, as during the preemption the kernel has to store the file system state to continue after higher priority tasks have finished, and the state being restored when returning depicts possibly outdated data. Due to the continued writing on the file system, restoring the prior state may no longer be valid, as ongoing writes may have changed file system metadata. OrcFS~\cite{2018-Yoo-OrcFS} implements \af{qpsc}, which sets a maximum time interval $T_{max}$ (default of 100ms). After cleaning a segment it checks if the timer has expired, and if so it checks if outstanding writes are present from the host. If there are outstanding writes, the locks are released and the write is executed, and if there are no outstanding writes the next segment can be cleaned and the timer is reset. This allows any host write command to not encounter a segment cleaning overhead higher than $T_{max}$.

A further drawback of segment cleaning with \as{lfs}, in particular F2FS, is that the modification of metadata during segment cleaning requires a checkpoint to be created after each segment clean. This constitutes to a significant overhead for the segment cleaning process. To avoid the excessive checkpointing after segment cleaning, \af{scj}~\cite{2021-Gwak-SCJ} adds support to F2FS to journal metadata updates made during segment cleaning, instead of creating a checkpoint. This journal is stored in a journal area, which delays the updating of the original metadata until the journal becomes large enough or the checkpointing time interval is reached. However, metadata still points to old invalid data blocks (referred to as \textit{pre-invalid blocks}), which requires that data only be invalidated once the metadata is updated by the \as{scj}. Therefore, \as{scj} implements an adaptive checkpointing that evaluates the cost of checkpointing to flush the metadata updates and the accumulation of pre-invalid blocks, and checkpoints if its cost is lower. 

In addition to utilizing \as{gc} policies to reduce its overheads, the policy can further be used to incorporate management of fragmentation. Park el at.~\cite{2016-Park-LFS_Defragmentation} propose to use a \af{vbq} in which valid blocks are sorted during garbage collection. Typically, a victim segment in the \as{lfs} is selected for cleaning, valid data is copied to the free space at the log head, and the old segment is erased. The \as{vbq} is added such that after a victim segment is selected, it is copied into the \as{vbq}, where the blocks it contains are sorted by the \as{inode} number. Then are the valid blocks written to a new segment and the old one is erased. This sorting allows maintaining of file associated blocks together based on their \as{inode} number. 

A different approach to mitigate \as{gc} overheads is to design the \as{gc} procedure such that accesses do not suffer from high tail latency when \as{gc} is running. TTFlash~\cite{2017-Yan-Tiny_tail} achieves this through several mechanisms. It implements \textit{plane-blocking GC}, which limits any resources that are blocking \as{io}s to only the affected planes on the flash. However, this leads to blocking of requests to the \as{gc} affected planes. Therefore, TTFlash implements \textit{rotating \as{gc}} that only runs at most one \as{gc} operation on a \textit{plane group}. The plane group assignment is based on the possible parallelism of the device, such as plane- and channel-level parallelism. This ensures that a plane group is never blocked for more than a single \as{gc} operation, implying that any request will not be blocked for more than one \as{gc} operation. 

On embedded devices, and in particular mobile devices, in order to save energy the device suspends all threads when not needed (e.g., when the mobile screen is turned off). This implies that all file system threads are also suspended, which means the file system cannot run the background \as{gc} during these inactive sessions. Therefore, \af{sac}~\cite{2015-Park-Suspend_Aware_cleaning-F2FS} is an addition to \as{lfs}, which is the time between the suspend initiation and the suspending of the file system threads, also referred to as the \textit{slack time}. It uses this slack time to run background \as{gc}, however it does not write any data on the flash storage, but instead selects a victim block and brings the still valid pages in the page cache and marks these as dirty. As a result, all pages in the victim block to be dirty, which allows it to be erased. This process is referred to as \textit{virtual segment cleaning}, which however is not run every time the screen is turned off, but rather based on the device utilization, which is called \textit{utilization based segment cleaning}. 

\subsection{Coordinating the Software Stack Layers}\label{sec:software_stack}
\begin{figure}[!t]
    \centering
    \includegraphics[width=0.45\textwidth]{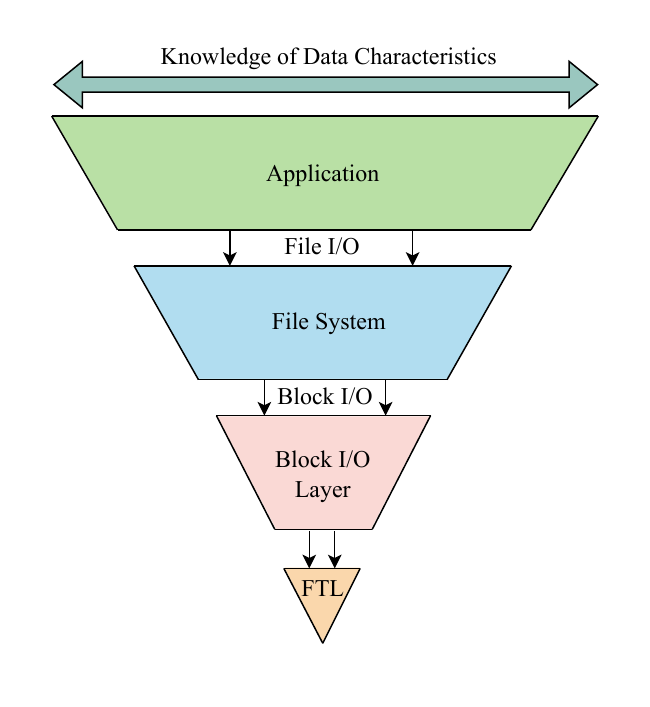}
    \caption{Layers in the storage software stack across which data specific knowledge decreases as \as{io} is passed downwards to the \as{ftl}, representing the semantic gap~\cite{zhang2017flashkv} with flash storage.}
    \label{fig:software_stack}
\end{figure}

The various layers in the storage software stack introduces several redundant operations, such as \as{gc} that is run in the \as{lfs} and \as{gc} run on the storage device. This duplicate work leads to significant performance impacts~\cite{2014-yang-dont_stack_log_on_log}, requiring a co-design of the storage device software and the file system. Qiu et al.~\cite{2013-Qiu-Codesign_FTL_FS} show that the co-design of \as{ftl} and the file system show benefits of reduced memory requirements for \as{l2p} mappings, and increased device parallelism that can be exploited with better file system knowledge of the storage characteristics. In addition to duplicated work, information about data characteristics are lost across the various, since in order to integrate communication across the layers, each utilizes an interface for the other layers to communicate with. However, the expressiveness of the interfaces limits the capability to communicate data characteristics across the layers, as \cref{fig:software_stack} illustrates. Applications have the highest knowledge of data characteristics, how it is best allocated on the flash and managed in order to reduce \as{gc}, however cannot forward all of this information to the file system due to the lack of such \as{api}s. Similarly, the file system may group specific data together, however cannot forward this information to the block layer, and similarly the \as{ftl} may take the submitted \as{io} requests and organize these different on the flash storage. This \textit{semantic gap} between the storage device and storage software is a result of the device integrating into the existing block \as{io} interface~\cite{zhang2017flashkv}, however failing to represent the flash-specific characteristics. Such mismatch between storage device and its accessing interface, requires storage software to enhance capabilities to pass information across the layers to avoid increasing the semantic gap across the layers.

DualFS~\cite{2020-Wu-DualFS} utilizes the custom integration with \as{ocssd} to merge the garbage collection of the file system with that of the \as{ftl}, and present this scheme as \textit{global garbage collection}. ParaFS~\cite{2016-Zhang-ParaFS} implements a similar coordination of file system \as{gc} and \as{ftl} \as{gc}. A different approach taken by Lee at al.~\cite{2016-Lee-AMF} is to modify the block interface with the flash characteristics, moving responsibility directly to the file system, or other application built on top of it. The resulting interface called \af{amf} exposes a block interface that does not allow overwriting unless an explicit erase is issued for the blocks. This matches the flash requirement that prior to overwriting data it has to be erased. The interface is implemented as a custom \as{ftl}, called AFTL, on top of which the ALFS file system is built. This avoids the duplicate garbage collection of the \as{ftl} and the file system, as the garbage collection of the ALFS erases blocks during garbage collection, informing the \as{ftl} to erase the physical block.

Co-designing the FTL and the file system allows removing uncoordinated duplicate work, and coordinate the flash management. To this end, Lee et al.~\cite{lee2014refactored} present a redesigned \as{io} architecture, called REDO, which avoids the duplicate operations from file system and \as{ftl} by implemented the new framework directly as the storage controller, and building the \af{rfs} on top of the new controller interface. By combining the file system operations with the storage controller, the file system is responsible for running \as{gc} and managing the storage by maintaining the \as{l2p} mappings.

\subsection{Summary}
With \as{gc} being a significant performance bottleneck for flash storage, adequate data grouping and effective \as{gc} policies allows to reduce and manage the \as{gc}. Furthermore, the complexity of storage system software lacks support for effective \as{api}s for coordination on data placement between the storage stack layers. Such coordination, allows to reduce the semantic gap and eliminate redundant and duplicated operations of the various layers.

\section{FIC-3: \as{io} Amplification}\label{sec:io_amplification}
Several of the applied mechanisms for dealing with asymmetric flash performance and garbage collection are key mechanisms to eliminate \as{io} amplification. \cref{tab:io_amplification} shows the different mechanisms that can be applied to lower the various types of \as{io} amplification, including \af{wa}, \af{ra}, and \af{sa}. These include the benefit of write buffering, which in the case of write requests that are smaller than the flash allocation unit, avoids the unnecessary \as{wa} to fill the flash allocation unit. Similar buffering as is employed in \as{wods}, such as the $\text{B}^\varepsilon$-trees, allows decreasing the \as{wa}. Another key mechanism is reducing the generated write traffic, which minimizes \as{wa}, \as{ra}, and \as{sa} through deduplication, compression, delta-encoding, and virtualization. Similarly, the discussed methods of grouping data, avoiding fragmentation, allows reducing the \as{ra} to locate data, and furthermore limits \as{gc} overheads and \as{gc} caused \as{wa}. Throughout this section we evaluate the additional methods for limiting the various types of \as{io} amplification.

\begin{table}[!t]
    \centering
    \crefformat{section}{\S#2#1#3}
    \crefformat{subsection}{\S#2#1#3}
    \crefformat{subsubsection}{\S#2#1#3}
    \begin{tabular}{||p{30mm}|p{20mm}|p{20mm}||}
        \hline 
        Mechanism & Amplification Type & File Systems \\
        \hline
        \cellcolor{lightgreen!75} Write Optimized Data Structures (\cref{sec:wods}) & \cellcolor{lightgreen!75}\as{wa} & \cellcolor{lightgreen!75}\cite{2020-Tu-URFS,hunter2008brief,rodeh2013btrfs,2022-Jiao-BetrFS,dubeyko2019ssdfs} \\
        \hline
        \cellcolor{lightgreen!75} Write Buffering (\cref{sec:write_buffering}) & \cellcolor{lightgreen!75}\as{wa} & \cellcolor{lightgreen!75}\cite{park2006cflru,jo2006fab,park2013enffis,2018-Kannan-DevFS,josephson2011direct,2019-Lu-Sync_IO_OCSSD,2010-Josephson-DFS,2011-Park-Multi_NAND} \\
        \hline
        \cellcolor{lightgreen!75} Reducing Write Traffic (\cref{sec:reduce_write_traffic}) & \cellcolor{lightgreen!75}\as{wa}, \as{ra}, \as{sa} & \cellcolor{lightgreen!75}\cite{huang2013improve,2011-Lim-DeFFS,woodhouse2001jffs,dubeyko2019ssdfs,2021-Ji-F2FS_compression,2007-Hyun-LeCramFS,ning2011design,2009-Zuck-NANDFS,2022-Lee-F2FS_Address_Remapping} \\
        \hline
        \cellcolor{lightgreen!75} Data Grouping (\cref{sec:data_grouping}) & \cellcolor{lightgreen!75}\as{wa}, \as{ra} & \cellcolor{lightgreen!75}\cite{2006-Lim-NAND_fs,2022-Zhang-ELOFS,2020-Zhang-LOFFS,2015-Changman-f2fs,2018-Rho-Fstream} \\
        \hline
        \cellcolor{lightgreen!75} \as{gc} Policies (\cref{sec:gc_policy}) & \cellcolor{lightgreen!75}\as{wa}, \as{ra} & \cellcolor{lightgreen!75}\cite{2012-Min-SFS,2018-Yoo-OrcFS,2021-Gwak-SCJ,2015-Park-Suspend_Aware_cleaning-F2FS} \\
        \hline
        Space Optimized Data Structure (\cref{sec:space_optimized_ds}) & SA & \cite{2014-Lu-ReconFS,2015-Changman-f2fs,2021-Liao-Max} \\
        \hline
        \as{wa} with Coarse Granularity Flash Mappings (\cref{sec:coarse_gran}) & \as{wa} & \cite{2018-Yoo-OrcFS,kim2006mnfs,2018-Yoo-OrcFS,lee2007log} \\
        \hline
        Reverse Indexing (\cref{sec:reverse_indexing}) & \as{wa} & \cite{2014-Lu-ReconFS}\\
        \hline
    \end{tabular}
    \crefformat{section}{Section #2#1#3}
    \crefformat{subsection}{Section #2#1#3}
    \crefformat{subsubsection}{Section #2#1#3}
    \caption{Mechanisms for file systems to deal with \as{io} amplification caused by flash storage integration, and the respective file systems that implement a particular mechanism. Green highlighted table cells depict previously discussed mechanisms with their respective section.}
    \label{tab:io_amplification}
\end{table}

\subsection{Space Optimized Data Structures}\label{sec:space_optimized_ds}
Similar to the design of a \as{wods}, data structures with particular focus on optimally utilizing available space are a mechanism to deal with \as{sa}. A commonly applied method for file systems to optimize space utilization is to possibly embed file data in the \as{inode}. Commonly the allocation of the file system \as{inode} occupies at least a block, such as 4KiB in F2FS, which however is more space than file metadata requires. Therefore, several bytes ($\sim$3.4KB in F2FS) are free, which are used to inline file data in the inode. This particularly allows for small files to entirely fit into the \as{inode}, avoiding writing the \as{inode} and leave the unused space empty, and additionally write an additional data block, which also has free space. Different to inline data, ReconFS~\cite{2014-Lu-ReconFS} uses an \as{inode} with a size of 128B, allowing to place numerous inodes in a single flash page. With such an \as{inode} size, writing each \as{inode} change directly requires filling the flash page with unnecessary data. Therefore, ReconFS implements a \textit{metadata persistent log}, in which metadata changes are logged and compacted to align with pages, and are only written back to the storage when evicted or checkpointed, in order for the file system to remain consistent.

Similar to effective tree-based \as{wods}, the radix tree is a space optimized tree variant, that is commonly used as the directory and inode tree for file systems~\cite{2015-Changman-f2fs,2021-Liao-Max}. The directory tree is commonly constructed and maintained in memory and written to the persistent flash storage. Its space optimization revolves around merging of nodes that have a single child with that child node. This eliminates the need for an individual node that is assigned to each child, lowering the space requirement. As a result, the radix tree is also referred to as a compressed tree, due to the compression of single child nodes.

\subsection{Coarse Granularity Flash Mappings}\label{sec:coarse_gran}
A similar goal of file systems is to reduce the amount of memory that is required for the mapping table to maintain the \as{l2p} mappings. \as{l2p} mappings are commonly persisted periodically, from the storage device \as{ram} to the flash storage, such that in the case of system shutdown or the device is unplugged, upon reconnection the mapping information can be recovered. Hence, the mapping information similarly requires flash pages to be stored. A common solution is to increase the granularity of the mapping table (e.g., block-level mapping instead of page-level mapping), requiring fewer mappings. MNFS~\cite{kim2006mnfs} manages flash storage with page-level and block-level mapping, depending on the update frequency. Metadata is updated more frequently and therefore utilizes a page-level mapping compared to larger mapping granularity for data. OrcFS~\cite{2018-Yoo-OrcFS} similarly utilizes a page-level for metadata, and a superblock-level mapping for data, which represents several flash blocks. The allocation unit is called a superblock as it consists of multiple blocks (not to be confused with the file system superblock). Furthermore, logical addresses are mapped to the same physical addresses in the data partition, requiring no mapping table, and file system sections are aligned to the superblock unit. Therefore, OrcFS only requires a block allocation information for each file in the superblock, which are stored in the inode block in the metadata area. 

However, this comes at the cost of having a larger allocation unit, and if a host write is smaller than the allocation unit it causes \as{wa}, due to the partial flash page write when the flash page size and the allocation unit are not aligned~\cite{2018-Yoo-OrcFS,lee2007log}. OrcFS~\cite{2018-Yoo-OrcFS,lee2007log} implements \textit{block patching} to solve this issue. It takes write requests that are smaller than the flash page size and pads the remaining space with dummy data to align the write request to flash page size. This mechanism avoids copying data if a flash page is partially written, and the next \as{lba} in the same flash page is written, which triggers a copy of all \as{lba}s in the flash page followed by writing the \as{lba} for the new write. For instance, for a flash page containing 4 \as{lba}s, if \as{lba}s 1-3 are written by one request, the first 3 \as{lba}s are mapped to the data and the fourth holds dummy data, such that the page is fully filled. If a second request to \as{lba} 4 is issued, it cannot fill the flash page as it has already been written. Therefore, to write the newly written data after the already written \as{lba}s, it must copy \as{lba}s 1-3, append the new write to \as{lba} 4, and write the 4 \as{lba}s to a new flash page. Adding of dummy data to fill pages reduces the \as{wa}, which would be caused by copying of all data in the flash page, as it now avoids copying the added dummy data on consecutive writes. While reduction in \as{wa} are presented, the adding of dummy data nonetheless adds \as{wa} to fill the flash page. However, as latter updates require less data written, and the importance of data grouping indicates, maintaining related data in the same flash page is more beneficial and possibly decreases future \as{wa}. Related data remains in the same flash page, as only the valid data in flash pages is copied on writes, as opposed to copying the entire flash page, introducing copied dummy data.

\subsection{Reverse Indexing}\label{sec:reverse_indexing}
As file system metadata is commonly maintained in a tree-based data structure, updates to metadata in the leaf nodes can propagate changes to the root node, known as the Wandering Tree Problem~\cite{bityutskiy2005jffs3}. Due to the update of leaf metadata, such that when file data is modified, the metadata points to the new location of the file data, causing new metadata to be written, which in turn requires its parent to be updated to point to the new location of the metadata. This propagates up to the root note, causing significant \as{wa}. F2FS utilizes a table based indexing, with the \as{nat}, such that only a table entry is required to change to update the data location, and metadata points to the table entry to locate the data. ReconFS~\cite{2014-Lu-ReconFS} utilizes an inverted indexing tree, which also avoids the wandering tree problem. With such a tree, each node points to its parent node, instead of the parent node pointing to a child node. Therefore, upon address change of a child node, the parent does not need to be modified, since the child node points upwards to the parent node. Similar mechanisms are utilized in \as{ftl} design~\cite{2013-Lu-Flash_Lifetime_Reduce_With_WA}, where indexing data is written in the \as{oob} space of the flash page from the data, in order to locate its metadata. In order to avoid increased scan times on failure recovery, which can no longer traverse the tree from the root, the updated pages are tracked to locate the most recently updated valid page, which is then periodically included in the checkpointing to ensure consistency. 

\subsection{Summary}
The introduced \as{io} amplification of flash storage, particularly a result of \as{gc}, requires careful consideration to reduce the write requests, such that the flash device lifetime can be extended. Several of the previously discussed mechanisms, such as reducing write traffic and utilizing effective \as{gc} policies, aid in reducing the \as{io} amplification, however furthermore particular data structures optimized for space utilization similarly provide efficient methods for reducing \as{io} amplification.

\section{FIC-4: Flash Parallelism}\label{sec:flash_parallelism}
With the capabilities of flash storage relying largely on increased parallelism, several existing mechanisms are leveraging these. Depending on the level of flash integration, different mechanisms are possible, where at the \as{ssd} integration the host has not control over the possible physical parallel utilization of flash, as the \as{ftl} controls this, however deeper flash integration at the custom and embedded levels provide more possibility. \cref{tab:flash_parallelism} depicts the various mechanisms to aid the utilization of flash parallelism and exploit the physical characteristics of flash.

\begin{table}[!t]
    \centering
    \crefformat{section}{\S#2#1#3}
    \crefformat{subsection}{\S#2#1#3}
    \crefformat{subsubsection}{\S#2#1#3}
    \begin{tabular}{||p{40mm}|p{35mm}||}
        \hline 
        Mechanism & File Systems \\
        \hline
        \hline
        \cellcolor{lightgreen!75}Aligning the Allocation Unit (\cref{sec:align_alloc}) & \cellcolor{lightgreen!75}\cite{2020-Tu-URFS,2009-Park-Multimedia_NAND_Fs} \\
        \hline
        Clustered Allocation \& Striping (\cref{sec:clustered_alloc}) & \cite{2012-Min-SFS,dubeyko2019ssdfs,2018-Yoo-OrcFS,2016-Zhang-ParaFS,2011-Park-Multi_NAND,manning2010yaffs,aleph2001yaffs,manning2002yaffs} \\
        \hline
        Concurrency (\cref{sec:concurrency}) & \cite{2021-Liao-Max,2018-Kannan-DevFS,2015-Kang-SpanFS,2019-Lee-Parallel_LFS,2016-Lee-AMF,2016-Zhang-ParaFS} \\
        \hline
    \end{tabular}
    \crefformat{section}{Section #2#1#3}
    \crefformat{subsection}{Section #2#1#3}
    \crefformat{subsubsection}{Section #2#1#3}
    \caption{Mechanisms for file systems to exploit flash parallelism capabilities, and the respective file systems that implement a particular mechanism. Green highlighted table cells depict previously discussed mechanisms with their respective section.}
    \label{tab:flash_parallelism}
\end{table}

\subsection{Clustered Allocation \& Striping}\label{sec:clustered_alloc}
As host software has no direct access to flash storage with \as{ssd}, the \as{ftl} implements and manages all device-level parallelism. The possibility for the host to utilize flash parallelism comes from aiding the \as{ftl} in providing large enough \as{io}s such that the \as{ftl} can stripe data across flash chips and channels. This is achieved with \textit{clustered blocks/pages}~\cite{2012-Kim-clustered_blocks}, where blocks or pages on different units (such as blocks on different planes) are accessed in parallel. The \as{ftl} can possibly stripe data across these clustered blocks, given that the \as{io} request is large enough to fill the clustered unit. Such a mechanism aligns with prior discussed aligning of the allocation unit (\cref{sec:align_alloc}) to a physical unit to reduce \as{io} amplification, such as making the file system segment unit a multiple of the flash allocation unit. SFS~\cite{2012-Min-SFS} takes advantage of the achieved device-level parallelism with clustered blocks by aligning segments to a multiple of the clustered block size. During garbage collection SFS ensures that cleaning of segments, that do not have enough blocks to fill the clustered block size, is delayed until enough data is present.

SSDFS~\cite{dubeyko2019ssdfs} utilizes the custom flash integration to map data allocation of segments to the unit of a \af{peb}, where the \as{peb} is split over the parallel unit on the device, such as previously mentioned parallel erasing of flash blocks over channels. Therefore, utilizing \as{peb}s over the parallel unit allows striping writes into a segment over the varying channels, increasing the device parallelism. In order to achieve this, \as{io} requests have to be large enough such that they can be striped across the channel and fill the \as{peb}s mapped to the segment. For this, SSDFS utilizes aggressive merging of \as{io} requests to achieve the larger \as{io}s that can be striped across the parallel units (solving \textbf{\as{fic}7}). Instead of merging \as{io} requests, OrcFS~\cite{2018-Yoo-OrcFS} increases its file system allocation unit to a \textit{superblock} (not to be confused with the file system superblock), which represents a set of flash blocks. These flash blocks are then split over the parallel units of the flash storage by the file system for increased parallelism. The file system utilizes a custom flash integration, and hence is capable of managing the parallelism of the flash storage.

The large allocation unit however introduces increased \as{gc} overheads, and block-level striping has lower performance than page-level striping~\cite{2016-Zhang-ParaFS}. Therefore, ParaFS~\cite{2016-Zhang-ParaFS} implements a 2-D allocation scheme, with page-level striping over the flash channels, where striping is also based on the data hotness, hence having a 2-dimensional allocation scheme. Different groups are assigned for the hotness levels, where writes are issued to the corresponding hotness group striped over the flash channels. Several other file systems implement variations of striping across different parallel units on flash storage~\cite{2011-Park-Multi_NAND,manning2010yaffs,aleph2001yaffs,manning2002yaffs}. These mechanisms are also present in the design of storage applications, such as key-value stores~\cite{zhang2017flashkv,wang2014efficient}.

\subsection{Concurrency}\label{sec:concurrency}
Similar to increasing the data allocation unit, concurrency is a mechanism to exploit the flash parallelism. \as{lfs} design relies on a single append point at the head of the log, in the simplistic implementations, depending on methods such as locking to ensure only one write is issued at the log head. This has a significant impact on the performance where other \as{io}s are idle while a single \as{io} completes. F2FS however suffers from severe lock contention overheads, where the performance of the multi-headed logging is nearly fully deprecated due to the serialization of data updates~\cite{2021-Liao-Max}. In particular, as data has to be written persistently before \as{inode} and other metadata can be written. Furthermore, F2FS suffers from contention of the in-memory data structures, for which it uses reader-writer semaphores for read and write operations from the user (termed \textit{external \as{io} operations}), and reader-writer locks for writing of checkpoints and other metadata (termed \textit{internal operations})~\cite{2021-Liao-Max}. As lock counters are shared among all cores, cache coherence adds a significant overhead that increases with more cores. Max~\cite{2021-Liao-Max} extends F2FS to increase the concurrency scalability with three main modifications.

Firstly, in order to eliminate cache coherence overheads, it introduces a \af{rps} that uses a per-process counter. Secondly, the shared data structures in memory are partitioned by the \as{inode}, such that concurrent accessing does not require locking on parts of the radix tree, but instead on an \as{inode} basis. Lastly, it utilizes multiple independent logs, called a \af{mlog}, which are accessed concurrently. The different between \as{mlog} and multi-headed logging in F2FS is that atomic data blocks are mapped to the same \as{mlog}, eliminating the need to ensure concurrency control across different logs. Ordering for persistence, ensuring data blocks are written before metadata, is delegated to the recovery mechanism using a global versioning number in each inode to identify ordering across \as{mlog}s, and recover the most recent version number in case of a system crash. These mechanisms eliminate much of the needed concurrency control, which sequentialized major parts of operations and hindered multicore scalability.

With this increased concurrency capabilities, the file system can issue more \as{io} requests to the device, allowing to leverage a higher degree of on-device parallelism. Similarly, DevFS~\cite{2018-Kannan-DevFS} utilizes the parallel capabilities by exploiting the high number of \as{io} queues supported by \as{nvme}. It maps \as{io} queues to individual files, allowing single file operations to submit \as{io}s concurrently without interfering on the \as{io} queue, therefore increasing the per-file concurrency as well. Likewise to the concept of \as{mlog}, SpanFS~\cite{2015-Kang-SpanFS} maps files and directories to different \textit{domains}, such that individual domains can be accessed in parallel. Such methods have a higher lock granularity, where concurrency below the lock granularity is not possible, as a single process is holding the lock. Therefore, instead of holding locks for individual inodes or files, preventing concurrent writing to the same inode or file, Lee et al.~\cite{2019-Lee-Parallel_LFS} extend F2FS to utilize \af{rl}, in which ranges of a file are locked, and different ranges can be written concurrently. Therefore, it provides the possibility for intra-file parallelism. 

ALFS~\cite{2016-Lee-AMF}, exploits the flash parallelism by mapping consecutive file system segments to the flash channels and utilizing different \as{io} queues for each flash channel. Similarly, ParaFS~\cite{2016-Zhang-ParaFS} implements \textit{parallelism-aware scheduling}, which also maintains different \as{io} queues for each channel. However, it extends this concept by using a \textit{dispatching phase} and a \textit{request scheduling phase}. The dispatching phase optimizes write \as{io}s by scheduling \as{io} requests to the respective channels based on the utilization of the channel, such that the least utilized channels receives more requests. All requests are assigned a weight, which indicates their priority in the queue, where read requests weight is lower than that of write requests, because of the asymmetric performance of flash storage. During the request scheduling phase the scheduler assigns slices to the read and write/erase operations in the individual queues, such that if the time from the slice of a read operation is up and the queue contains no other read requests, a write or erase is scheduled, based on a fairness formula that incorporates the amount of free space in the block and concurrently active erase operations on other channels. This allows to minimize the erase operations on the flash, giving always free channels to utilize and maintain a fair schedule between write and erase operations.

\subsection{Summary}
Due to the architecture of flash storage providing a high degree of parallelism, numerous methods are employed to leverage these parallel units in order to maximize the performance. Depending on the level of flash integration, particular design choices can be made, such as clustered allocation and striping can be achieved with flash \as{ssd} integration, by providing large write \as{io} requests such that the \as{ftl} can stripe the data across parallel units. With a higher degree of control over the flash storage, file systems can directly rely on utilizing concurrency to leverage the parallelism of flash storage.

\section{FIC-5: Wear Leveling}\label{sec:wear_leveling}
As flash cells wear out over time, it is important to utilize the flash evenly to avoid burning out particular flash cells faster than others. The possibility for ensuring even wear at the different levels of integration is limited, as at the \as{ssd} flash integration level, the \as{ftl} handles all wear leveling, without host considerations. However, similar to prior flash integration challenges, several mechanisms are nonetheless applicable. In particular, reducing the write traffic to the flash device, as less writing incurs less flash wear, and particularly flash-specific data structures inherently provide a degree of wear leveling. Based on the sequential write requirement, data structures, such as the \as{lfs} must write sequentially in an append-only fashion, which evenly writes the space. At closer to flash integrations, where host systems have more control over the flash management, there are particular mechanisms to ensure better flash wear. \cref{tab:wl} depicts the methods we discuss in this section for enabling increased wear leveling for file systems.

\begin{table}[!t]
    \centering
    \crefformat{section}{\S#2#1#3}
    \crefformat{subsection}{\S#2#1#3}
    \crefformat{subsubsection}{\S#2#1#3}
    \begin{tabular}{||p{50mm}|p{25mm}||}
        \hline 
        Mechanism & File Systems \\
        \hline
        \hline
        \cellcolor{lightgreen!75} Write Optimised Data Structures (\cref{sec:wods}) & \cellcolor{lightgreen!75} \cite{2020-Tu-URFS,hunter2008brief,rodeh2013btrfs,2022-Jiao-BetrFS,dubeyko2019ssdfs} \\
        \hline
        \cellcolor{lightgreen!75} Reducing Write Traffic (\cref{sec:reduce_write_traffic}) & \cellcolor{lightgreen!75} \cite{huang2013improve,2011-Lim-DeFFS,woodhouse2001jffs,dubeyko2019ssdfs,2021-Ji-F2FS_compression,2007-Hyun-LeCramFS,ning2011design,2009-Zuck-NANDFS,2022-Lee-F2FS_Address_Remapping} \\
        \hline
        \cellcolor{lightgreen!75} \as{gc} Policies (\cref{sec:gc_policy}) & \cellcolor{lightgreen!75} \cite{2012-Min-SFS,2018-Yoo-OrcFS,2021-Gwak-SCJ,2015-Park-Suspend_Aware_cleaning-F2FS} \\
        \hline
        Write \& Read Leveling (\cref{sec:write_leveling,sec:read_leveling}) & \cite{2006-Lim-NAND_fs,2016-Lee-AMF,2020-Wu-DualFS,2009-Sungjin-FlexFS} \\
        \hline
    \end{tabular}
    \crefformat{section}{Section #2#1#3}
    \crefformat{subsection}{Section #2#1#3}
    \crefformat{subsubsection}{Section #2#1#3}
    \caption{Mechanisms for file systems to deal with wear leveling of the flash storage, and the respective file systems that implement a particular mechanism. Green highlighted table cells depict previously discussed mechanisms with their respective section.}
    \label{tab:wl}
\end{table}

\subsection{Write Leveling}\label{sec:write_leveling}
Several flash integration challenges proved data grouping to be an effective method for dealing with \as{gc} overheads and \as{io} amplification. However, this can have an effect on the flash storage. In particular the hot data, such as file system metadata which is more frequently updated and written, must be moved across the flash space more than cold data. CFFS~\cite{2006-Lim-NAND_fs} therefore switches the allocation areas for metadata and data blocks, such that an erased metadata block becomes a free data block. Therefore, cold data should be placed in blocks that have been written more frequently, whereas hot data should be placed in blocks with a lower write count history. The principle of migrating cold data from less written blocks, which are also referred to as \textit{younger blocks}, to more frequently written, \textit{older blocks}, is referred to as \textit{cold-data migration}, and similarly moving hot data from old blocks to younger blocks is known as \textit{hot-data migration}~\cite{chang2007efficient}. These methods are commonly used in \as{ftl} implementations due to their simplicity and effectiveness, and similarly in file systems such as ALFS~\cite{2016-Lee-AMF}.

Wear leveling is an increasingly vital concern on file systems that utilize flash dual mode~\cite{2020-Wu-DualFS,2009-Sungjin-FlexFS}, where it switches the flash level to increase performance for critical \as{io} requests. Due to the lowering in flash cell level, the same amount of written data requires a larger amount of space, where a switch from \as{mlc} to \as{slc} divides the capacity in half, requiring double the space for the same \as{io} request. Therefore, these file systems include a \textit{write budget} that is maintained for the areas, and dynamically resizes the available lower cell level area, such as decreasing the space if the wear is reaching a threshold. This switches the cell level back to a higher number, allowing to write more with less wear. Additionally, the file systems utilize the wear budget in order to identify if an \as{io} request should be redirected to the larger cell area, instead of being written to the lower cell level area.

\subsection{Read Leveling}\label{sec:read_leveling}
While write operations are the major cause of flash cells burning out, read operations also pay a toll on flash cells, as the current flash cell technology utilizes flash cells that are only capable of holding very few electrons (determining the charge of the gate) due to their size~\cite{lu2009future,shin2005non}. This makes the cells increasingly susceptible to \textit{read disturbance}~\cite{2015-Liu-Read_leveling}, where reading of a page results in shifting of voltages in nearby cells (typically in the same block), requiring frequent rewriting to ensure the charge stays consistent. In order to control read disturbance, Liu et al.~\cite{2015-Liu-Read_leveling} propose to read-leveling mechanisms in the FTL. While their proposal is aimed at \as{ftl} implementations, the ideas are applicable to file systems for flash storage devices. With the proposed read-leveling, the read-hot data, that is read more frequently, is isolated from other data pages by placing the hot pages into \textit{shadow blocks}, which contain no valid data, in order to avoid disturbing that data. However, this requires to identify the read-hot pages, where a tracking of read counters for each page would require significant resources. Therefore, a \textit{second-chance monitoring strategy} is proposed, which initially tracks the reads for each block, therefore requiring counters at a higher block granularity, and upon reaching a threshold indicating the block contains read-hot pages, the individual pages in the block are tracked on their read counters. Finally, the pages in these blocks that reach a certain threshold are copied to the shadow blocks. Therefore, this avoids the tracking of read counters for individual pages and only copies read-hot pages into the shadow blocks. While this strategy requires copying of read-hot pages to shadow blocks, it minimizes read disturbance which in turn minimizes the \as{wa} it causes.

\subsection{Summary}
In addition to previously discussed mechanisms to reduce \as{io} amplification and \as{gc}, resulting in decreased wear of the flash storage, write leveling, ensuring that write requests are spread across the available storage space, and read leveling are important mechanisms for ensuring the longevity of the flash storage.

\section{FIC-6: \as{io} Management}\label{sec:io_sched}
Given that flash storage has the capabilities to achieve single digit $\mu$-second latency, whereas overheads in the software stack, such as context switching in the kernel caused by system calls, can already require $\mu$-seconds to complete~\cite{soares2010flexsc}, making software the dominating factor in overheads~\cite{2010-Caulfield-Moneta,2012-Caulfield-Fast_User_Space_Access,foong2010towards,seppanen2010high,vasudevan2012using}. In addition, \textit{interrupts} cause significant overheads for systems. Aimed at slow storage devices, the \as{io} request is submitted to the device, the context is switched, such that the process can continue with other work, and upon completion of the \as{io}, the host is interrupted, and the context is switched again. Any added interrupt on the \as{io} path can cause significant delays~\cite{2014-Shin-OS_IO_Path}. Cache effects are another drawback of context switching, since other work is continued, replacing data in the caches, it requires bringing the replaced data back into the caches after the interrupt and resuming of the prior context. Similarly, it also causes \af{tlb} pollution on the host system. A different approach to submitting \as{io} requests is with \textit{polling}, which eliminates the need for context switches. With polling, the \as{io} request is submitted and instead of continuing other work, the process regularly checks the \as{io} for completion. Using polling for \as{io} requests has been shown to be a favored method of building application for fast storage devices~\cite{kourtis2019reaping,yang2012poll,didona2022understanding}.

Particular for fast storage devices, with the utilization of \textit{synchronous \as{io}}, to saturate the storage device, additional threads are required, which each submit \as{io} requests to the numerous \as{io} queues in the of the storage device. However, this mechanism does not scale efficiently, where each thread must wait until the \as{io} request is completed, and thread scheduling overheads are introduced. Therefore, with asynchronous \as{io} the threads do not wait for completion, but instead submit a larger number of \as{io} requests each, allowing to fill the device \as{io} queues more effectively. The \as{io} requests for which a thread as submitted a request, but have not completed, are referred to as \textit{outstanding} or \textit{in-flight} \as{io} requests. \cref{tab:io_sched} shows the mechanisms for host systems to better leverage the flash storage performance and minimize overheads. In addition to file systems implementing particular mechanisms, we discuss more general methods applicable to all applications for benefiting from flash storage and enhancing performance.

\begin{table}[!t]
    \centering
    \crefformat{section}{\S#2#1#3}
    \crefformat{subsection}{\S#2#1#3}
    \crefformat{subsubsection}{\S#2#1#3}
    \begin{tabular}{||p{40mm}|p{35mm}||}
        \hline 
        Mechanism & File Systems \\
        \hline
        \hline
        \as{io} Operations (\cref{sec:io_ops}) & \cite{2020-Tu-URFS} \\
        \hline
        \as{io} Scheduler (\cref{sec:io_scheduler}) & \cite{2021-Qin-Atomic_Writes} \\
        \hline
        \as{io} Path - User-Space File Systems (\cref{sec:user_space_fs}) & \cite{2020-Tu-URFS,2019-Yoshimura-EvFS,2018-Kannan-DevFS,2019-Liu-fs_as_process} \\
        \hline
    \end{tabular}
    \crefformat{section}{Section #2#1#3}
    \crefformat{subsection}{Section #2#1#3}
    \crefformat{subsubsection}{Section #2#1#3}
    \caption{\as{io} scheduling mechanisms to exploit performance capabilities of flash storage, and the respective file systems that implement a particular mechanism.}
    \label{tab:io_sched}
\end{table}

\subsection{\as{io} Operations}\label{sec:io_ops}
Given that particular \as{io} patterns can have degrading affects on the \as{ssd} performance, such as mixing read and write operations, as they share resources on the device, including the mapping table and \as{ecc} engine, and furthermore possibly invalidating the cached data in the \as{ssd} \as{ram}. Similarly, mixing \as{io} operations with different block sizes can result in increased fragmentation~\cite{2020-Tu-URFS}. As the Linux kernel relies on a submission and completion queue for \as{io}, user-space frameworks such as \as{spdk} and NVMeDirect provide more flexibility for user-space file systems to design different queues, depending on the requirements. URFS~\cite{2020-Tu-URFS} utilizes this possibility to create adaptive queues that can better optimize \as{io} submissions to the device. Based on the workload characteristics URFS dynamically creates flexible \as{io} queues (e.g., group by size, read/write operation) to increase \as{ssd} performance. Similarly, Borge et al.~\cite{2019-Borge-SSD_read_variability} show with a case study on HDFS performance with \as{ssd}, that in order to leverage the capabilities of flash \as{ssd}, direct \as{io}, and increased parallel requests with buffered \as{io} are needed. 

\subsection{\as{io} Scheduler}\label{sec:io_scheduler}
With the possibility for asynchronous \as{io} to merge and reorder requests, the Linux kernel implements several schedulers, such as \textit{NOOP}, \textit{deadline}, and \textit{CFQ}~\cite{sun2014exploring,pratt2004workload,moallem2008study,heger2010linux}. NOOP being the least intrusive scheduler only merges \as{io} requests in its \as{fifo}, but does not reorder them, which is beneficial on devices such as \as{ocssd}, that require consecutive LBAs. Son et al.~\cite{2015-Son-Optimizing_FS} showcase the benefits of merging random write requests, regardless of contiguity of the LBAs, in order to better enhance performance with fast storage devices. The deadline scheduler adds to NOOP by utilizing merging and reordering, however also applies a deadline for each \as{io} request to ensure requests are submitted to the device eventually. Two separate queues, one for read requests and an additional one for write requests are utilized, which are both ordered by the deadline of the request. Another scheduler variant of deadline exists, called \textit{mq-deadline}, which is aimed at multi-queue devices, such as \as{nvme} \as{ssd}. \af{cfq} implements a round-robin based queue that assigns time slices to each in order to prevent starvation and provide fairness. While these are the common schedulers in the Linux kernel, to see details on all schedulers present in the Linux kernel consult~\cite{2019-ubuntu_wiki_io_schedulers}. Such scheduling configuration begs the question on which scheduler is best suited for file systems on flash storage. Several studies into performance of the schedulers exist~\cite{sun2014exploring,yu2014optimizing}, showcasing that merging of read \as{io} requests in synchronous \as{io} provides beneficial performance gains, and similarly the merging of write \as{io}s in asynchronous \as{io} shows performance gains.

Qin et al.~\cite{2021-Qin-Atomic_Writes} argue that \as{io} ordering limits exploiting the parallelism of flash devices. Especially as the Linux block layer does not guarantee particular ordering, flags such as \af{fua}, indicating that \as{io} completion is signaled upon arrival of data on the storage, and \textit{PREFLUSH}, which before completing the request flushes the volatile caches, have to be set in order to ensure a specific ordering~\cite{2021-Qin-Atomic_Writes}. With file systems, the \as{io} of metadata and data has a particular ordering, such that metadata can only point to data that exists, needing to ensure that data is written prior to metadata. Removing of \as{io} ordering allows eliminating this need and better utilize the flash parallelism. Utilizing the \as{oob} area on flash pages, the file system developed by Qin et al., called NBFS, maintains versioning in order to identify out of order updates. Furthermore, updates are done using atomic writes (discussed in Section~\ref{sec:failure_consistency}). The issuing of \as{fua} requests further implies that its \as{io}s cannot be merged in the scheduler~\cite{2021-Qin-Atomic_Writes}, implying that if a smaller than flash page size \as{fua} \as{io} request is issued, it is padded to the page size, causing \as{wa}. NBFS solves this with its atomic writes that imply that the \as{fua} request does not immediately have to be written to the flash, but instead wait for all data blocks to arrive, which are then used to fill the pages, allowing to reduce the \as{wa} (solving \textbf{\as{fic}3}).

\subsection{\as{io} Path - User-Space File Systems}\label{sec:user_space_fs}
A mechanism that is gaining significant attention in the research community is the utilization of user-space file systems, bypassing the kernel layers and avoid its associated overheads. These file systems run only in the user-space, as opposed to commonly used file systems (e.g., F2FS) running in kernel space. In addition to the benefit of avoiding kernel overheads, user-space file systems are easier to develop, have increased reliability and security by avoiding kernel bugs and vulnerabilities, and provide increased portability~\cite{2015-Tarasov-User_space_fs_practicality}. A widely adopted framework for building user-space storage applications is \af{fuse}~\cite{szeredi2010fuse}. It is implemented over a kernel module with which it exports a virtual file system from the kernel, where data and metadata are provided by a user-space process, hence allowing user-space applications to interact with it. Since \as{fuse} is implemented with a kernel module, \as{fuse} based file systems suffer significant performance penalties, requiring more \as{cpu} cycles than file systems in kernel space. Particularly contributing to overheads is the need to copy memory between user-space and kernel-space, caused by the way \as{fuse} handles \as{io} requests~\cite{2019-Vangoor-Fuse_performance,vangoor2017fuse}. Furthermore, \as{fuse} still suffers from context switching~\cite{vangoor2017fuse,2019-Vangoor-Fuse_performance,rajgarhia2010performance} overheads and \af{ipc} between the \as{fuse} kernel module and \as{fuse} user-space daemon~\cite{zhu2018direct}. 

A similar framework for building user-space applications with direct storage access is \as{nvme}Direct~\cite{2016-Kim-NVMe_Direct}. However, it also relies on a kernel driver to provide enhanced \as{io} policies. \as{spdk}~\cite{2017-Yang-SPDK} is another framework for building user-space storage applications, however it provides the mechanisms to bypass the kernel and submit \as{io} directly to the device, by implementing a user-space driver for the storage device. Such a framework allows building high performance storage applications in user-space, which eliminate the overheads coming from the kernel \as{io} stack. 

URFS~\cite{2020-Tu-URFS} provides increased concurrency performance by implementing a multi-process shared cache in memory, in order to avoid the kernel overheads of copying data as is present in \as{fuse}. It furthermore helps avoid contention on the storage device. Eliminating of data copy is also addressed in ZUFS (zero-copy user-mode file system)~\cite{2019-Harrosh-zufs}, which is a
user-space file system for persistent memory, which completes \as{io} requests by requesting exact data locations instead of copying data into the own address space. A similar user-space file system that implements a shared cache for process is EvFS~\cite{2019-Yoshimura-EvFS}, which is \as{spdk}-based. While this file system can also support multiple readers/writers in the page cache, it only supports these for a single user process. User-space frameworks often provide capabilities to either expose the storage device as a block device, which the user-space application then accesses, or build a custom block device module (e.g., with \as{spdk}, which also has default driver modules such as \as{nvme}). For \as{nvme} devices that support \as{nvme} controller memory buffer management, the file system can manage parts of the device memory. DevFS~\cite{2018-Kannan-DevFS} utilize such an integration to manage the device memory for file metadata and \as{io} queues.

Different from prior discussed development frameworks, \af{fsp}~\cite{2019-Liu-fs_as_process} provides a storage architecture design for user-space file systems. The emphasis of \as{fsp} is to scale with the arrival of faster storage, and similarly to other user-space frameworks, minimize the software stack. For this it bases development on running file systems as processes, providing safer metadata integrity, data sharing coordination, and consistency control, as the process running the file system is in control of everything, instead of trusting libraries. Furthermore, \as{fsp} relies on \as{ipc} for fast communication, which unlike FUSE has a low overhead since it does not require context switching. Inter-core message passing comes at a low overhead and cache-to-cache transfers on multi-core systems can complete in double-digit cycles~\cite{soares2010flexsc}. DashFS~\cite{2019-Liu-fs_as_process} is built with \as{fsp}, providing a safe user-space file system with isolation of different user processes, and efficient \as{ipc}.

\subsection{Summary.}
With the complexity of the storage software stack and the high performance of flash storage, the storage software stack dominates the \as{io} latency~\cite{2010-Caulfield-Moneta,2012-Caulfield-Fast_User_Space_Access}, requiring careful \as{io} management to enhance the performance. In addition to \as{io} scheduling, allowing to merge consecutive \as{io} requests and reduce the overall issued \as{io} requests, user-space file systems, bypassing the Linux storage stack, avoid the storage stack layer overheads.

\section{Failure Consistent Operations}\label{sec:failure_consistency}
An important aspect of file systems is to ensure that in the case of power failure or system crashes, the file system state and its data remain in a consistent state, and can be recovered upon a subsequent mount. While we do not classify it as a challenge of flash storage, we discuss the methods for ensuring failure consistent operations for flash. Most commonly used mechanisms for ensuring this are journaling, CoW, and checkpointing. However, journaling suffers from having to write data twice, once for the metadata log and again for the data log, to ensure full consistency~\cite{pillai2017application,xu2016nova}, also referred to as the double-write problem~\cite{2018-Kannan-DevFS}. Therefore, file systems commonly only provide metadata consistency by logging only the metadata. Furthermore, power failure is a concern with flash storage, as it has capacitors that can flush parts of the on-device memory buffers, unless more expensive capacitors are used that can flush the entirety of the memory buffers. In the case of power failure, partially written data or metadata updates can result in the file system being in an inconsistent state.

While checkpointing, \as{cow}, and other consistency mechanisms aim to handle these failures and provide recovery after failure, other methods exist for providing interfaces that eliminate partial operations. Particularly, the design of flash storage, such as the available \as{oob} space of pages, provides beneficial options for such implementations. \cref{tab:consistency} depicts the methods for ensuring failure consistency with flash storage, and the file systems implementing such methods.

\begin{table}[!t]
    \centering
    \crefformat{section}{\S#2#1#3}
    \crefformat{subsection}{\S#2#1#3}
    \crefformat{subsubsection}{\S#2#1#3}
    \begin{tabular}{||p{40mm}|p{35mm}||}
        \hline 
        Mechanism & File Systems \\
        \hline
        \hline
        Atomic Writes (\cref{sec:atomic_writes}) & \cite{2021-Qin-Atomic_Writes,cheon2017exploiting,f2fs-atomic-write} \\
        \hline
        Transactions (\cref{sec:transactions}) & \cite{2022-Oh-exF2FS} \\
        \hline
    \end{tabular}
    \crefformat{section}{Section #2#1#3}
    \crefformat{subsection}{Section #2#1#3}
    \crefformat{subsubsection}{Section #2#1#3}
    \caption{Failure consistency mechanisms to for flash storage, and the respective file systems that implement a particular mechanism.}
    \label{tab:consistency}
\end{table}


\subsection{Atomic Writes}\label{sec:atomic_writes}
A possible method for ensuring that operations are not completed partially is through atomic writes. This is important to file systems as an update of data requires its respective metadata to also be updated. Failure of one should result in the other not being completed. A variety of flash devices support atomic writes through mechanisms such as in \as{ftl}~\cite{ouyang2011beyond} and through user-programmable \as{ssd}~\cite{Seshadri-2014-Willow}, however it can also be implemented in the file system itself. F2FS supports multi-block atomic writing, which allows updating multiple file system blocks in a single \textit{ioctl} command~\cite{2021-Qin-Atomic_Writes,cheon2017exploiting,f2fs-atomic-write}. ReconFS~\cite{2014-Lu-ReconFS} provides multi-page atomicity by using a flag in each page to indicate if it is valid. This is achieved by writing a 1 in the head of the last flash page of the metadata updated, and all other pages have a 0 in the head. Therefore, in the case of power failure, when the file system is reconstructed, any metadata pages in between two pages with a 1 in the head (depicting two ending page updates) are valid pages, and in the case when the log has no page with the flag set to 1, the update failed and all pages after the last 1 flag are invalid. 

Similarly, Qin et al.~\cite{2021-Qin-Atomic_Writes} showcase the utilization of the \as{oob} area on flash with \as{ocssd} in order to store metadata for ensuring data integrity in the No-Barrier File System (NBFS), which is extending F2FS. The file system uses \af{dnchain} to ensure consistency measurements. \as{dnchain} is a linked list of all the blocks in an atomic operation, which are stored through a pointer in the \as{oob} space, thus representing a linked list of pointers in an atomic operation. The last block in the \as{dnchain} points to itself to indicate the end of the linked list and the atomic operation. Furthermore, each block contains a checkpoint version, allowing the recovery after a failure to traverse the \as{dnchain} and identify if an invalid checkpoint version number is present, which indicates a failed atomic operation. As the initial writing of journal or \as{cow} is done in memory and later flushed to the flash storage, mechanisms such as failure-atomic \textit{msync()}~\cite{park2013failure}, provide atomicity in such operations.

\subsection{Transactions}\label{sec:transactions}
An additional approach for ensuring consistency is to use transactions, enforcing an ``all-or-nothing`` mechanism that either writes all data or no data at all if there are any failures during the transaction. Transactional support can stem from transactional block devices, such as TxFlash (Transactional Flash)~\cite{prabhakaran2008transactional} LightTX with embedded transaction support in the flash storage FTL~\cite{2013-Lu-LightTX,lu2015high}, and similar block device implementations that expose transaction support~\cite{huang2012bvssd,kang2013x,2013-Coburn-SSD_Txs}. Additions to file systems can similarly implement transactions for block devices without transaction support. exF2FS~\cite{2022-Oh-exF2FS} is an extension on F2FS that provides it with support for transactions. In order to support transactions, exF2FS implements several features. Firstly, with transactions relying on either all updates being present or no updates at all, the \as{gc} policy is adapted such that the \as{gc} module cannot reclaim flash pages that contain data from a transaction that is in progress. If such a page is garbage collected without the transaction having finished, it can be recovered and thus violates the all-or-nothing mechanism. 

For this, exF2FS implements \textit{shadow garbage collection}, which prohibits the \as{gc} module from using pages involved in pending transactions. Secondly, in order to provide large scale transactions with multiple exF2FS maintains \textit{transaction file groups}, which are kernel objects that identify the set of files involved in a transaction. Lastly, it implements \textit{stealing}, which allows dirty pages of pending transactions to be committed to the flash storage, however do not allow it to be garbage collected until the transaction completed. This mechanism is referred to as \textit{delayed invalidation and relocation record}, and the process of allowing dirty pages to be evicted in uncommited transactions is referred to as \textit{stealing}~\cite{2022-Oh-exF2FS}.

\subsection{Flash Failures}
Flash failures are another important aspect of file systems on how they manage and recover from these failures. A study into file system reliability showed that 16\% of injected faults resulted in an unmountable file system, which furthermore was not fixable by a file system checker~\cite{2020-Jaffer-SSD_Errors}. Especially, as the density of flash is increasing, where more layers of NAND is stacked on top of each other, resulting in a decrease in the flash reliability~\cite{meza2015large}. While flash employs \as{ecc} to handle correcting of data, there are errors that are not fixable such as the \textit{uncorrectable bit corruption}~\cite{boboila2010write,belgal2002new,brand1993novel}, which can be caused by flash wear, read disturbance, and other factors~\cite{2020-Jaffer-SSD_Errors}. We do not go into detail of flash induced errors, for a detailed evaluation on flash errors recovery mechanisms consult \cite{2020-Jaffer-SSD_Errors,gatla2018towards}. Jaffer et al.~\cite{2020-Jaffer-SSD_Errors} provide several guidelines for file systems to enhance reliability with flash errors, including adding more sanity checks, especially of metadata, and finding better tradeoffs between checksums and the checksum granularity, where a large checksum granularity can result in significant data loss if it has an unrecoverable invalid checksum, and a small checksum adds overheads on the needed space and performance. While file system checkers aim to solve a degree of faults, they are no panacea to fixing corrupted file systems~\cite{gatla2018towards} and require better failure handling in the file system itself.

\subsection{Summary}
The importance of failure consistent operations for storage systems has resulted mechanisms, such as atomic writes, to utilize the available \as{oob} area of flash pages for metadata to ensure consistency in the case of failures. Similarly, in order to avoid flash errors, checksum methods have been introduced to alleviate errors as a result of flash failures.

\section{Discussion}
With the multitude of methods for dealing with flash integration, there are several that are of key concern. In particular, \as{gc} and mechanisms of dealing with \as{gc} have shown to be applicable, and enhance other flash integration challenges. However, there is no panacea for solving all flash integration challenges. Incorporating a particular mechanism causes difficulty of integration with other flash challenges. Especially, depending on the level of integration of flash storage, there is limited possibility of integrating particular mechanisms. Therefore, the closer integration is largely necessary in order to more optimally integrate flash, however coming with the additional flash management complexity. Therefore, a tradeoff between the level of integratin and the required complexity must be made. A higher level of management allows the application to more optimally be designed for the flash. This however comes at the expense of increased complexity. Furthermore, generic file systems working on numerous integration levels, require more generic interface management, limiting their customizability for different integrations. 


Clear trends in the storage community are becoming apparent, focusing on eliminating the hiding of flash management idiosyncrasies, and exposing its characteristics to the host. Therefore, allowing the host to integrate and optimize the software for its access patterns with increased knowledge of the underlying storage device characteristics~\cite{bjorling2017lightnvm,2012-asplos-moneta,kang2014multi,bhimani2017enhancing,2010-Josephson-DFS,2021-Bjorling-ZNS}. As stated in the CompSys Manifesto~\cite{2022-compsys-manifesto}, ''the grand challenge in storage systems is to combine heterogeneous storage layers, leveraging their programmability and capabilities to deliver a new class of cost, data, and performance efficiency for all kinds of applications''. Reducing the semantic gap between storage hardware and software must be a driving concern in future storage system design, leveraging the existing hardware capabilities and enhancing integration for more efficient, effective, sustainable, and reliable storage systems.

The recent addition of \af{zns} \as{ssd}~\cite{2021-Bjorling-ZNS,2022-Tehrany-Understanding_ZNS}, standardized in the \as{nvme} 2.0 specification~\cite{2022-nvme-spec}, similarly provides host the control over \as{gc} management on the storage device, and matches the interface to the underlying flash characteristics. 
Leveraging such device interfaces allows the file system development to partially take control of on-device operations. In addition to eliminating duplicate work of file system and \as{gc} on the device, where the increased coordination provides better performance capabilities, the higher-level control of the file system allows it to apply its data knowledge, such as particular grouping based on file characteristics, without a significant increase in complexity through the zone interface of \as{zns} \as{ssd}. Such integration presents a well-defined interface aimed at eliminating the interface mismatch between flash storage and storage software, leaving a plethora of possibility for storage software to enhance flash integration. Exposing more storage device control to the host system, particularly in the configuration of \as{zns}, allows not only data placement to be better integrated into storage software, but furthermore allows to fully utilize the parallel units of the storage device. With \as{zns} devices, the parallelism unit of the storage device is clearly defined~\cite{2022-Bae-ZNS_Parallelism}, and can be leveraged by the storage software.


We imagine interfaces exposing an increasing level of flash hardware characteristics to the host software, to begin appearing, in order to better coordinate and integrate the storage. Similarly, we envision future efforts to aim at further decreasing the semantic gap between storage hardware and software, and leverage a higher degree of coordination across storage software/hardware layers. The gain in popularity of user-space based applications, as we discussed in \cref{sec:user_space_fs}, presents prominent possibility for future development and limiting kernel involvement, which has become an increasing overhead in the software stack. Therefore, we furthermore picture an increase in user-space applications, and particularly file systems.

\section{Related Work}\label{sec:related_work}
With the growing adoption of flash based storage systems~\cite{daim2008forecasting,2008-Lee-Flash_in_enterprise_db}, there has been a plethora of proposed systems to optimize for flash characteristics. We focus on the existing file systems for flash storage, but other aspects such as key-value stores are another popular use case for flash storage.

\noindent \textbf{Flash Optimized File Systems.} Egger~\cite{egger2010file} provide a survey on the file systems for flash storage at the time (2010), comparing the file systems in key features important for flash storage. This includes the feature set of the file system, time complexity of operations such as mount time, and space requirements for memory. However, evaluated file systems are not limited to flash specific file systems. While the survey presents an insightful summary of flash file systems, it was published in 2010, thus limiting the number of available file systems significantly. Similarly, Gal and Toledo~\cite{gal2005algorithms} present a survey of algorithms and data structures for flash storage, which encompasses flash mappings and flash-specific file systems.

Jaffer~\cite{jafferevolution} provide a comparison of five file systems for flash storage, analyzing the feature set of each and discussing limitations. In particular, the author focuses on the evolution of design trends for flash storage over the past decades, where the earliest file system included in the comparison was presented in 1994. The author additionally includes a discussion on file system optimizations for data management with Streams~\cite{kang2014multi} and \as{ocssd}~\cite{picoli2020open,bjorling2017lightnvm}. While the author presents an insightful analysis into design trends for flash storage, the literature review is limited to only the five discussed file systems and does not differentiate in the file system application domain.

Dubeyko~\cite{dubeyko2019ssdfs} presents SSDFS, a file system designed for SSDs. Albeit not being a literature review of flash file systems, the author presents an extensive comparison of related work that proposes flash file systems. Including a discussion on flash-friendly and flash-oriented file systems, and summarizing the available methods for optimizing flash specific operations and storage management. While the discussion of flash file systems provides a comparison of flash file systems, it is similarly not differentiating between application domain of the file system. Munegowda et al.~\cite{munegowda2014evaluation} showcase a study into several file systems on Windows and Linux for SSD and flash devices. Furthermore, the authors present a comparison of features for the varying file systems, including flash support and FTL integration, as well as a high level performance comparison. However, the study is focused on the main adopted file systems, lacking the inclusion of less adopted flash file systems.

Ramasamy and Karantharaj~\cite{2014-Ramasamy-Flash_FS_Challenges} survey the challenges of building file systems for flash storage, including the performance implications, caching techniques, and implications of the FTL. The authors additionally discuss available solutions to the presented design challenges for storage systems on flash memory. Similarly, Di Carlo et al.~\cite{2011-DiCarlo-Design_Issues_FS_Flash} present a study into the design issues and challenges of flash memory file systems. The authors analyze the inherent implications of the flash storage from the type of cell type used and required wear leveling, error correction, and bad block management. A comparison of several available flash file systems at the time of publication (2011) is presented, with focus on the discussed design challenges.

\noindent \textbf{Flash Optimized Applications.} Not focusing of file systems for flash storage, there have been several surveys into flash characteristics. Luo and Carey~\cite{2018-Luo-LSM_Survey} present a survey into log-structured merge-tree (LSM-tree) design techniques for storage systems. With LSM-trees being a widely adopted and popular choice of database and key-value store design to optimize for flash storage, the presented survey showcases relevant flash storage optimizations for data management. Similarly, Doekemeijer and Trivedi~\cite{2022-Doekemeijer-KV_Flash_Survey} present a study into key-value stores optimized for flash storage, showcasing several techniques that can likewise be applied to file system design for flash storage.

\noindent \textbf{Flash Translation Layer.} Chung et al.~\cite{chung2009FTL_survey} provide a survey into the various FTL algorithms, discussing the design issues of various algorithms. Flash file system performance will largely depend on the FTL implementation for SSDs, making optimal FTL design an important aspect of optimizing file system performance. A similar survey on FTL algorithms is presented by Kwon et al.~\cite{kwon2011ftl}. As embedded devices and possible custom integrations require management of flash at the file system level, the concepts for efficient and performant FTL algorithms are applicable to file system level management of flash storage.

\section{Conclusion}
The move from \as{hdd} to flash \as{ssd}, has been one of the most fundamental shifts in the storage hierarchy. The increased performance of flash \as{ssd} over \as{hdd}, capable of achieving single several GB/s bandwidth with millions of \as{iops}, however required adaptations in the software stack, changing the design of file systems and storage software to integrate with these devices. In this literature study we evaluate the current state of file systems for flash storage devices, how these file systems design to align with introduced flash characteristics, and how the integration of flash storage into file system design has affected storage stack design. We evaluate the findings of this literature study to each of the proposed \textit{survey research questions (SRQs)}.

\textbf{SRQ1: What are the main challenges arising from NAND flash characteristics and its integration into file system
design?}\\
The architecture of flash-based storage introduces six key challenges for file system and storage software developers to consider during software design and development. The first key challenge of flash storage is the (1) \textbf{asymmetric read and write performance}, for which file systems developers commonly resort to methods and data structures for enhancing write performance. 
Due to the architecture of flash storage, the lack of in-place updates introduces required (2) \textbf{garbage collection}, presenting a significant performance implication for flash storage. Furthermore, the implications of \as{gc} extend to introducing (3) \textbf{\as{io} amplification}, which additionally increases the wear of flash cells, requiring effective (4) \textbf{wear leveling} methods to employed. In order to leverage the performance capabilities of flash storage, the (5) \textbf{flash parallelism} must be exploited with particular methods that are capable of increasing the concurrency on the flash storage. The last key challenge of flash storage arises from the storage software stack into which the devices integrate, where the performance of the storage stack becomes the bottleneck. Therefore, (6) \textbf{\as{io} management} for flash storage is a vital aspect at limiting software overheads, and maximizing the utilization of the flash storage.


\textbf{SRQ2: How has NAND flash storage influenced the design and development of file system and the storage software
stack?}\\
The main challenges of integrating with NAND flash storage resulted in file system design to utilize specific data structures, algorithms, and mechanisms. Log-based data structures are widely adopted for flash-based file system, due to the matching characteristics to the flash storage. In addition to log-based data structures, file system development has focused on several key methods to exploit the parallelism capabilities of flash storage, including clustered allocation, data striping, and increasing the \as{io} sizes. Similarly, the design of \as{io} management has propagated out of the file system design, into the \as{io} scheduler architecture, to optimize \as{io} activity for fast flash-based storage.

\textbf{SRQ3: How will NAND flash storage and newly introduced NAND flash-based storage devices and interfaces affect
future file system design and development?}\\
Given the increasing rise in adoption for flash storage, and the introduction of new flash-based storage devices and interfaces, future implications of flash storage \textbf{(RQ3)} provide a promising ground of better integrating the flash storage with software, leveraging the increased performance capabilities further. Through new interfaces exposing a larger amount of flash characteristics, the host software gets an increasing level of possibility to design application specific flash management, integrating the application design with flash management. Future storage software developments are likely to continue integrating the closer integration of flash storage into host storage software design, optimizing the flash utilizing for full leveraging of flash performance. We envision the semantic gap between storage hardware and software to slowly decrease over time, allowing to build more performant, efficient, reliable, and responsible storage systems. Such efforts align with the grand challenges of future storage~\cite{2022-compsys-manifesto}, particularly with increasing demand of systems due to the digitalization of the world, and the push towards a more sustainable future. 


\printglossaries

\bibliographystyle{plain}
\bibliography{main}

\end{document}